\documentstyle[floats,prd,aps,epsf]{revtex}

\draft
\begin{document}

\newcommand{\beq}{\begin{equation}}
\newcommand{\eeq}{\end{equation}}
\newcommand{\beqn}{\begin{eqnarray}}
\newcommand{\eeqn}{\end{eqnarray}}
\newcommand{\pa}{\partial}
\newcommand{\vp}{\varphi}


\begin{center}
{\large\bf{Black hole tidal problem in the Fermi normal coordinates}} 
~~\\
~~\\
Masaki Ishii$^1$, Masaru Shibata$^1$, and Yasushi Mino$^2$
~~\\
~~\\
{\em $^1$ Graduate School of Arts and Sciences, 
University of Tokyo, Komaba, Meguro, Tokyo 153-8902, Japan \\
$^2$ Center for Gravitational Wave Astronomy, 
The University of Texas at Brownsville,
80 Fort Brown, Brownsville TX, 78520-4993 
\footnote{The present address: 
Mail Code 130-33, Caltech, Pasadena, CA 91125
}
}
\end{center}

\begin{abstract}
~\\
We derive a tidal potential for a self-gravitating fluid star
orbiting Kerr black hole along a timelike geodesic extending previous
works by Fishbone and Marck. In this paper, the tidal potential is
calculated up to the third and fourth-order terms in $R/r$, where $R$
is the stellar radius and $r$ the orbital separation, in the
Fermi-normal coordinate system following the framework developed by
Manasse and Misner.  The new formulation is applied for determining
the tidal disruption limit (Roche limit) of corotating Newtonian stars
in circular orbits moving on the equatorial plane of Kerr black holes. 
It is demonstrated that the third and fourth-order terms
quantitatively play an important role in the Roche limit for close
orbits with $R/r \agt 0.1$. It is also indicated that the Roche limit
of neutron stars orbiting a stellar-mass black hole near the
innermost stable circular orbit may depend
sensitively on the equation of state of the neutron star. 
\end{abstract}

\section{Introduction}

Black hole is the most compact object in the universe and 
can tidally disrupt ordinary stars and compact stars such as 
white dwarfs and neutron stars. A star of mass $m$ and 
radius $R$ will be tidally disrupted by a black hole of mass $M$ for the case 
\beqn
\mu \equiv {m \over M} \biggl({r \over R}\biggr)^3
< \mu_{\rm crit}, \label{eq1.1}
\eeqn
where $r$ denotes the orbital separation and $\mu_{\rm crit}$ is a
constant $\approx 10$--$20$ which depends on $r/M$, $R/M$, spin of the
black hole, and equations of state (see Sec. V). If the condition
(\ref{eq1.1}) is satisfied outside a minimum orbital radius (e.g.,
$r=6GM/c^2$ for a star in a circular orbit around a Schwarzschild black
hole where $c$ and $G$ are the speed of light and gravitational constant),
a star will be tidally disrupted. According to
Eq. (\ref{eq1.1}), (i) an ordinary star of mass $\sim M_{\odot}$
plunging inside a tidal radius of a supermassive black hole of mass
smaller than $\sim 10^8 M_{\odot}$ will be tidally disrupted, (ii) a white
dwarf of mass $\sim 0.7M_{\odot}$ and radius $\sim 10^4$ km plunging
into a massive black hole of mass smaller than $\sim 10^5 M_{\odot}$
will be tidally disrupted, and (iii) an inspiraling neutron star of mass
$\sim 1.4M_{\odot}$ and radius $\sim 10$ km in a circular orbit around
a stellar-mass black hole of mass smaller than $\sim 5 M_{\odot}$ will
be tidally disrupted before the neutron star reaches the innermost
stable circular orbit (ISCO). During the tidal interaction of the
ordinary star by a supermassive black hole, an ultraviolet flare of 
a characteristic light-curve may be emitted at the center of galaxy
and be observed (e.g., \cite{ayal,obs} for a review). White dwarfs or
neutron stars tidally disrupted by a stellar-mass black hole will form
a massive disk which may be a possible candidate of the central engine
of gamma-ray bursts \cite{fryer}. Detection of gravitational waves at
a tidal disruption of neutron stars by a stellar-mass black hole may
constrain the equation of state of neutron stars
\cite{cutler,micheal}. These examples show the tidal disruption of
ordinary stars and compact objects by a black hole is likely to
happen frequently and be an interesting phenomenon in the
universe. This fact stimulates the theoretical study for the tidal
disruption of stars by a black hole. In this paper, we present a new
general relativistic formulation with higher-order corrections of the
tidal potential which can be used to clarify the criterion of the
tidal disruption for close orbits more accurately than that by
previous works.

Numerical studies for the tidal disruption of a star by a black hole
have been extensively performed assuming the Newtonian (e.g.,
\cite{EK,UE,Max,CL,Novikov0,MLB}) or post-Newtonian \cite{ayal}
gravity with a point particle approximation for the black
hole. However, such an approximation is not quantitatively appropriate
for analyzing the tidal disruption near the black hole, since general
relativistic effects are essential for such close orbits. In
\cite{Laguna}, a numerical result for a general relativistic
simulation was presented, but the authors ignored the self-gravity of
the fluid star assuming that it is much weaker than the tidal force of
the black hole. Relativistic tidal problems have been widely studied
in the so-called tidal approximation (e.g.,
\cite{Fishbone,mash,Marck,novikov2,shibata}). In this approximation,
one assumes that the mass of a star $m$ is much smaller than the black
hole mass $M$ and that the stellar radius $R$ is smaller than the
orbital radius $r$ (or the curvature radius of the black hole
spacetime). As a result of these assumptions, one may assume that (i)
the center of mass of the star moves around the black hole along a
timelike geodesic in the black hole spacetime and (ii) the tidal field
from the black hole is computed from the Riemann tensor of the black
hole spacetime in terms of the geodesic deviation equation. In
addition to (i) and (ii), one often assumes that the self-gravity of
the star is described by the Newtonian gravity. This approach has been
used for studies of the tidal disruption limit of ordinary stars and
compact objects \cite{CL,MLB,Fishbone,shibata}, and for hydrodynamic
simulations of the tidal disruption of ordinary stars and white
dwarfs \cite{novikov2}. The purpose of this paper is to improve
a calculation of the 
tidal potential in this framework, since the improvement is necessary
for some problems as explained in the following.

In the tidal approximation, one assumes $R \ll r$ and then
expands the tidal potential in terms of $R/r$. This approximation
is illustrated for the Newtonian potential $\phi$ as follows: 
Considering $\phi$ of a point particle of mass $M_p$ at 
$(r, 0, 0)$ and expanding it around the origin, one obtains 
\beqn
\phi &=& -{G M_p \over \sqrt{(x-r)^2 + y^2 + z^2}} \nonumber \\
& = & -{G M_p \over r}\biggl[
1 + {x \over r} + {2x^2 - y^2 -z^2 \over 2r^2}
+ {x(2x^2 - 3y^2 - 3z^2) \over 2r^3} \nonumber \\
&& ~~~~~ +{8x^4 - 24x^2 (y^2 + z^2) +3y^4 +3z^4 +6y^2z^2
\over 8r^4} +O(r^{-5})\biggr],
\eeqn
where we assume that $|x|, |y|, |z| \ll r$ (i.e., $R \ll r$).
In the standard tidal approximation, one takes into account
terms up to the second order in $R/r$ 
neglecting the terms of $O[(R/r)^3]$, resulting in 
\beqn
\phi_{\rm tidal~approximation} =  -{G M_p \over r}\biggl[
1 + {x \over r} + {2x^2 - y^2 -z^2 \over 2r^2}\biggr]. 
\eeqn

The standard tidal approximation works for determining the
tidal disruption limit in many problems, but does not in some problems: 
Equation (\ref{eq1.1}) implies that a tidal disruption occurs 
if the following condition is satisfied; 
\beqn
{R \over r} \agt \mu_{\rm crit}^{-1/3} Q^{1/3}.  
\eeqn
Here, $Q \equiv m/M$. 
Thus, with increasing the mass ratio $Q$, the critical
value of $R/r$ for the tidal disruption increases: 
For $Q \agt 10^{-3}$, the tidal disruption sets in for 
$R/r \agt 0.05$, indicating that neglecting more than third-order 
terms in $R/r$ yields an error of $\agt 5\%$ and 
may not be a very good approximation 
for the computation of the tidal disruption limit. For example, 
the tidal disruption of white dwarfs by 
an intermediate mass black hole of $\sim 10^3 M_{\odot}$ is the case.
Even for $Q < 10^{-3}$, the star will be elongated 
during the close-encounter with a black hole in parabolic or 
highly elliptical orbits. In such cases,
$R$ will increase with decreasing $r$, and hence, 
the higher-order terms may be important. 
For binaries of a black hole of 
mass smaller than $\sim 5M_{\odot}$ and a 
neutron star in close quasicircular orbits,
the higher-order terms are also important 
since $R/r$ $\agt 0.2$ at the ISCO of radius $\sim 6GM/c^2$.
Our numerical results in the framework of Newtonian gravity indeed 
suggest that the third- and fourth-order terms play an important 
role for a system of $R/r \agt 0.1$ \cite{IS}. 
For a rapidly rotating black hole, the radius of the ISCO can be
as small as $\sim GM/c^2$. In this case, the higher-order terms in $R/r$ 
may become quite important. These facts illustrate that 
it is often necessary to take into account the higher-order corrections
of the tidal potential for computation of quantitatively improved
results in the tidal problem. 

In this paper, we derive a general relativistic
tidal potential induced by a black hole 
in which higher-order corrections are taken into account. 
In contrast to the previous works \cite{Fishbone,Marck}, 
we do not use the geodesic deviation equation to calculate 
the tidal potential. Instead, we derive the tidal metric for an
observer moving along a timelike geodesic on a black hole spacetime in
the Fermi normal coordinates following Manasse and Misner \cite{MM}. 
In this method, the tidal potential can be 
calculated from the tidal metric in a straightforward manner. 

Using the new formulation, we numerically compute equilibrium states
and a tidal disruption limit (Roche limit)
\footnote{The Roche limit is defined as the tidal disruption limit
for a star of corotating velocity fields.} for corotating stars of
polytropic equations of state in a close circular orbit around a Kerr
black hole. Comparing the numerical results with those computed by 
the standard tidal approximation, we illustrate that the
higher-order terms of the tidal potential play a quantitatively
important role for $R/r=O(0.1)$. The numerical results in this
framework will be also used to calibrate accuracy of a numerical
result of fully general relativistic quasiequilibrium states for a
binary of a black hole and a neutron star which will be computed in
near future. (It should be noted that there has been no comprehensive
work about the tidal disruption of stars by a black hole in full
general relativity, although there are primitive works for binaries of
a black hole and a neutron star \cite{miller,bss}.)  The solution
obtained in this work is valid for $r \gg R$ and $M \gg m$, and hence,
the calibration can be carried out for the case of low-mass neutron
stars. In addition, dependence of the tidal disruption limit on the
equations of state for the star is investigated. It is indicated that
the tidal disruption limit of a neutron star by a stellar-mass black
hole depends sensitively on the equation of state of the neutron star.

The paper is organized as follows. In Sec. II, we derive a general
expression of the metric for an observer moving along timelike
geodesics in the Fermi normal coordinates. In Sec. III, the results of
Sec. II are applied to the Kerr spacetime. In Sec. IV, the tidal
potentials are computed for equatorial circular orbits. In Sec. V, the
tidal disruption limit (Roche limit) of corotating stars of equatorial
circular orbits around a Kerr black hole is presented for a wide range
of the spin parameter of the black hole and the equations of state for
the star. Sec. VI is devoted to a summary.

In the following, we adopt the geometrical units $c=G=1$. 
Latin and Greek indices denote spatial and spacetime components
with $\tau=x^0$ as the time coordinate in the Fermi normal frame. 
$g_{\mu\nu}$, $\Gamma^{\mu}_{\nu\sigma}$, and 
$R_{\mu\nu\sigma\delta}$ denote the spacetime metric,
Christoffel symbols, and Riemann tensor, respectively. 
$\delta_{ij}(=\delta^{ij})$ denotes the Kronecker delta.
The comma and semi-colon denote the ordinary and covariant derivatives. 
For simplicity, we often use the notations \cite{Wald}
\beqn
&& A_{(ij)}={1 \over 2}(A_{ij}+A_{ji}),\\
&& A_{(ijk)}={1 \over 6}(A_{ijk}+A_{jik}+A_{jki}+A_{kji}+A_{kij}+A_{ikj}),\\
&& A_{(a_1 \cdots a_l)}
={1 \over l!}\sum_l A_{a_{f(1)} \cdots a_{f(l)}},
\eeqn
where in the last equation, the sum is taken for all permutations.

\section{External metric in the Fermi normal coordinates}

\subsection{Outlines and definition of the Fermi normal coordinates}

We derive the metric in an observer frame which moves along
timelike geodesics around a Kerr black hole.
As Manasse and Misner \cite{MM} show, 
this can be done by finding the relation between the 
Riemann tensor and the metric in the Fermi normal coordinate system 
which is one of the local inertial frames. 
Then, our goal is to write the metric in the 
neighborhood of an observer in this coordinate system as
\beq
g_{\mu\nu}=\eta_{\mu\nu}+{1 \over 2} g_{\mu\nu,ij}x^i x^j
+{1 \over 6} g_{\mu\nu,ijk}x^i x^j x^k
+{1 \over 24} g_{\mu\nu,ijkl}x^i x^j x^k x^l+O(x^5), \label{goal}
\eeq
where $\eta_{\mu\nu}$ is the flat metric and
$x^i$ denotes a spatial coordinate in the Fermi normal
coordinate system. The coefficients 
$g_{\mu\nu,ij\cdots}$ are related to the Riemann tensor 
of the spacetime. Using the definition of the Riemann tensor and
the geodesic deviation equations, 
Manasse and Misner \cite{MM} derived the quadratic term in Eq. (\ref{goal}). 
Our purpose is to derive the third and fourth terms following the 
method developed by them. 

Since the Fermi normal coordinate system is a local inertial
frame, $g_{\mu\nu}$ is the flat metric along the observer frame, and 
the Christoffel symbols vanish 
\beqn
&& \Gamma^{\mu}_{\nu\sigma}=0, \label{eq1} \\
&& \Gamma_{\mu\nu\sigma}=g_{\mu\alpha}\Gamma^{\alpha}_{\nu\sigma}=0. 
\label{eq2}
\eeqn
In addition, the time direction in the Fermi normal coordinates 
is chosen as the direction of a timelike geodesic ${\cal G}$.
Since Eqs. (\ref{eq1}) and (\ref{eq2}) are preserved
along the timelike geodesic ${\cal G}$, one obtains 
\beqn
&& \Gamma^{\mu}_{\nu\sigma,0}
=\Gamma^{\mu}_{\nu\sigma,00}
=\Gamma^{\mu}_{\nu\sigma,0 \cdots 0}=0,\label{eq3}\\
&& \Gamma_{\mu\nu\sigma,0}
=\Gamma_{\mu\nu\sigma,00}
=\Gamma_{\mu\nu\sigma,0 \cdots 0}=0. \label{eq4}
\eeqn
{}From the definition of the Christoffel symbols,
\beq
g_{\mu\nu,\alpha}=\Gamma_{\mu\nu\alpha}+\Gamma_{\nu\mu\alpha},
\eeq
one finds along ${\cal G}$, 
\beqn
g_{\mu\nu,\alpha 0}=g_{\mu\nu,\alpha 00}=
g_{\mu\nu,\alpha 0 \cdots 0}=0. \label{eq000}
\eeqn

\subsection{Relation obtained from the definition of
the Riemann tensor}

In this subsection, we present relations necessary for
deriving the tidal potential up to the fourth order 
using the definition of the Riemann tensor 
\beq
R_{\mu\nu\rho\sigma}=
\Gamma_{\sigma\mu\rho,\nu}-\Gamma_{\sigma\nu\rho,\mu}
+\Gamma^{\alpha}_{\nu\rho}\Gamma_{\alpha\mu\sigma}
-\Gamma^{\alpha}_{\mu\rho}\Gamma_{\alpha\nu\sigma}.\label{defrie}
\eeq
In the following, the components of the Riemann tensor along ${\cal G}$ 
are computed. For $\mu=0$ in Eq. (\ref{defrie}), 
we find that the nonzero components are 
\beqn
R_{0i\nu j}=\Gamma_{j0\nu,i}=-\Gamma_{\nu 0j,i},\label{gamrie0}
\eeqn
where Eqs. (\ref{eq1})--(\ref{eq4}) are used.  
For $\nu=0$ in Eq. (\ref{gamrie0}), one finds 
\beq
g_{00,ij}=-2\Gamma_{j00,i}=-2R_{0i0j}, \label{gamrie00}
\eeq
where Eq. (\ref{eq000}) is used.
This gives one of the second-order terms in Eq. (\ref{goal}). 

Since the Christoffel symbols are vanishing,
the covariant derivative of the Riemann tensor is equal to
the ordinary derivative along ${\cal G}$ as
\beq
R_{\mu\nu\rho\sigma;\alpha}=R_{\mu\nu\rho\sigma,\alpha}. 
\eeq
Thus, from Eq. (\ref{defrie}) and $\Gamma^{\mu}_{\nu\rho}=0$,
one obtains a relation along ${\cal G}$ 
\beq
R_{\mu\nu\rho\sigma;\alpha}
=\Gamma_{\sigma\mu\rho,\nu\alpha}-\Gamma_{\sigma\nu\rho,\mu\alpha}, 
\label{rieder}
\eeq
and thus, 
\beqn
R_{i0j0;k}
&=&\Gamma_{0ij,0k}-\Gamma_{00j,ik} \nonumber \\
&=&{1 \over 2}\biggl(g_{0i,jk0} + g_{0j,ik0} - g_{00,ijk}\biggr). 
\label{rieder2}
\eeqn
Equation (\ref{rieder2}) leads to
\beq
R_{0i0j;k}+R_{0j0k;i}+R_{0k0i;j}
=(g_{0i,jk} + g_{0j,ki} + g_{0k,ij})_{,0}
- {3 \over 2} g_{00,ijk}. \label{eq101}
\eeq
In Sec. \ref{sec24}, 
this equation is used for deriving the third-order terms
in Eq. (\ref{goal}). 

The second covariant derivative of the Riemann tensor along ${\cal G}$ is
written as
\beqn
R_{\mu\nu\rho\sigma;\alpha\beta}=R_{\mu\nu\rho\sigma,\alpha\beta}
-\Gamma^{\gamma}_{\mu\beta,\alpha}R_{\gamma\nu\rho\sigma}
-\Gamma^{\gamma}_{\nu\beta,\alpha}R_{\mu\gamma\rho\sigma}
-\Gamma^{\gamma}_{\rho\beta,\alpha}R_{\mu\nu\gamma\sigma}
-\Gamma^{\gamma}_{\sigma\beta,\alpha}R_{\mu\nu\rho\gamma}. 
\eeqn
Thus, 
\beqn
R_{\mu\nu\rho\sigma;(\alpha\beta)}
=R_{\mu\nu\rho\sigma,\alpha\beta}
-\Gamma^{\gamma}_{\mu(\beta,\alpha)}R_{\gamma\nu\rho\sigma}
-\Gamma^{\gamma}_{\nu(\beta,\alpha)}R_{\mu\gamma\rho\sigma}
-\Gamma^{\gamma}_{\rho(\beta,\alpha)}R_{\mu\nu\gamma\sigma}
-\Gamma^{\gamma}_{\sigma(\beta,\alpha)}R_{\mu\nu\rho\gamma},
\eeqn
and 
\beqn
R_{0i0j;(kl)}=R_{0i0j,kl}
-\Gamma^{\gamma}_{i(k,l)}R_{\gamma 0j0}
-\Gamma^{\gamma}_{0(k,l)}(R_{i\gamma j0}+R_{j\gamma i0})
-\Gamma^{\gamma}_{j(k,l)}R_{\gamma 0i0}. \label{rieddr}
\eeqn
Using Eq. (\ref{defrie}), one obtains 
\beqn
R_{0i0j,kl}&=&{1 \over 2}\biggl(g_{0i,0jkl} + g_{0j,0ikl} - g_{00,ijkl}
-g_{ij,kl00}\biggr) \nonumber \\
&&
+\Gamma^{\alpha}_{0i,k}\Gamma_{\alpha 0j,l}
+\Gamma^{\alpha}_{0i,l}\Gamma_{\alpha 0j,k}
-\Gamma^{\alpha}_{00,l}\Gamma_{\alpha ij,k}
-\Gamma^{\alpha}_{00,k}\Gamma_{\alpha ij,l}, 
\eeqn
and hence, 
\beqn
&&R_{0i0j,kl}+R_{0i0k,jl}+R_{0i0l,jk}+R_{0j0k,il}+R_{0j0l,ik}+R_{0k0l,ij}
\nonumber \\
&&={3 \over 2}(g_{0i,jkl}+g_{0j,ikl}+g_{0k,ijl}+g_{0l,ijk})_{,0}
-3g_{00,ijkl} \nonumber \\
&&~~~-{1 \over 2}(g_{ij,kl}+g_{ik,jl}+g_{il,jk}+g_{jk,il}+g_{jl,ik}
+g_{kl,ij})_{,00} \nonumber \\
&&~~~
+4(\Gamma^{\alpha}_{0(i,j)}\Gamma_{\alpha 0(k,l)}
+\Gamma^{\alpha}_{0(i,k)}\Gamma_{\alpha 0(j,l)}
+\Gamma^{\alpha}_{0(i,l)}\Gamma_{\alpha 0(j,k)}) \nonumber \\
&&~~~
-3(\Gamma^{\alpha}_{00,i}\Gamma_{\alpha (jk,l)}
+\Gamma^{\alpha}_{00,j}\Gamma_{\alpha (ik,l)}
+\Gamma^{\alpha}_{00,k}\Gamma_{\alpha (ij,l)}
+\Gamma^{\alpha}_{00,l}\Gamma_{\alpha (ij,k)}). \label{g00ijkl0}
\eeqn
In Sec. \ref{sec24}, 
this equation is used to derive the fourth-order terms in Eq. (\ref{goal}). 

\subsection{Relations obtained from the geodesic deviation equations}

In this subsection, relations necessary for calculating 
the tidal potential up to the fourth order are derived 
using the geodesic deviation equations. 

For the tangent of a geodesic, $u^{\mu}$, and
the displacement vector to an infinitesimally nearby geodesic 
$z^{\mu}$, the geodesic deviation equation is written as 
\beq
u^{\mu} \nabla_{\mu} (u^{\nu} \nabla_{\nu} z^{\sigma})
=-R_{\mu\nu\gamma}^{~~~~\sigma} u^{\mu}u^{\gamma}z^{\nu}. \label{eqdev}
\eeq
Note here that the geodesic that we consider is not restricted
to ${\cal G}$. Using an affine parameter $\lambda$ for the geodesic
(i.e., $u^{\mu}=(\pa/\pa \lambda)^{\mu}$), 
Eq. (\ref{eqdev}) is rewritten to
\beq
{d^2 z^{\mu} \over d\lambda^2}
+2 {d z^{\nu} \over d\lambda}\Gamma^{\mu}_{\nu\sigma} u^{\sigma}
+(\Gamma^{\mu}_{\nu\alpha,\sigma}
+ \Gamma^{\mu}_{\nu\beta} \Gamma^{\beta}_{\alpha\sigma}
- \Gamma^{\mu}_{\alpha\beta} \Gamma^{\beta}_{\nu\sigma})
u^{\nu} u^{\sigma} z^{\alpha}
=-R_{\nu\alpha\sigma}^{~~~~\mu}u^{\nu} u^{\sigma} z^{\alpha},
\label{eqgeo0}
\eeq
or 
\beq
{d^2 z^{\mu} \over d\lambda^2}
+2 {d z^{\nu} \over d\lambda}\Gamma^{\mu}_{\nu\sigma} u^{\sigma}
+\Gamma^{\mu}_{\nu\sigma,\alpha} u^{\nu} u^{\sigma} z^{\alpha}=0.
\label{eqgeo}
\eeq

Now, we consider the family of spacelike geodesics in the Fermi
normal coordinates of the form 
\beq
x^0={\rm const}.~~{\rm and}~~x^i=\alpha^i \lambda, 
\eeq
where $\alpha^i$ is a constant three-component.
Using its definition, $u^{\mu}$ is written to 
\beq
u^{\mu}=\biggl( {\pa \over \pa \lambda} \biggr)^{\mu}
=\delta^{\mu}_{~i} \alpha^i. \label{equ}
\eeq
As the displacement vector $z^{\sigma}$, 
we choose the spatial vector defined by
\beq
z^{\mu}=\biggl( {\pa \over \pa \alpha^i} \biggr)^{\mu}
=\lambda \delta^{\mu}_{~i}. \label{eqz}
\eeq
Substituting Eqs. (\ref{equ}) and (\ref{eqz}) into Eq. (\ref{eqgeo})  
leads to
\beq
2\Gamma^{\mu}_{ij} \alpha^j
+ \lambda \Gamma^{\mu}_{jk,i} \alpha^j \alpha^k=0. \label{eqgeo2}
\eeq

Our purpose here is to derive the relations among the derivatives of
the Christoffel symbols along ${\cal G}$ which will be useful in the subsequent
calculations. To obtain the relations, Eq. (\ref{eqgeo2}) should be
evaluated for $\lambda=0$. However, Eq. (\ref{eqgeo2}) is trivial
because of the vanishing Christoffel symbols in the Fermi normal 
coordinates. Thus, we carry out a Taylor expansion of Eq. (\ref{eqgeo2})
assuming that $\lambda$ is small, and 
the first-, second-, and third-order terms in $\lambda$ provide 
\beqn
&& 2\Gamma^{\mu}_{ij,k} \alpha^j \alpha^k +
\Gamma^{\mu}_{jk,i}\alpha^j \alpha^k =0,\label{chris1} \\
&& \Gamma^{\mu}_{ij,kl} \alpha^j \alpha^k \alpha^l 
+ \Gamma^{\mu}_{jk,il}\alpha^j \alpha^k \alpha^l =0,\label{chris2} \\ 
&& {1 \over 3} \Gamma^{\mu}_{ij,kln} \alpha^j \alpha^k \alpha^l \alpha^n
+{1 \over 2} \Gamma^{\mu}_{jk,iln}\alpha^j \alpha^k \alpha^l \alpha^n =0.
\label{chris3}
\eeqn
Equations (\ref{chris1})--(\ref{chris3}) lead to cyclic relations among 
the derivatives of the Christoffel symbols along ${\cal G}$ as
\beqn
&& \Gamma_{\mu(ij,k)} = 0, \label{cyclic1}\\
&& \Gamma_{\mu(ij,kl)} = 0, \label{cyclic2}\\
&& \Gamma_{\mu(ij,kln)} = 0. \label{cyclic3}
\eeqn
Setting $\mu=0$ in Eqs. (\ref{cyclic1})--(\ref{cyclic3}),
the following useful relations are derived; 
\beqn
&& g_{0(i,jk)}=0, \label{eqgg1}\\
&& 3(g_{0i,jkl}+g_{0j,ikl}+g_{0k,ijl}+g_{0l,ijk})
=(g_{ij,kl}+g_{ik,jl}+g_{il,jk}+g_{jk,il}+g_{jl,ik}+g_{kl,ij})_{,0}, 
\label{eqgg2}\\
&& 4(g_{0i,jkln}+g_{0j,ikln}+g_{0k,ijln}+g_{0l,ijkn}+g_{0n,ijkl})
\nonumber \\
&&~~~=(g_{ij,kln}+g_{ik,jln}+g_{il,jkn}+g_{in,jkl}
+g_{jk,iln}+g_{jl,ikn}+g_{jn,ikl}+g_{kl,ijn}+g_{kn,ijl}+g_{ln,ijk})_{,0}. 
\label{eqgg3}
\eeqn

Using Eq. (\ref{eqgeo0}) with the same strategy as that for
deriving Eqs. (\ref{chris1})--(\ref{chris3}), one can also obtain
relations among the Christoffel symbols and Riemann tensor as
\beqn
&& R_{ijk}^{~~~~\mu}\alpha^j \alpha^k
=3 \Gamma^{\mu}_{ij,k} \alpha^j \alpha^k,\label{rie11}\\
&& R_{ijk~~,l}^{~~~~\mu} \alpha^j \alpha^k \alpha^l 
=2 \Gamma^{\mu}_{ij,kl} \alpha^j \alpha^k \alpha^l,\\
&& {1 \over 2} R_{ijk~~,ln}^{~~~~\mu} \alpha^j \alpha^k \alpha^l \alpha^n 
={5 \over 6} \Gamma^{\mu}_{ij,kln} \alpha^j \alpha^k \alpha^l \alpha^n
+(\Gamma^{\sigma}_{ij,l} \Gamma^{\mu}_{\sigma k,n}
-\Gamma^{\mu}_{i\sigma,l} \Gamma^{\sigma}_{jk,n})
\alpha^j \alpha^k \alpha^l \alpha^n.  \label{rie13}
\eeqn
Here, $\Gamma^{\sigma}_{jk,n}\alpha^j\alpha^k\alpha^n=0$ 
because of Eq. (\ref{cyclic1}). Thus, the last term of
Eq. (\ref{rie13}) vanishes.

{}From Eqs. (\ref{rie11})--(\ref{rie13}) together with 
Eqs. (\ref{cyclic1})--(\ref{cyclic3}), one obtains the relations between
the Christoffel symbols and the Riemann tensor along ${\cal G}$ as 
\beqn
&&\Gamma^{\mu}_{ij,k}+\Gamma^{\mu}_{ik,j}
=-{1 \over 3}\biggl(R_{jik}^{~~~~\mu}+R_{kij}^{~~~~\mu}\biggr),
\label{gamrie1}\\
&&\Gamma^{\mu}_{ij,kl}+\Gamma^{\mu}_{ik,jl}+\Gamma^{\mu}_{il,jk}
=-{1 \over 4}\biggl(
 R_{jik~~,l}^{~~~~\mu}+R_{jil~~,k}^{~~~~\mu}
+R_{kij~~,l}^{~~~~\mu}+R_{lij~~,k}^{~~~~\mu}
+R_{lik~~,j}^{~~~~\mu}+R_{kil~~,j}^{~~~~\mu} \biggr),\label{gamrie2}\\ 
&&\Gamma^{\mu}_{ij,kln}+\Gamma^{\mu}_{ik,jln}
+\Gamma^{\mu}_{il,jkn}+\Gamma^{\mu}_{in,jkl} \nonumber \\
&&~~~={2 \over 5}\biggl(
 R_{i(jk)~,ln}^{~~~~~\mu}+R_{i(jl)~,kn}^{~~~~~\mu}
+R_{i(jn)~,kl}^{~~~~~\mu}+R_{i(kl)~,jn}^{~~~~~\mu}
+R_{i(kn)~,jl}^{~~~~~\mu}+R_{i(ln)~,jk}^{~~~~~\mu}\biggr) \nonumber \\
&& ~~~~
-{4 \over 5}\biggl(
 \Gamma^{\nu}_{i(j,k)}\Gamma^{\mu}_{\nu(l,n)}
+\Gamma^{\nu}_{i(j,l)}\Gamma^{\mu}_{\nu(k,n)}
+\Gamma^{\nu}_{i(j,n)}\Gamma^{\mu}_{\nu(k,l)}
+\Gamma^{\nu}_{i(k,l)}\Gamma^{\mu}_{\nu(j,n)}
+\Gamma^{\nu}_{i(k,n)}\Gamma^{\mu}_{\nu(j,l)}
+\Gamma^{\nu}_{i(l,n)}\Gamma^{\mu}_{\nu(j,k)}
\biggr)
\label{gamrie3} 
\eeqn
Using Eq. (\ref{cyclic1}), Eq. (\ref{gamrie1}) is rewritten to
\beqn
\Gamma_{\mu jk,i} = -{2 \over 3} R_{i(jk)\mu}. \label{gamijkl}
\eeqn
For $\mu=0$ in Eq. (\ref{gamijkl}), 
\beq
-{2 \over 3}R_{i(jk)0}=
\Gamma_{0jk,i}={1 \over 2}(g_{0j,ik}+g_{0k,ij})=-{1 \over 2}g_{0i,jk},
\eeq
where Eqs. (\ref{eqgg1}) and $g_{jk,i0}=0$ are used. Thus, 
\beqn
g_{0k,ij}={2 \over 3} \biggl( R_{0ijk} + R_{0jik} \biggr).\label{g0kij}
\eeqn
This gives one of the second-order terms in  Eq. (\ref{goal}).

Equation (\ref{gamijkl}) is also used to derive the relation for
$g_{kl,ij}$ along ${\cal G}$ 
\beqn
&& g_{kl,ij}=\Gamma_{kli,j}+\Gamma_{lki,j}=
{1 \over 3} \biggl( R_{iklj} + R_{ilkj} \biggr). \label{gijkl}
\eeqn
This also gives one of the second-order terms in Eq. (\ref{goal}).

{}From Eqs. (\ref{gijkl}), one finds the following relations
along ${\cal G}$; 
\beqn
&& g_{i(j,kl)}=0,\label{g41} \\
&& g_{(kl,j)i}=0. \label{g42}
\eeqn
Equations (\ref{eqgg1}), (\ref{g41}), and (\ref{g42})
are preserved along ${\cal G}$, i.e.,
\beqn
&& [g_{0(k,ij)}]_{,0}=0,\label{g400} \\
&& [g_{i(j,kl)}]_{,0}=0,\\
&& [g_{(kl,j)i}]_{,0}=0. 
\eeqn
Substituting the last two relations into Eq. (\ref{eqgg2}), one obtains
\beq
g_{0(i,jkl)}=0. \label{eqgg22}
\eeq
These relations are useful for deriving the third-order terms in Eq.
(\ref{goal}). 

\subsection{Deriving the third- and fourth-order terms}\label{sec24}

{}From Eqs. (\ref{eq101}), (\ref{eqgg1}), (\ref{gamrie2}),
(\ref{g400}), and (\ref{eqgg22}), 
third derivatives of $00$ and $0i$ components of the metric are found to be 
\beqn
&& g_{00,ijk}=-{2 \over 3}\biggl(R_{0i0j;k}+R_{0j0k;i}+R_{0k0i;j}\biggr)
=-2 R_{0(i|0|j;k)},\label{g00ijk} \\
&& g_{0i,jkl}={1 \over 4} \biggl(
R_{ijk0;l} + R_{ikj0;l} + R_{ilk0;j}
+R_{ikl0;j} + R_{ijl0;k} + R_{ilj0;k}\biggr)
={3 \over 2}R_{i(jk|0|;l)}. \label{g0ijkl}
\eeqn
For $ij$ components, 
\beqn
g_{ij,kln}
&=&\Gamma_{ijk,ln}+\Gamma_{jik,ln} \nonumber \\
&=&{1 \over 3}\biggl(
 \Gamma_{ijk,ln}+\Gamma_{jil,kn}+\Gamma_{ijn,kl}+\Gamma_{jik,ln}
+\Gamma_{jil,kn}+\Gamma_{jin,kl}
\biggr) \nonumber \\
&=&-{1 \over 6}\biggl(
R_{kjli;n} + R_{kjni;l} + R_{ljki;n}
+R_{njki;l} + R_{njli;k} + R_{ljni;k}\biggr),\label{gijkln}
\eeqn
where we use Eqs. (\ref{cyclic2}) and (\ref{gamrie2}).

Using Eqs. (\ref{rieddr}), (\ref{g00ijkl0}), and (\ref{eqgg2}), the 
fourth derivative of $00$ component of the metric is written as 
\beqn
g_{00,ijkl}
&=&
-{1 \over 3} \biggl[ R_{0i0j;(kl)} + R_{0i0k;(jl)} + R_{0i0l;(jk)}
+R_{0j0k;(il)} + R_{0j0l;(ik)} + R_{0k0l;(ij)}\biggr] \nonumber \\
&&~~ + {8 \over 3}\biggl[
\Gamma^{\mu}_{0(k,l)} \Gamma_{\mu 0(i,j)}
+\Gamma^{\mu}_{0(j,l)} \Gamma_{\mu 0(i,k)}
+\Gamma^{\mu}_{0(i,l)} \Gamma_{\mu 0(j,k)}
\biggr] \nonumber \\
&=&
-{1 \over 3} \biggl[ R_{0i0j;(kl)} + R_{0i0k;(jl)} + R_{0i0l;(jk)}
+R_{0j0k;(il)} + R_{0j0l;(ik)} + R_{0k0l;(ij)}\biggr] \nonumber \\
&&~~ + {8 \over 3}\biggl[
R^{\mu}_{~(kl)0} R_{\mu (ij)0}
+R^{\mu}_{~(jl)0} R_{\mu (ik)0}
+R^{\mu}_{~(jk)0} R_{\mu (il)0}\biggr] \nonumber \\
&=&-2R_{0(i|0|j;kl)}+8 R^{\mu}_{~(kl|0|} R_{|\mu|ij)0}. \label{g00ijkl}
\eeqn

{}From Eq. (\ref{gijkln}) and $R_{(ijk)l}=0$, it is found
\beq
g_{(ij,kln)}=g_{i(j,kln)}=0, \label{g51}
\eeq
and hence,
\beqn
[g_{(ij,kln)}]_{,0}=[g_{i(j,kln)}]_{,0}=0. \label{g510}
\eeqn
Using Eqs. (\ref{g510}) and (\ref{eqgg3}), one obtains
\beq
g_{0(i,jkln)}=0. \label{g52}
\eeq
By a straightforward calculation, one finds 
\beqn
\Gamma_{\mu ij,kln}+\Gamma_{\mu ik,jln}
+\Gamma_{\mu il,jkn}+\Gamma_{\mu in,jkl}
=2[g_{\mu i,jkln}+g_{\mu(j,kln)i}-g_{i(j,kln)\mu}]. \label{eq2.58}
\eeqn
For $\mu=0$ in Eq. (\ref{eq2.58}),
\beqn
\Gamma_{0ij,kln}+\Gamma_{0ik,jln}
+\Gamma_{0il,jkn}+\Gamma_{0in,jkl}={3 \over 2} g_{0i,jkln}, 
\eeqn
where Eqs. (\ref{g51}) and (\ref{g52}) are used. Thus, 
\beqn
g_{0i,jkln}&=&{2 \over 3}\biggl(
\Gamma_{0ij,kln}+\Gamma_{0ik,jln}+\Gamma_{0il,jkn}+\Gamma_{0in,jkl} \biggr)
\nonumber \\
&=&{4 \over 15}\biggl(
 R_{i(jk)0,ln}+R_{i(jl)0,kn}+R_{i(jn)0,kl}
+R_{i(kl)0,jn}+R_{i(kn)0,jl}+R_{i(ln)0,jk} \biggr) \nonumber \\
&& ~~-{8 \over 15}\biggl(
 \Gamma^{\nu}_{i(j,k)}\Gamma_{0\nu(l,n)}
+\Gamma^{\nu}_{i(j,l)}\Gamma_{0\nu(k,n)}
+\Gamma^{\nu}_{i(j,n)}\Gamma_{0\nu(k,l)} \nonumber \\
&&~~~~~~~~+\Gamma^{\nu}_{i(k,l)}\Gamma_{0\nu(j,n)}
+\Gamma^{\nu}_{i(k,n)}\Gamma_{0\nu(j,l)}
+\Gamma^{\nu}_{i(l,n)}\Gamma_{0\nu(j,k)}
\biggr) \nonumber \\
&=&{4 \over 15}\biggl(
 R_{i(jk)0,ln}+R_{i(jl)0,kn}+R_{i(jn)0,kl}
+R_{i(kl)0,jn}+R_{i(kn)0,jl}+R_{i(ln)0,jk} \biggr) \nonumber \\
&&-{8 \over 45}\biggl(
 R_{i(jk)}^{~~~~~0} R_{0(ln)0}
+R_{i(jl)}^{~~~~~0} R_{0(kn)0}
+R_{i(jn)}^{~~~~~0} R_{0(kl)0}
+R_{i(kl)}^{~~~~~0} R_{0(jn)0}
+R_{i(kn)}^{~~~~~0} R_{0(jl)0}
+R_{i(ln)}^{~~~~~0} R_{0(jk)0}
\biggr) \nonumber \\
&&-{8 \over 135}\biggl(
 R_{i(jk)}^{~~~~~m} R_{m(ln)0}
+R_{i(jl)}^{~~~~~m} R_{m(kn)0}
+R_{i(jn)}^{~~~~~m} R_{m(kl)0} \nonumber \\
&&~~~~~~~~
+R_{i(kl)}^{~~~~~m} R_{m(jn)0}
+R_{i(kn)}^{~~~~~m} R_{m(jl)0}
+R_{i(ln)}^{~~~~~m} R_{m(jk)0} \biggr) \nonumber\\
&=&{8 \over 5}R_{i(jk|0|,ln)}
-{16 \over 15}R_{i(jk}^{~~~~0}R_{|0|ln)0}
-{16 \over 45}R_{i(jk}^{~~~~m}R_{|m|ln)0}, \label{gi61}
\eeqn
where we use Eq. (\ref{gamrie3}). 

The fourth derivative of $ij$ components is derived as follows: 
\beqn
g_{ij,klmn}&=&\Gamma_{ijk,lmn}+\Gamma_{jik,lmn} \nonumber \\
&=&{1 \over 5}\biggl[
R_{i(kl)j,mn}+R_{i(km)j,ln}+R_{i(kn)j,lm}+R_{i(lm)j,kn}+R_{i(ln)j,km}+
R_{i(mn)j,kl} \biggr] \nonumber \\
&&~~~~-{1 \over 5}\biggl[
 \Gamma^{\nu}_{i(k,l)}\Gamma_{j\nu(m,n)}
+\Gamma^{\nu}_{i(k,m)}\Gamma_{j\nu(l,n)}
+\Gamma^{\nu}_{i(k,n)}\Gamma_{j\nu(l,m)} \nonumber \\
&&~~~~~~~~
+\Gamma^{\nu}_{i(l,m)}\Gamma_{j\nu(k,n)}
+\Gamma^{\nu}_{i(l,n)}\Gamma_{j\nu(k,m)}
+\Gamma^{\nu}_{i(m,n)}\Gamma_{j\nu(k,l)} \nonumber \\
&&~~~~~~~~
+\Gamma^{\nu}_{j(k,l)}\Gamma_{i\nu(m,n)}
+\Gamma^{\nu}_{j(k,m)}\Gamma_{i\nu(l,n)}
+\Gamma^{\nu}_{j(k,n)}\Gamma_{i\nu(l,m)} \nonumber \\
&&~~~~~~~~
+\Gamma^{\nu}_{j(l,m)}\Gamma_{i\nu(k,n)}
+\Gamma^{\nu}_{j(l,n)}\Gamma_{i\nu(k,m)}
+\Gamma^{\nu}_{j(m,n)}\Gamma_{i\nu(k,l)}
\biggr] \nonumber \\
&=&{1 \over 5}\biggl[
R_{i(kl)j,mn}+R_{i(km)j,ln}+R_{i(kn)j,ln}+R_{i(lm)j,kn}+R_{i(ln)j,km}+
R_{i(mn)j,kl} \biggr] \nonumber \\
&&-{2 \over 15}\biggl[
 R_{i(kl)}^{~~~~~0}R_{j(mn)0}
+R_{i(km)}^{~~~~~0}R_{j(ln)0}
+R_{i(kn)}^{~~~~~0}R_{j(lm)0} \nonumber \\
&&~~~~~
+R_{i(lm)}^{~~~~~0}R_{j(kn)0}
+R_{i(ln)}^{~~~~~0}R_{j(km)0}
+R_{i(mn)}^{~~~~~0}R_{j(kl)0}
\biggr] \nonumber \\
&&-{2 \over 45}\biggl[
 R_{i(kl)}^{~~~~~p}R_{j(mn)p}
+R_{i(km)}^{~~~~~p}R_{j(ln)p}
+R_{i(kn)}^{~~~~~p}R_{j(lm)p} \nonumber \\
&&~~~~
+R_{i(lm)}^{~~~~~p}R_{j(kn)p}
+R_{i(ln)}^{~~~~~p}R_{j(km)p}
+R_{i(mn)}^{~~~~~p}R_{j(kl)p}
\biggr] \nonumber \\
&=&{6 \over 5}R_{i(kl|j|,mn)}
-{4 \over 5} R_{i(kl}^{~~~~0}R_{|j|mn)0}
-{4 \over 15} R_{i(kl}^{~~~~p}R_{|j|mn)p}. \label{gi62}
\eeqn
We note that the second ordinary derivatives of the Riemann
tensor in Eqs. (\ref{gi61}) and (\ref{gi62}) are
transformed to the covariant derivatives using
\beqn
&&R_{i(jk|0|;ln)}=
R_{i(jk|0|,ln)}-{4 \over 3} R_{i(nl}^{~~~~\sigma} R_{|\sigma|jk)0},\\
&&R_{i(jk|m|;ln)}=
R_{i(jk|m|,ln)}-{2 \over 3}R_{i(nl}^{~~~~\sigma} R_{|\sigma|jk)m}. 
\eeqn
This implies that $g_{0i,jklm}$ and $g_{ij,klmn}$ can be written
in the covariant form. 

To summarize this section, we derive the coefficients
$g_{\mu\nu,ij\cdots}$ in Eq. (\ref{goal}) which denotes the metric
at $\tau=x^0=$constant in the neighborhood of the timelike geodesic
${\cal G}$. This is used as the tidal field from a black hole 
for a star moving along ${\cal G}$. 
The equation numbers for the relations between 
the derivatives of the metric and the Riemann tensor 
in the Fermi normal coordinates are summarized in Table I. 

\begin{table}[t]
\begin{center}
\begin{tabular}{c|cc|cc|cc} \hline
& $00$ components & equation & $0i$ components & equation &
  $ij$ components & equation \\ \hline
2nd order &$g_{00,ij}$ & (\ref{gamrie00}) &
 $g_{0i,jk}$ & (\ref{g0kij}) &
 $g_{ij,kl}$ & (\ref{gijkl}) \\ 
3rd order &$g_{00,ijk}$ & (\ref{g00ijk}) &
 $g_{0i,jkl}$ & (\ref{g0ijkl}) &
 $g_{ij,klm}$ & (\ref{gijkln}) \\ 
4th order &$g_{00,ijkl}$ & (\ref{g00ijkl}) &
 $g_{0i,jklm}$ & (\ref{gi61}) &
 $g_{ij,klmn}$ & (\ref{gi62}) \\ 
\end{tabular}
\caption{Equation numbers from which one finds the 
relation between $g_{\mu\nu,ijk\cdots}$ and the corresponding Riemann 
tensor in the Fermi normal coordinates. }
\end{center}
\end{table}

\section{Components of the Riemann tensor for a Kerr spacetime}

To compute components of the Riemann tensor and
its covariant derivative for the
Kerr metric in the Fermi normal coordinates, we adopt the 
method developed by Marck \cite{Marck}. Namely, we first calculate 
the components in a standard tetrad frame of the
Boyer-Lindquist coordinate system \cite{chandra}, and
then, perform a coordinate transformation to the Fermi normal coordinates. 

The Kerr metric in the Boyer-Lindquist coordinate system is written as
\beqn
ds^2 = -\biggl(1 - {2 Mr \over \Sigma}\biggr) dt^2
-{4 M r a \sin^2\theta \over \Sigma} dt d\varphi
+{(r^2+a^2)^2 - \Delta a^2 \sin^2\theta \over \Sigma}
\sin^2\theta d\varphi^2 
+{\Sigma \over \Delta} dr^2 + \Sigma d\theta^2, 
\eeqn
where
\beqn
\Sigma = r^2+a^2\cos^2\theta,~~~\Delta = r^2 + a^2 - 2Mr, 
\eeqn
and $M$ and $a$ denote the mass and spin parameter.
A standard tetrad for the Kerr metric is defined by 
\beqn
&& (e^{(0)})_{\mu}=\biggl(\sqrt{{\Delta \over \Sigma}}, 0, 0,
-a\sin^2\theta \sqrt{{\Delta \over \Sigma}}\biggr),\\
&& (e^{(1)})_{\mu}=\biggl(0, \sqrt{{\Sigma \over \Delta}}, 0, 0 \biggr),\\
&& (e^{(2)})_{\mu}=\biggl(0, 0, \sqrt{\Sigma}, 0 \biggr),\\
&& (e^{(3)})_{\mu}=\biggl(-{a\sin\theta \over \sqrt{\Sigma}}, 0, 0,
{(r^2 +a^2)\sin\theta \over \sqrt{\Sigma}} \biggr). 
\eeqn
Note that the sign convention for $(e^{(3)})_{\mu}$ is 
different from that of \cite{Marck}. 

In the first step, we compute the components of the
Riemann tensor and its covariant derivative in the
standard tetrad. The nonvanishing components of the Riemann tensor
in this tetrad frame are 
\beqn
&& R_{(1)(2)(1)(2)}=R_{(1)(3)(1)(3)}
={1 \over 2}R_{(1)(0)(1)(0)}=-{1 \over 2}R_{(2)(3)(2)(3)} \nonumber \\
&&~~~~~~~~~=-R_{(2)(0)(2)(0)}=-R_{(3)(0)(3)(0)}
={M r(3a^2 \cos^2\theta - r^2) \over \Sigma^3},\\
&& R_{(1)(2)(3)(0)}=-R_{(1)(3)(2)(0)}=-{1 \over 2}R_{(1)(0)(2)(3)}
={a M \cos\theta (3r^2 -a^2 \cos^2\theta) \over \Sigma^3}. 
\eeqn
The tetrad components associated with the first and
second covariant derivatives are
defined as
\beqn
&& Q_{(a)(b)(c)(d)(e)} \equiv \nabla_{\mu} R_{\nu\rho\sigma\lambda}
(e_{(a)})^{\mu}(e_{(b)})^{\nu}(e_{(c)})^{\rho}
(e_{(d)})^{\sigma}(e_{(e)})^{\lambda}, \\
&& P_{(a)(b)(c)(d)(e)(f)} \equiv \nabla_{(\alpha} \nabla_{\beta)} 
R_{\nu\rho\sigma\lambda}
(e_{(a)})^{\alpha}(e_{(b)})^{\beta}(e_{(c)})^{\nu}(e_{(d)})^{\rho}
(e_{(e)})^{\sigma}(e_{(f)})^{\lambda}. 
\eeqn
The nonvanishing components of $Q_{(a)(b)(c)(d)(e)}$ are
\beqn
&& Q_{(a)(1)(2)(1)(2)}=Q_{(a)(1)(3)(1)(3)}={1 \over 2}Q_{(a)(1)(0)(1)(0)}
=-{1 \over 2}Q_{(a)(2)(3)(2)(3)}=-Q_{(a)(2)(0)(2)(0)}
=-Q_{(a)(3)(0)(3)(0)} \nonumber \\
&& \hskip 5cm =\biggl({3 M J_1\Delta^{1/2}  \over \Sigma^{9/2}},
-{12 M a^2 r J_2 \sin\theta\cos\theta \over \Sigma^{9/2}}, 0, 0\biggr),\\
&& Q_{(a)(1)(2)(1)(0)}=Q_{(a)(2)(3)(3)(0)}
=\biggl(0,0,{-12 M a r J_2 \Delta^{1/2} \cos\theta \over \Sigma^{9/2}},
-{12 M a^2 r J_2 \sin\theta\cos\theta \over \Sigma^{9/2}}\biggr),\\
&& Q_{(a)(1)(2)(2)(3)}=-Q_{(a)(1)(0)(3)(0)}
=\biggl(0,0,{3 M J_1 \Delta^{1/2} \over \Sigma^{9/2}},
{3 M a J_1 \sin\theta \over \Sigma^{9/2}}\biggr),\\
&& Q_{(a)(1)(2)(3)(0)}=-Q_{(a)(1)(3)(2)(0)}
=-{1 \over 2}Q_{(a)(1)(0)(2)(3)}
=\biggl(-{12 M a r \Delta^{1/2} J_2 \cos\theta \over \Sigma^{9/2}},
-{3 M a J_1 \sin\theta \over \Sigma^{9/2}}, 0, 0\biggr),\\
&& Q_{(a)(1)(3)(1)(0)}=-Q_{(a)(2)(3)(2)(0)}
=\biggl(-{3 M a J_1 \sin\theta \over \Sigma^{9/2}},
{12 M a r J_2 \Delta^{1/2} \cos\theta \over \Sigma^{9/2}}, 0, 0\biggr),\\
&& Q_{(a)(1)(3)(2)(3)}=Q_{(a)(1)(0)(2)(0)}
=\biggl(-{12 M a^2 r J_2 \sin\theta\cos\theta \over \Sigma^{9/2}},
-{3 M J_1 \Delta^{1/2} \over \Sigma^{9/2}}, 0, 0\biggr),
\eeqn
where
\beqn
&& J_1=r^4 - 6a^2 r^2 \cos^2\theta + a^4 \cos^4\theta,\\
&& J_2=r^2 - a^2 \cos^2\theta . 
\eeqn
For $P_{(a)(b)(c)(d)(e)(f)}$, the explicit form of the components in 
a general orbit is very complicated. Thus, we only write the
nonvanishing components in the equatorial plane setting $\theta=\pi/2$;
\beqn
&& P_{(a)(b)(1)(2)(1)(2)}=-P_{(a)(b)(3)(0)(3)(0)} 
={3M \over r^7}
\left(
\begin{array}{cccc}
-(4r^2-9 r M+5a^2) & 0 & 0 & 0 \\
\ast & \Delta + 4a^2 & 0 & 0\\
\ast & \ast & 3 \Delta & 3a \Delta^{1/2}  \\
\ast & \ast & \ast & -(M r-3a^2)  \\
\end{array}
\right),\\
&& P_{(a)(b)(1)(2)(1)(3)}=P_{(a)(b)(2)(0)(3)(0)} 
=-{3M \over r^7}
\left(
\begin{array}{cccc}
0 & 0 & 0 & 0 \\
\ast & 0 & \Delta & a\Delta^{1/2}\\
\ast & \ast & 0 & 0  \\
\ast & \ast & \ast & 0  \\
\end{array}
\right),\\
&& P_{(a)(b)(1)(2)(1)(0)}=P_{(a)(b)(2)(3)(3)(0)} 
={3M \over 2r^7}
\left(
\begin{array}{cccc}
0 & 0 & 0 & 0 \\
\ast & 0 & 8a\Delta^{1/2} & -(Mr-8a^2)\\
\ast & \ast & 0 & 0  \\
\ast & \ast & \ast & 0  \\
\end{array}
\right),\\
&& P_{(a)(b)(1)(2)(2)(3)}=-P_{(a)(b)(1)(0)(3)(0)} 
=-{3M \over 2r^7}
\left(
\begin{array}{cccc}
0 & 0 & 8\Delta-Mr & 8a\Delta^{1/2} \\
\ast & 0 & 0 & 0 \\
\ast & \ast & 0 & 0  \\
\ast & \ast & \ast & 0  \\
\end{array}
\right),\\
&& P_{(a)(b)(1)(2)(2)(0)}=-P_{(a)(b)(1)(3)(3)(0)} 
={3M \over r^7}
\left(
\begin{array}{cccc}
0 & 0 & a\Delta^{1/2} & a^{2} \\
\ast & 0 & 0 & 0 \\
\ast & \ast & 0 & 0  \\
\ast & \ast & \ast & 0  \\
\end{array}
\right),\\
&& {1 \over 15} P_{(a)(b)(1)(2)(3)(0)}=-{1 \over 21} P_{(a)(b)(1)(3)(2)(0)}
= -{1 \over 36} P_{(a)(b)(1)(0)(2)(3)}
={M \over r^7}
\left(
\begin{array}{cccc}
0 & a \Delta^{1/2} & 0 & 0 \\
\ast & 0 & 0 & 0 \\
\ast & \ast & 0 & 0  \\
\ast & \ast & \ast & 0  \\
\end{array}
\right),\\
&& P_{(a)(b)(1)(3)(1)(3)}=-P_{(a)(b)(2)(0)(2)(0)} 
={3M \over r^7}
\left(
\begin{array}{cccc}
-(4r^2-9 r M+7a^2) & 0 & 0 & 0 \\
\ast & (3\Delta + 4a^2) & 0 & 0\\
\ast & \ast &  \Delta & a \Delta^{1/2}  \\
\ast & \ast & \ast & -M r+a^2  \\
\end{array}
\right),\\
&& P_{(a)(b)(1)(3)(1)(0)}=-P_{(a)(b)(2)(3)(2)(0)} 
={3M \over 2r^7}
\left(
\begin{array}{cccc}
16 a \Delta^{1/2} & 0 & 0 & 0 \\
\ast & -16 a \Delta^{1/2} & 0 & 0\\
\ast & \ast & 0 & -Mr  \\
\ast & \ast & \ast & 0  \\
\end{array}
\right),\\
&& P_{(a)(b)(1)(3)(2)(3)}=P_{(a)(b)(1)(0)(2)(0)} 
={3M \over 2r^7}
\left(
\begin{array}{cccc}
0 & 8r^2-17Mr+16a^2 & 0 & 0 \\
\ast & 0 & 0 & 0 \\
\ast & \ast & 0 & 0  \\
\ast & \ast & \ast & 0  \\
\end{array}
\right),\\
&& P_{(a)(b)(1)(0)(1)(0)}=-P_{(a)(b)(2)(3)(2)(3)} 
={6M \over r^7}
\left(
\begin{array}{cccc}
-(4r^2 - 9Mr +6a^2)& 0 & 0 & 0 \\
\ast & 2(\Delta + 2a^2) & 0 & 0\\
\ast & \ast &  2\Delta & 2 a\Delta^{1/2}  \\
\ast & \ast & \ast & -(M r - 2a^2)  \\
\end{array}
\right).
\eeqn
Here, the components in the 2-matrix form of
the subscripts $(a)$ and $(b)$ are shown 
in the order (1), (2), (3), and (0).

To compute $R_{abcd}$, $Q_{abcde}$, and $P_{abcdef}$ in the Fermi
normal coordinates, we need to prepare the transformation matrix
from the standard tetrad frame to the Fermi normal coordinate frame.
Denoting it as $\Lambda_a^{(a)}$, we have 
\beqn
R_{abcd}&=&R_{(a)(b)(c)(d)}\Lambda_a^{(a)}\Lambda_b^{(b)}
\Lambda_c^{(c)}\Lambda_d^{(d)},\\
Q_{abcde}&=&Q_{(a)(b)(c)(d)(e)}\Lambda_a^{(a)}\Lambda_b^{(b)}
\Lambda_c^{(c)}\Lambda_d^{(d)}\Lambda_e^{(e)},\\
P_{abcdef}&=&P_{(a)(b)(c)(d)(e)(f)}\Lambda_a^{(a)}\Lambda_b^{(b)}
\Lambda_c^{(c)}\Lambda_d^{(d)}\Lambda_e^{(e)}\Lambda_f^{(f)}. 
\eeqn
$\Lambda_a^{(a)}$ is constructed from an orthonormal set of vectors
$\lambda_a^{\mu}~(a=0,1,2,3)$, 
which are parallel-propagated along a timelike geodesic ${\cal G}$ as
\beq
\Lambda_a^{(a)}=\lambda_a^{\mu} (e^{(a)})_{\mu}.
\eeq
Since the tangent vector of the timelike geodesic,
$u^{\mu}\equiv(\pa/\pa \tau)^{\mu}$, 
is parallel-propagated along ${\cal G}$, $\lambda_0^{\mu}$ should be
equal to $u^{\mu}$. Here, the geodesic equations are integrated
to give (e.g., \cite{Marck})
\beqn
u^t&=&{1 \over \Delta \Sigma}[(r^2+a^2)\Sigma E
-2Mar B],\\
(\Sigma u^r)^2&=&A^2-\Delta (r^2 + K),\\
(\Sigma u^{\theta})^2&=&K-a^2\cos^2\theta-{B^2 \over \sin^2\theta},\\
u^{\varphi}&=&{1 \over \Delta\Sigma\sin^2\theta}
[-2Mr B + L\Sigma],
\eeqn
where
\beqn
A=E(r^2+a^2) -a L,~~~B= L-a E \sin^2\theta. 
\eeqn
$E=-u_t$ and $L=u_{\varphi}$ are the specific
energy and angular momentum for a particle of mass $\mu$
moving around a Kerr black hole. 
$K=(L-aE)^2+C_K$ is the so-called Carter constant with $C_K$ a
constant \cite{carter}.
The first integral of the geodesic equations leads to
\beqn
&& \Lambda_0^{(0)}={A \over \sqrt{\Delta \Sigma}},\\
&& \Lambda_0^{(1)}={\sqrt{\Sigma} u^r \over \sqrt{\Delta}},\\
&& \Lambda_0^{(2)}=\sqrt{\Sigma}u^{\theta},\\
&& \Lambda_0^{(3)}={B \over \sqrt{\Sigma}\sin\theta}.
\eeqn
For other components of $\lambda_{a}^{(a)}$, 
we follow Marck \cite{Marck}, and thus, we choose 
\beqn
&& \Lambda_1^{(0)}=
{\alpha\sqrt{\Sigma}r u^r \over \sqrt{K\Delta}}\cos\Psi
-{\alpha A \over \sqrt{\Delta\Sigma}}\sin\Psi,\\
&& \Lambda_1^{(1)}=
{\alpha r A \over \sqrt{K\Delta\Sigma}}\cos\Psi
-{\alpha \sqrt{\Sigma} u^r \over \sqrt{\Delta}}\sin\Psi,\\
&& \Lambda_1^{(2)}=
-{\beta a B \cos\theta  \over \sqrt{K\Sigma}\sin\theta}\cos\Psi
-\beta \sqrt{\Sigma} u^{\theta} \sin\Psi,\\
&& \Lambda_1^{(3)}=
{\beta \sqrt{\Sigma}a u^{\theta} \cos\theta \over \sqrt{K}}\cos\Psi
-{\beta B \over \sqrt{\Sigma}\sin\theta}\sin\Psi,\\
&& \Lambda_2^{(0)}=
{\sqrt{\Sigma}a\cos\theta u^r \over \sqrt{K\Delta}},\\
&& \Lambda_2^{(1)}=
{a A\cos\theta \over \sqrt{K\Delta\Sigma}},\\
&& \Lambda_2^{(2)}=
{r B \over \sqrt{K\Sigma}\sin\theta},\\
&& \Lambda_2^{(3)}=
-{\sqrt{\Sigma}r u^{\theta} \over \sqrt{K}},\\
&& \Lambda_3^{(0)}=
{\alpha\sqrt{\Sigma}r u^r \over \sqrt{K\Delta}}\sin\Psi
+{\alpha A \over \sqrt{\Delta\Sigma}}\cos\Psi,\\
&& \Lambda_3^{(1)}=
{\alpha r A \over \sqrt{K\Delta\Sigma}}\sin\Psi
+{\alpha \sqrt{\Sigma} u^r \over \sqrt{\Delta}}\cos\Psi,\\
&& \Lambda_3^{(2)}=
-{\beta a B \cos\theta  \over \sqrt{K\Sigma}\sin\theta}\sin\Psi
+\beta \sqrt{\Sigma} u^{\theta} \cos\Psi,\\
&& \Lambda_3^{(3)}=
{\beta \sqrt{\Sigma}a u^{\theta} \cos\theta \over \sqrt{K}}\sin\Psi
+{\beta B \over \sqrt{\Sigma}\sin\theta}\cos\Psi,
\eeqn
where $\alpha$ and $\beta$ are normalization constants defined by 
\beqn
\alpha=\sqrt{{K-a^2\cos^2\theta \over r^2 + K}},~~~~
\beta={1 \over \alpha}.
\eeqn
We note that the direction of the components 1, 2, and 3
[not (1), (2), and (3)] adopted by Marck 
are approximately equal to $x$, $-z$, and $y$
of the Cartesian coordinates in the comoving frame. 
The time evolution of the rotation angle $\Psi$ is computed by
\beq
{d\Psi \over d\tau}={\sqrt{K} \over \Sigma}
\biggl({A \over r^2+K}+{a B \over K-a^2\cos^2\theta}\biggr).
\eeq

\section{Formulation for equilibrium Newtonian stars in equatorial circular 
orbits in the black hole tidal field} 

\subsection{Basic equations}

In this section, we give a formulation for computing 
equilibrium states of a fluid star 
in circular orbits around equatorial plane of a Kerr spacetime.
Here, we assume that the self-gravity of the star is described by
the Newtonian gravity. In this case, the gravitational potential 
associated with the tidal potential can be linearly superposed
\cite{Fishbone}. Using this property, we write 
the hydrodynamic equation for the fluid body as
\beqn
\rho {\pa v_i \over \pa \tau}+\rho v^j {\pa v_i \over \pa x^j}
=-{\pa P \over \pa x^i}
-\rho {\pa (\phi + \phi_{\rm tidal}) \over \pa x^i}
+\rho\biggl[v_j\biggl({\pa A_j \over \pa x^i}-{\pa A_i \over \pa x^j}
\biggr) -{\pa A_i \over \pa \tau}\biggr], \label{euler}
\eeqn
where $\rho$ is the mass density, $v^i$ the three-velocity $(dx^i/d\tau)$, 
$P$ the pressure, and $\phi$ the Newtonian potential produced 
by a star which obeys the Poisson equation 
\beq
\Delta \phi=4\pi \rho. \label{poisson}
\eeq
$A_i$ is a vector potential defined in Eq.(\ref{Aeq}), which
is associated with the so-called gravitomagnetic force \cite{Membrane}. 
$\phi_{\rm tidal}$ denotes the tidal potential associated
with the background Kerr spacetime, which is
related to the metric computed in previous sections as
\beqn
\phi_{\rm tidal}&=&-{1 \over 2}(g_{00}+1) \nonumber \\
&=&-{1 \over 4}g_{00,ij}x^i x^j-{1 \over 12}g_{00,ijk}x^i x^j x^k
-{1 \over 48}g_{00,ijkl}x^i x^j x^k x^l +O(x^5) \nonumber \\
&=&{1 \over 2} C_{ij}x^ix^j+{1 \over 6}C_{ijk}x^i x^j x^k
+{1 \over 24} \bigl[C_{ijkl}+4C_{(ij}C_{kl)}-4B_{(kl|n|}B_{ij)n} \bigr]
x^i x^j x^k x^l+ O(x^5),
\eeqn
where
\beqn
&&C_{ij}=R_{0i0j},\\
&&C_{ijk}=R_{0(i|0|j;k)},\\
&&C_{ijkl}=R_{0(i|0|j;kl)},\\
&&B_{ijk}=R_{k(ij)0},\\
&&A_k={2 \over 3}B_{ijk} x^i x^j. \label{Aeq}
\eeqn

In the equations of motion (\ref{euler}),
we include the lowest-order gravitomagnetic term
associated with $A_k$, although it is a first post-Newtonian term
and does not appear in Newtonian order from the point of
view of post-Newtonian approximations \cite{MTW}. 
The reason we add it is that the order of magnitude of
this term is as large as that of the fourth-order terms in 
$\phi_{\rm tidal}$ if the spin angular velocity
of the fluid star is of order $\Omega$ as in the corotational velocity
field (see a discussion in the final paragraph of this section).
On the other hand, for the irrotational velocity field in which 
the spin of the fluid star is negligible in the frame of 
the Fermi normal coordinates,
the magnitude of this term will be much smaller than
the fourth-order term in $\phi_{\rm tidal}$. 
In the following calculation, we neglect the gravitomagnetic
terms for most of
calculations, but to clarify the quantitative 
effect of this term, we perform a few computations including it. 

In this paper, we restrict out attention to circular orbits in the
equatorial plane, i.e., $\theta=\pi/2$, $u^r=0$, $u^{\theta}=0$, and 
$C_K=0$. Then \cite{BPT}
\beqn
E&=&{r^2-2Mr+a\sqrt{Mr} \over r D},\\
L&=&{\sqrt{Mr} (r^2 - 2a \sqrt{Mr} + a^2) \over r D},\\
D&\equiv & \sqrt{r^2-3Mr+2a\sqrt{Mr}}, 
\eeqn
where $r$ denotes the orbital radius. 
In this case, the evolution equation for $\Psi$ becomes
\beq
{d \Psi \over d \tau}= \sqrt{{M \over r^3}},~{\rm and~hence},~
\Psi=\sqrt{{M \over r^3}} \tau. 
\eeq
For the equatorial circular orbits, 
the transformation matrix $\Lambda_a^{(a)}$
reduces to a simple form as
\beqn
&& \Lambda_0^{(a)}=\biggl(\sqrt{1+{B^2 \over r^2}}, 0, 0, {B \over r}
\biggr),\\
&& \Lambda_1^{(a)}=\biggl(-{B \over r}\sin\Psi, \cos\Psi, 0,
-\sqrt{1+{B^2 \over r^2}}\sin\Psi \biggr),\\
&& \Lambda_2^{(a)}=(0, 0, 1, 0 ),\\
&& \Lambda_3^{(a)}=\biggl({B \over r}\cos\Psi, \sin\Psi, 0,
\sqrt{1+{B^2 \over r^2}}\cos\Psi
\biggr),
\eeqn
where $B=L-aE=r(\sqrt{Mr}-a)/D$. Note that $1+B^2/r^2$ may be written as
$\Delta/D^2$. A known interesting property is that
independent of the value of $a$,
$B/r=1/\sqrt{3}$ and $\Delta/D=4/3$ at ISCOs \cite{Fishbone} at which
$r$ satisfies 
\beqn
r^2 - 6Mr+8M^{1/2}a r^{1/2}-3a^2=0. 
\eeqn
This implies that at the ISCO, 
$r^3 C_{ij}$ and $r^3 B_{ijk}$ are independent of $a$ (see below).

To derive the tidal tensors 
$C_{ij}$, $C_{ijk}$, $C_{ijkl}$, and $B_{ijk}$, as a first step, 
it is better to calculate the components in a tetrad 
frame defined by 
\beqn
&&\tilde \Lambda_0^{(a)}=\Lambda_0^{(a)},\\
&&\tilde \Lambda_1^{(a)}=\Lambda_1^{(a)}\cos\Psi + \Lambda_3^{(a)}\sin\Psi,\\
&&\tilde \Lambda_2^{(a)}=\Lambda_2^{(a)},\\
&&\tilde \Lambda_3^{(a)}=-\Lambda_1^{(a)}\sin\Psi + \Lambda_3^{(a)}\cos\Psi. 
\eeqn
We refer to this frame as the tilde frame in the following. 
In the tilde frame, $\tilde \Lambda_a^{(a)}$ is independent of $\Psi$, but
the coordinate basis of this frame is not parallel-transported
along the timelike geodesic. 
In the second step, we should perform the coordinate
transformation to the parallel-transported frame. 

The nonvanishing components of
$\tilde C_{ij}$, $\tilde C_{ijk}$, $\tilde C_{ijkl}$, and $\tilde B_{ijk}$, 
which denote the tidal tensor in the tilde frame, are
\beqn
&&\tilde C_{11}={M \over r^3}\biggl(1 - 3{r^2 + B^2 \over r^2}\biggr),\\
&&\tilde C_{22}={M \over r^3}\biggl(1 + 3{B^2 \over r^2}\biggr),\\
&&\tilde C_{33}={M \over r^3},\\
&&\tilde B_{131}=\tilde B_{311}=-\tilde B_{232}=-\tilde B_{322}
=- {1 \over 2}\tilde B_{113}={1 \over 2} \tilde B_{223}
=-{3M B \over 2r^4}\sqrt{1+{B^2 \over r^2}},\\
&&\tilde C_{111}={3M \Delta^{1/2} \over D r^7}[2D r^2 -2a B r + 3B^2D],\\
&&\tilde C_{122}=\tilde C_{212}=\tilde C_{221}
=-{M \Delta^{1/2} \over D r^7}[3Dr^2 - 8a B r+7B^2D],\\
&&\tilde C_{133}=\tilde C_{313}=\tilde C_{331}=
-{M \Delta^{1/2} \over D r^7}[3 D r^2 +2a B r +2B^2D],\\
&& \tilde C_{1111}={3M \over Dr^9}
\bigl[-8 D r^4+18 D M r^3 +16 a B r \Delta - 12(a^2+B^2)Dr^2 
+27B^2 D M r - 19 a^2 B^2 D\bigr],\\
&& \tilde C_{1122}=\tilde C_{1212}=\tilde C_{1221}
=\tilde C_{2112}=\tilde C_{2121}=\tilde C_{2211}\nonumber \\
&&={M \over 2Dr^9}
\bigl[24Dr^4 -51D M r^3 -108 a B r \Delta + 51(a^2+B^2)Dr^2
-109 B^2 D M r + 102 a^2 B^2 D\bigr],\\
&& \tilde C_{1133}=\tilde C_{1313}=\tilde C_{1331}
=\tilde C_{3113}=\tilde C_{3131}=\tilde C_{3311} \nonumber \\
&&={M \over 2Dr^{11}}
\bigl[24 D r^6 -51 D M r^5 +20 a B r^3 \Delta + 25(a^2+B^2)Dr^4
-56 B^2 D M r^3 \nonumber \\
&& \hskip 2.5cm
+10 a B^3 r \Delta 
+5B^4 D r^2 +25a^2 B^2 D r^2-15B^4 D M r + 10a^2 B^4 D\bigr],\\
&& \tilde C_{2222}=-{3 M \over D r^9}
\bigl[3 D r^4 -6 D M r^3 -16 a B r \Delta + 7(a^2+B^2)Dr^2
-14 B^2 D M r +19 a^2 B^2 D\bigr],\\
&& \tilde C_{2233}=\tilde C_{2323}=\tilde C_{2332}
=\tilde C_{3223}=\tilde C_{3232}=\tilde C_{3322} \nonumber \\
&&={M \over 2Dr^{11}}
\bigl[-6 D r^6 +12 D M r^5 +10 a B r^3 \Delta -10(a^2+B^2)Dr^4
+21 B^2 D M r^3 \nonumber \\
&& \hskip 2.5cm
-10 a B^3 r \Delta 
-5 B^4 D r^2 +5 a^2 B^2 D r^2+15 B^4 D M r -10a^2 B^4 D\bigr],\\
&& \tilde C_{3333}={3 M \over D r^9}
\bigl[-3 D r^4 +6 D M r^3 - 6 a B r \Delta -3(a^2+B^2)Dr^2
+7 B^2 D M r - 6 a^2 B^2 D\bigr]. 
\eeqn
The expression for $\tilde C_{ij}$
agrees with that derived by Marck \cite{Marck}. 
The nonvanishing components of 
$C_{ij}$, $C_{ijk}$, $C_{ijkl}$, and $B_{ijk}$ are derived by 
the coordinate transformation from $\tilde x^i$ to $x^i$. 

In the tilde frame, the tidal potential up to the
fourth order is written as
\beqn
\phi_{\rm tidal}
={1 \over 2} \tilde C_{ij}\tilde x^i \tilde x^j
+{1 \over 6}\tilde C_{ijk}\tilde x^i \tilde x^j \tilde x^k
+{1 \over 24} \bigl[\tilde C_{ijkl} + 4\tilde C_{(ij} \tilde C_{kl)}
-4\tilde B_{(kl|n|}\tilde B_{ij)n} \bigr]
\tilde x^i \tilde x^j \tilde x^k \tilde x^l,
\eeqn
where
\beqn
&& \tilde x^1=x^1 \cos\Psi+ x^3 \sin\Psi,\\
&& \tilde x^2=x^2,\\
&& \tilde x^3=-x^1\sin\Psi + x^3\cos\Psi. 
\eeqn
In the Newtonian limit $r \gg M ( > a)$,
it is written as
\beqn
\phi_{\rm tidal}
&=&{M \over 2r^3}
\Big[ -2(\tilde x^1)^2 + (\tilde x^2)^2 + (\tilde x^3)^2 \Big]
-{M \over 2r^4} \tilde x^1
\Big[-2(\tilde x^1)^2 + 3\{(\tilde x^2)^2 + (\tilde x^3)^2\}\Big]
\nonumber \\
&&-{M \over 8r^5} \Big[8 (\tilde x^1)^4+3(\tilde x^2)^4+3(\tilde x^3)^4
-24\{(\tilde x^1)^2 (\tilde x^2)^2+(\tilde x^1)^2 (\tilde x^3)^2\}
+6(\tilde x^2)^2 (\tilde x^3)^2 \Big]. \label{eq168}
\eeqn
This agrees with the expansion form of the Newtonian tidal potential
from a point source of mass $M$ at a distance $r$ as
\beq
-{M \over \sqrt{(\tilde x^1+r)^2 + (\tilde x^2)^2 + (\tilde x^3)^2}}.
\eeq

The orders of the magnitude of the second-, third-, and fourth-order 
tidal forces in Eq. (\ref{eq168})
are $O(MR/r^3)$, $O(MR^2/r^4)$, and $O(MR^3/r^5)$, respectively. 
On the other hand, the order of magnitude of the gravitomagnetic 
force in the Fermi normal coordinates 
is $O(M^{3/2}R v/r^{7/2})$ where $v$ denotes the characteristic 
magnitude of $v^i$. For the corotational velocity field, 
$v =O(M^{1/2}R/r^{3/2})$, and hence, the gravitomagnetic tidal force 
is of $O(M^{2}R^2/r^{5})$ which is the same 
as the order of the fourth-order tidal potential for 
stars with $R=O(M)$. For close corotating orbits with $r=O(M)$, it is
as larger as the third-order term. 
For rapidly rotating stars with $v \alt c$, the gravitomagnetic 
tidal force is always larger than the third-order term.
On the other hand, for the irrotational velocity field,
it will be negligible. 

\subsection{Hydrostatic equations for 
corotational and irrotational equilibria}

Now, we turn our attention to the hydrostatic equations. First, 
we consider the case in which the velocity field is corotational, and 
assume 
\beq
v^i=[-\{x^3-x_c \sin(\Omega\tau)\}, 0, \{x^1-x_c \cos(\Omega\tau)\}]
\label{vcorot}
\eeq 
where $\Omega = d \Psi / d \tau =$const. 
$x_c$ is a correction constant which is much smaller than the stellar 
radius and nonzero only when we take into account the third-order 
terms in $\phi_{\rm tidal}$ or the gravitomagnetic terms. 
For $x_c \not=0$, the rotational axis deviates from the $x^2$ axis. 
Also, the center of mass 
of a fluid star is different from the origin slightly (see Sec. V). 

In this velocity field, Eq. (\ref{euler}) is integrated to give 
\beqn
{\Omega^2 \over 2}[(\tilde x^1-x_g)^2 + (\tilde x^3)^2]
=h + \phi + \phi_{\rm tidal}+\phi_{\rm mag}+C, \label{coro}
\eeqn
where $x_g=2x_c$, $C$ is an integration constant, and 
\beqn
&&h=\int {dP \over \rho}, \\
&&\phi_{\rm mag}=2{M B \over r^4}\sqrt{1+{B^2 \over r^2}} \Omega
\biggl[-(\tilde x^1)^3+\tilde x^1 \{(\tilde x^2)^2-(\tilde x^3)^2\}
+{3 \over 4}x_g\{(\tilde x^1)^2-(\tilde x^2)^2\}\biggr]. 
\eeqn
The first integral of the Euler equation is independent of $\tau$ 
in the tilde frame. Equations (\ref{poisson}) and (\ref{coro})
constitute the basic equations for the corotational binary. 

For the irrotational velocity field, we should set the
three-components of the four-velocity as 
\beq
u_i \equiv v_i+A_i = {\pa \psi \over \pa x^i}, 
\eeq
where $\psi$ denotes the velocity potential which is time-independent 
in the tilde frame. Then, Eq. (\ref{euler}) is integrated to give 
\beq
-{\pa \psi \over \pa \tau}
-{1 \over 2}\delta_{ij} {\pa\psi \over \pa x^i}{\pa\psi \over \pa x^j}
=h + \phi + \phi_{\rm tidal}-{\pa \psi \over \pa x^j} A^j+C. \label{irro1}
\eeq
In the tilde frame, the first term is written as
\beq
-{\pa \psi \over \pa \tau}
=\Omega \tilde x^i{\pa \psi \over \pa \tilde x^i}. 
\eeq
Also, we have a relation
\beq
\delta_{ij} {\pa \psi \over \pa x^i}{\pa \psi \over \pa x^j}
=\delta_{ij} {\pa \psi \over \pa \tilde x^i}
{\pa \psi \over \pa \tilde x^j}. 
\eeq
Thus, the first integral of the Euler equation is also independent of $\tau$ 
in the tilde frame. 

$\psi$ is determined by solving the continuity equation rewritten as 
\beq
\rho \tilde \Delta \psi+\delta_{ij}{\pa \psi \over \pa \tilde x^i}
{\pa \rho \over \pa \tilde x^j}=0, \label{irro2}
\eeq
where $\tilde \Delta$ is the Laplacian in the tilde coordinates. 
Equations (\ref{poisson}), (\ref{irro1}), and (\ref{irro2}) constitute 
the basic equations for irrotational binaries. 


\section{Roche limit in equatorial circular orbits}

To quantitatively illustrate the importance of the higher-order terms
in the tidal potential $\phi_{\rm tidal}$ as well as to clarify the
dependence of the tidal disruption limit of a star on the equations of
state and on general relativistic effects of the black hole, we
numerically compute corotating equilibria and determine the tidal
disruption limit (Roche limit) as an extension of a previous work by
Fishbone \cite{Fishbone}.  For some case such as binaries of a black
hole and a neutron star, the irrotational velocity field is more
realistic. However, we know that in the incompressible case, the tidal
disruption limit depends weakly on the velocity profile as far as the
spin angular velocity of the star is of order $M^{1/2}/r^{3/2}$
\cite{shibata}.  We expect that this will be also the case for
compressible stars, and hence, even in the assumption of the
corotational velocity field, we can obtain an approximate result of
the tidal disruption limit for the irrotational velocity field.

\subsection{Setting and numerical method}

We adopt polytropic equations of state for the star as 
\beqn
P=\kappa \rho^{1+{1\over n}},~{\rm and~thus},~h=\kappa(n+1)\rho^{1 \over n},
\eeqn
where $\kappa$ is the polytropic constant and $n$ the polytropic index. 
In this paper, we choose $n=0.5$, 1, and 1.5 to approximately
model neutron stars or white dwarfs. 

The basic equations in this problem are Eqs. (\ref{poisson}) and (\ref{coro}).
Numerical solutions are obtained by iteratively solving these
coupled equations. To achieve a convergence in the iteration, we 
rescale the coordinates as $\tilde x^i = p q^i$ where $p$ is a
constant and $q^i$ dimensionless coordinates.
Then, the basic equations are written in the form 
\beqn
&& \Delta_q \bar\phi= 4\pi \rho,\label{poissonq}\\
&& {\Omega^2 \over 2}p^2[(\tilde q^1-q_g)^2 + (\tilde q^3)^2]
=h + p^2 \bar\phi + p^2 (\bar\phi_{\rm tidal}+\bar\phi_{\rm mag})+C,
\eeqn
where $\Delta_q$ is the Laplacian in the coordinates of $q^i$,
$\bar \phi=p^{-2}\phi$, 
$\bar\phi_{\rm tidal}=p^{-2}\phi_{\rm tidal}$,
$\bar\phi_{\rm mag}=p^{-2}\phi_{\rm mag}$, and $q_g=q^{-2}x_g$. 
Thus, six free constants $M$, $a$, $\kappa$, $q_g$, $p$, and $C$
are contained in the equations. 
In the following, $M$ is fixed adopting the units $c=G=M=1$. 
An equilibrium configuration is computed for fixed values of 
$\kappa$, $\rho_c$ (the central density), $a$, and $r (>R)$.
Sequences of the equilibria are computed varying these parameters. 
$\kappa$ and $\rho_c$ determine the mass $m$ and the radius $R$ of a star,
and so do the ratios $Q=m/M$ and $R/r$. 
Note that for a given value of $r$, $\Omega$ is determined
to be $(M/r^3)^{1/2}$ in this problem.

Three remained constants $q_g$, $p$, and $C$ are
parameters determined at each step of the iteration
from the following conditions: 
During the iteration, we require that $\rho=\rho_c$ and
$\pa \rho/\pa q^1=0$ at the origin $(q^1, q^2, q^3)=(0, 0, 0)$. 
In addition, we fix the coordinates of 
the stellar surface along the $\tilde x^1$ axis (the axis connecting 
the origin and the center of a black hole) on the black hole side as
$(q_s, 0, 0)$ where $q_s <0$. From these three conditions,
the three free parameters are determined.

If we include the third-order terms in the tidal potential,
the center of mass of a star may be deviated from the origin,
although for consistency, it should be located approximately there.
To check that the deviation is much smaller than the stellar
radius, we calculate the value of $q^1$ coordinate for
a center of mass defined by 
\beqn
\langle q^1\rangle \equiv {p^3 \over m} \int d^3q \rho q^1,
\eeqn
where
\beqn
m \equiv p^3 \int d^3 q \rho. 
\eeqn
We found that $|\langle q^1 \rangle|$
is indeed much smaller than the stellar radius 
(e.g., $|\langle q^1 \rangle| \sim 0.005 |q_s|$ at ISCOs 
for $a=0$--$0.9M$ and the value is smaller for smaller orbital radii). 
 
The Poisson equation (\ref{poissonq}) is solved in the Cartesian coordinates 
with the uniform grid of size $(2N+1, N+1, 2N+1)$ for $(q^1, q^2, q^3)$
which covers the region with $-L \leq q^1 \leq L$, $0 \leq q^2 \leq L$,
and $-L \leq q^3 \leq L$ 
(the reflection symmetry with respect to the $q^2=0$ plane is assumed). 
The numerical method is essentially the same as that in \cite{shibapn}: 
We make the second-order finite-differencing equation and solve
it in a preconditioned conjugate gradient method. 
Typically, $N$ and the grid spacing $\Delta$ are set to be 50 and $|q_s|/40$.
To check the convergence, we varied the values of ($N$, $|q_s|/\Delta$)
as $(60, 48)$, $(50, 30)$, and $(40, 32)$.
It is found that the numerical results converge at the second order
and the error is within 0.1\% with $(50, 40)$. 

Following Fishbone \cite{Fishbone}, 
we often refer to a nondimensional parameter defined by 
\beq
\zeta \equiv {\Omega^2 \over \pi \rho_c}={M \over \pi \rho_c r^3}. 
\eeq
The order of magnitude of this parameter is
\beq
\zeta \sim {M R^3 \over m r^3} ={MR/r^3 \over m/R^2}. 
\eeq
Thus, it denotes the ratio of the tidal force by a black hole
to the self-gravity of a star. In particular, we focus on the value of 
$\zeta$ at the Roche limit, which is denoted by $\zeta_{\rm crit}$
in the following. $\zeta_{\rm crit}$ is a function of $r/M$
and depends on the equations of state and the black hole spin. 
We can also compute the minimum allowed mass of a star that can escape
tidal disruption for a given radius. The critical mass ratio associated
with such minimum mass is defined by $Q_{\rm crit}=(m/M)_{\rm minimum}$.
$Q_{\rm crit}$ is also a function of $r$ and depends on the
equations of state and the black hole spin. 
A star will be tidally disrupted for $Q < Q_{\rm crit}$. 

Numerical computations for determining $\zeta_{\rm crit}$ and
$Q_{\rm crit}$ are performed approximately fixing the value of
an averaged stellar radius for a given set of $r$, $a$, and $n$. 
The stellar radius of a spherical polytrope $R_0$ is written by \cite{ST} 
\beqn
R_0=\biggl[ {(n+1)\kappa \rho_c^{(1-n)/n} \over 4\pi} \biggr]^{1/2}\xi_1
=\biggl({h_c \over 4\pi\rho_c}\biggr)^{1/2} \xi_1, 
\eeqn
where $h_c=\kappa(n+1)\rho_c$ and
$\xi_1$ denotes the Lane-Emden coordinate at the stellar surface, 
which is $2.75270$, $\pi$, and 3.65375 for 
$n=0.5$, 1, and 1.5, respectively. 
Note that for $n=1$, fixing the value of $\kappa$ is equivalent to
fixing the value of $R_0$.
For other values of $n$, $\kappa$ varies 
along a sequence of a fixed value of $R_0$. 

To determine the Roche limit for a given set of
$r$, $a$, and $n$, we compute a sequence of solutions by varying 
$\rho_{c}$ from a large value to a small value 
until an inner edge of the star forms a cusp on the black hole side.
A configuration with such a cusp can be identified as the Roche limit, i.e., 
the self-gravity of the star is small enough to form 
a saddle point of the total gravitational potential at a stellar surface. 
Specifically, the Roche limit is determined monitoring the
following quantity at $(q^1, q^2, q^3)=(q_s, 0, 0)$: 
\beqn
H \equiv \Omega^2 (q_s-q_g)
- {\pa (\bar\phi + \bar\phi_{\rm tidal}) \over \pa q^1}. \label{eq187}
\eeqn
Here, $H$ is proportional to $\pa h/\pa q^1$. Thus, 
the value of $H$ for $q_s <0$ is positive for stable stars, 
and becomes zero for the marginally stable configuration against
tidal disruption. 

In this paper, we are in particular interested in 
making approximate models of binaries composed
of a stellar-mass black hole of mass 
$M \sim 3$--$30 M_{\odot}$ and a neutron star of mass $m \sim 1.4M_{\odot}$.
It is appropriate to assume that 
the radius of neutron stars is between 10 and 15 km. Then, $R_0$
in the present units should be between $\sim 3M$ and $\sim 0.2M$. 
In the following calculation, we adopt the values of $R_0/M$ as 
0.5, 1, and 2. For consistency in the framework of this paper, 
(i) $R_0$ should be much smaller than $r$, 
(ii) $m$ should be much smaller than $M$, and 
(iii) $r$ should be larger than the orbital radius of 
an ISCO (hereafter $r_{\rm ISCO}$) around the black hole. Thus, 
the computations are performed only for $m < M$ and 
for $r \geq r_{\rm ISCO}$. 
It would not be quantitatively appropriate to model 
a neutron star by Newtonian gravity since neutron stars are
compact with $m/R_0 \sim 0.2$ and thus general relativistic objects.
The following numerical results could contain a systematic error of 
magnitude $\sim m/R_0$. However, the quantitative importance 
of the higher-order terms as well as
the general relativistic effects in the tidal potential can be clarified
even if we neglect the general relativistic corrections of neutron stars. 
Also, the present study will be useful for qualitatively 
clarifying the dependence of the Roche limit 
on the equations of state.

\subsection{Tidal potential}

\begin{figure}[tb]
\vspace*{-6mm}
\begin{center}
\epsfxsize=2.85in
\leavevmode
(a)\epsffile{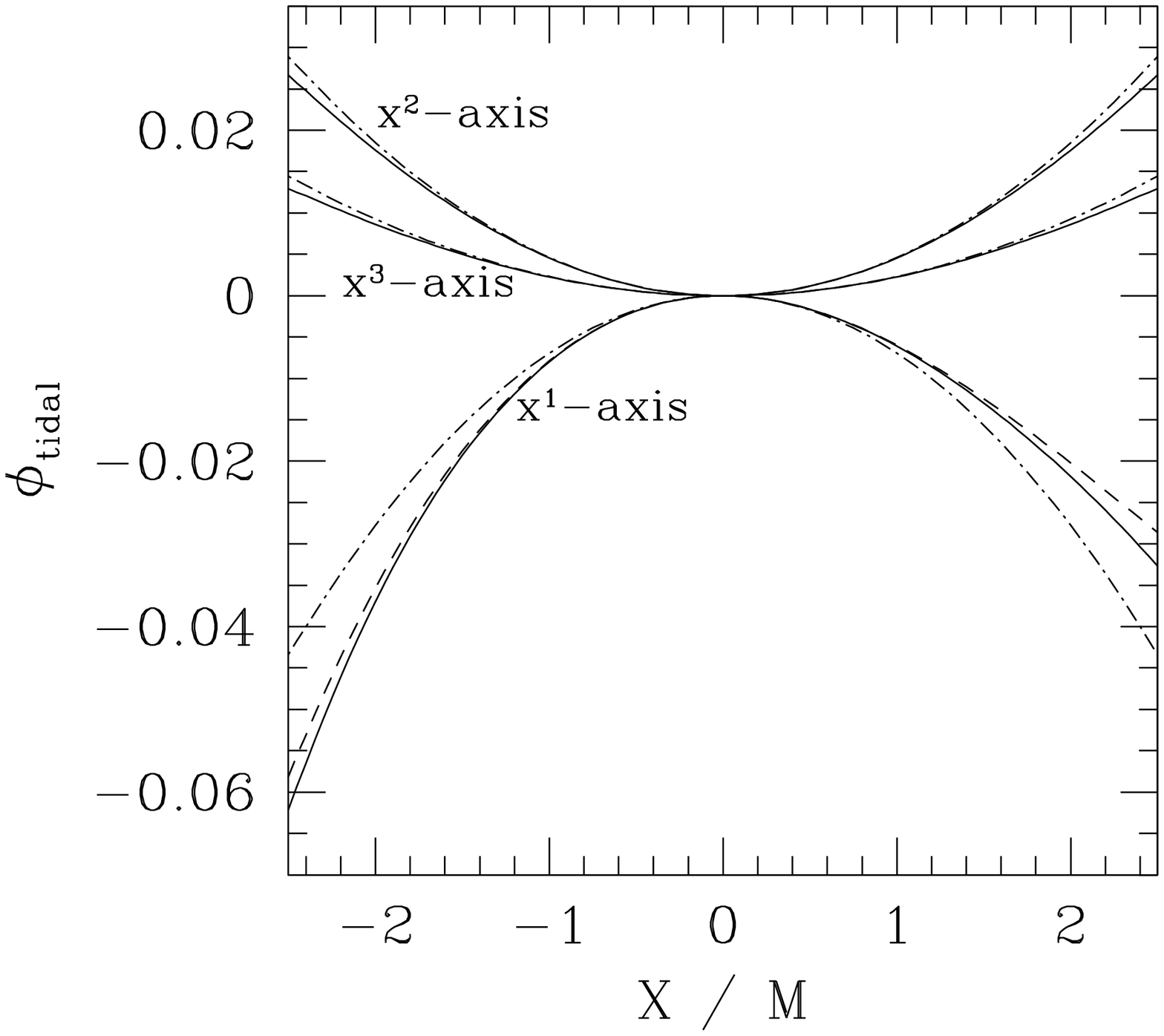}
\epsfxsize=2.85in
\leavevmode
~~(b)\epsffile{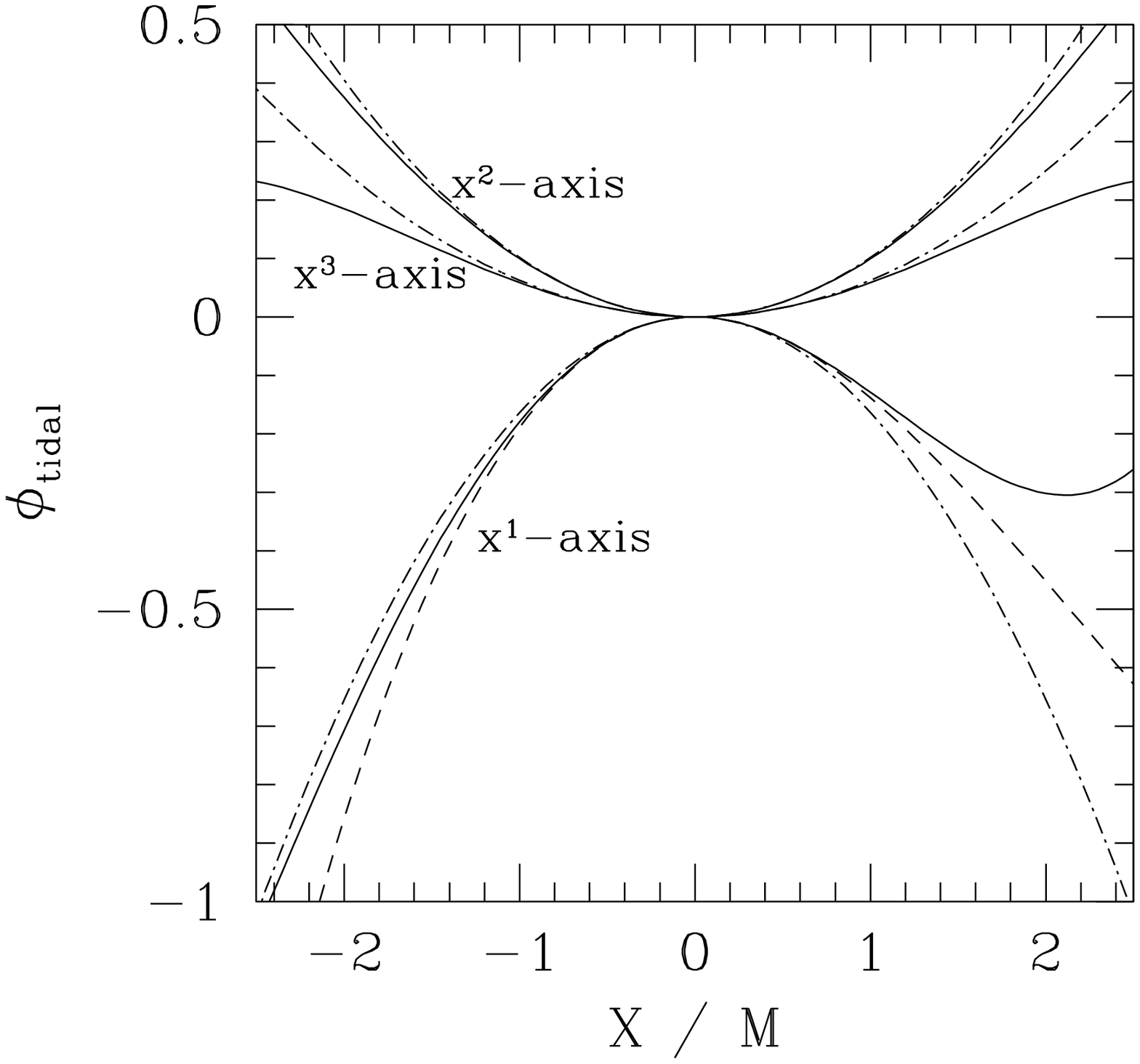} \\
\epsfxsize=2.85in
\leavevmode
(c)\epsffile{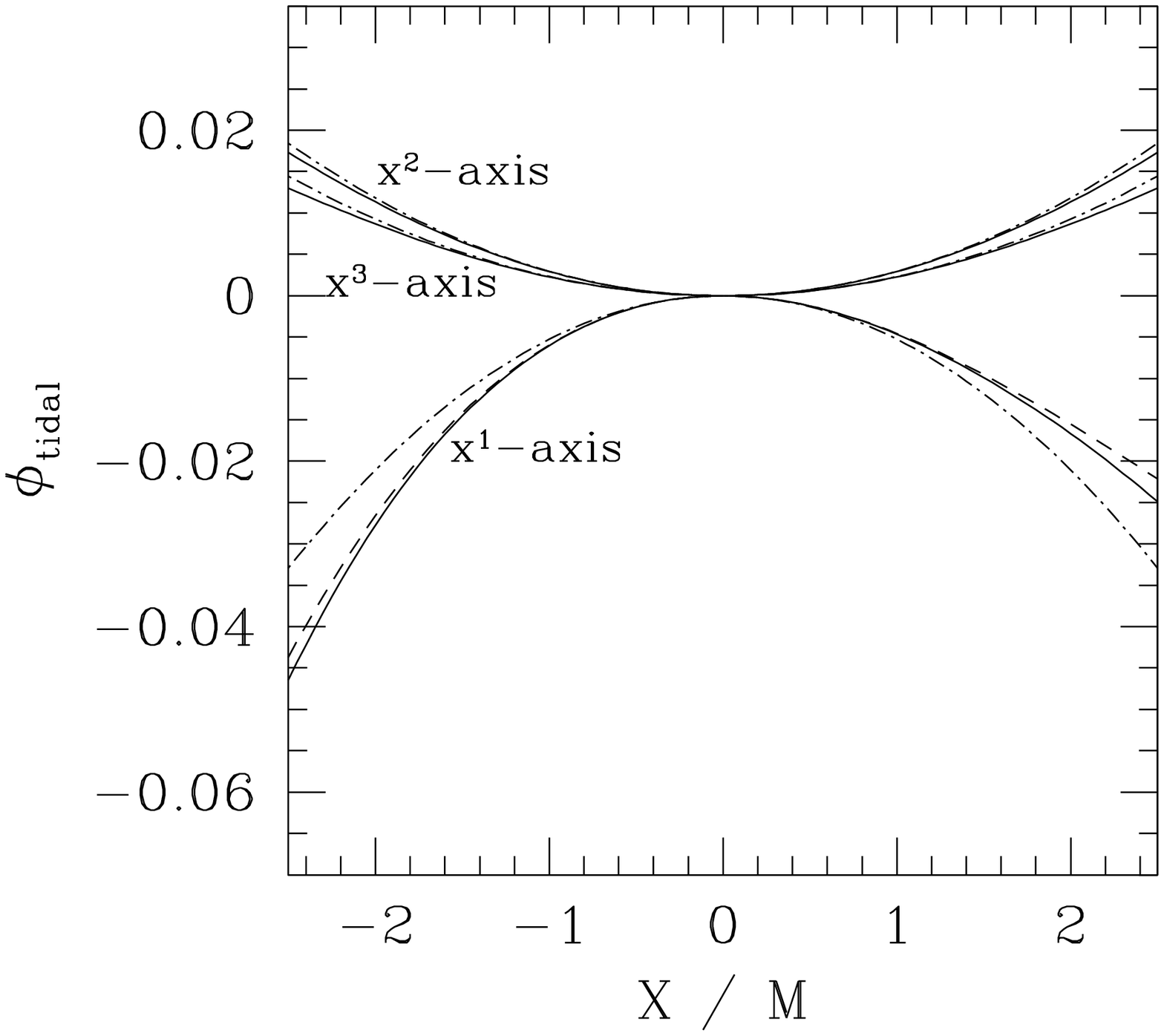}
\epsfxsize=2.85in
\leavevmode
(d)\epsffile{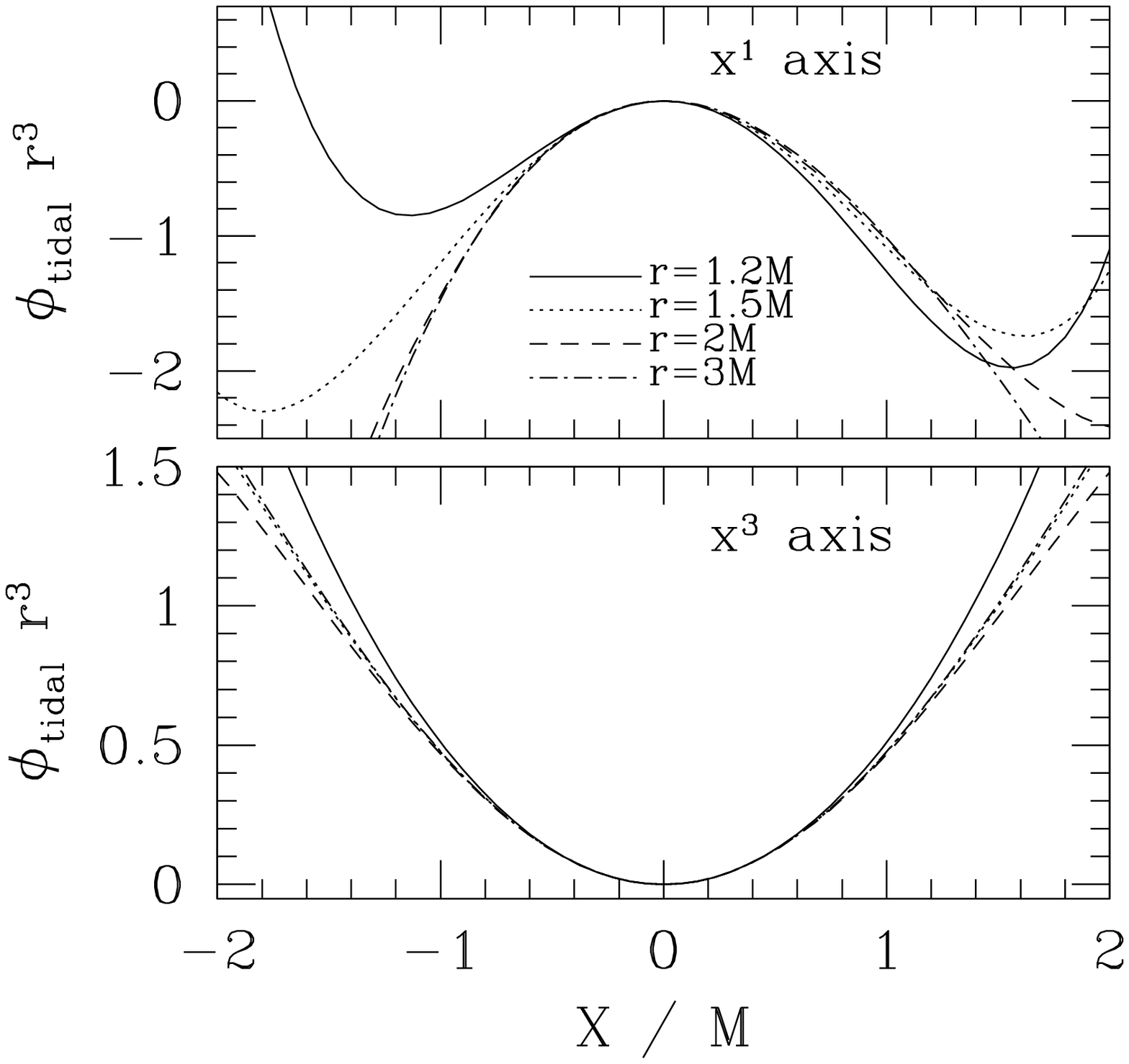}
\end{center}
\vspace*{-2mm}
\caption{The profiles of the tidal potential $\phi_{\rm tidal}$ 
along $x^1$, $x^2$, and $x^3$ axes
(a) for $r=6M$ and $a=0$, 
(b) for $r=2M$ and $a=M$,
(c) for $r=6M$ and $a=M$,
and (d) for $r/M=1.2$, 1.5, 2, and 3 and $a=M$. 
The solid, dashed, and dotted-dashed curves
in panels (a)--(c) denote $\phi_{\rm tidal}$ 
in the fourth-, third-, and second-order approximations,
respectively. In panel (d), the tidal potential in the fourth-order
approximation along $x^1$ and $x^3$ axes are shown. 
The units of $G=M=1$ are adopted in these figures. 
\label{FIG1} }
\end{figure}

In Fig. 1, we display the tidal potential $\phi_{\rm tidal}$ along 
$\tilde x^1$, $\tilde x^2$, and $\tilde x^3$ axes. 
Figures 1(a)--(c) show $\phi_{\rm tidal}$ 
for $(r/M, a/M)=(6, 0)$, $(2, 1)$, and $(6, 1)$, respectively.
Figure 1(d) is $\phi_{\rm tidal}$ in the fourth-order approximation 
for $a=M$ and $r/M=1.2$, 1.5, 2, and 3. 
To clarify the convergence with increasing the order, 
$\phi_{\rm tidal}$ in the second- (dotted-dashed curves),
third- (dashed curves), and fourth-order (solid curves) approximations are
shown together in Figs. 1(a)--(c). $\phi_{\rm tidal}$ in the second- and
third-order approximations are identical along $\tilde x^2$ and
$\tilde x^3$ axes. Thus, only the second-order results are presented.
At the second order, $\phi_{\rm tidal}$ is symmetric with respect to
the origin along all the axial directions. The third-order terms 
induce an asymmetry in the $\tilde x^1$ direction.

The difference in the magnitude of $\phi_{\rm tidal}$ 
is about 20--30\% between the second- and fourth-order
approximations for $|\tilde x^i| \sim R_0 \sim M$ near the ISCOs.
For $|\tilde x^i|\alt R_0 \sim M$, the difference in the magnitude of
the third- and fourth-order tidal potentials is $\alt 1\%$, and hence,
the convergence is approximately achieved. $\phi_{\rm tidal}$ 
in the third- and fourth-order approximations are larger than that
in the second-order approximation in the black hole side ($\tilde x^1 < 0$).
This indicates that a star becomes prone to 
tidal disruption in the higher-order approximations. 

Comparing Figs. 1(a) and (c) for which the results with the
same value of $r$ are shown, it is found that the spin effect reduces
the magnitude of $\phi_{\rm tidal}$ along $\tilde x^1$ and $\tilde x^2$
axes. This illustrates the property 
that for the larger value of $a/M$, the tidal effect
is weaker for a given value of $r/M$. 
Figure 1(d) shows that the magnitude of $r^3\phi_{\rm tidal}$
in the black hole side is smaller for the
smaller values of $r$. This suggests that a star is less prone to
the tidal disruption for an extremely high value of $a \approx M$
near the ISCOs. Such a special feature is not outstanding for
other values of $a$. 

\subsection{Roche limits}

\begin{figure}[tb]
\vspace{-6mm}
\begin{center}
\epsfxsize=2.85in
\leavevmode
(a)\epsffile{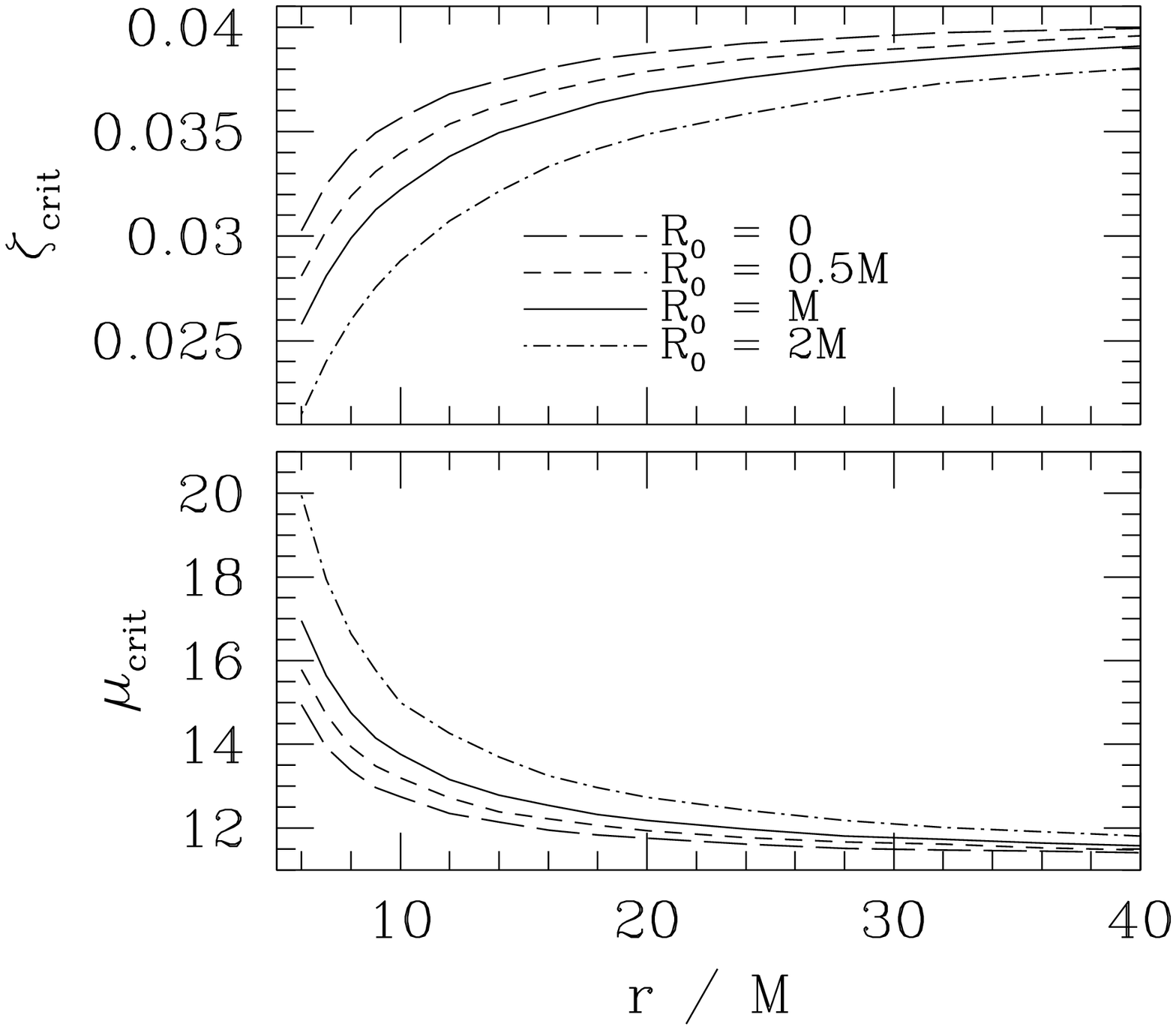}
\epsfxsize=2.85in
\leavevmode
~~(b)\epsffile{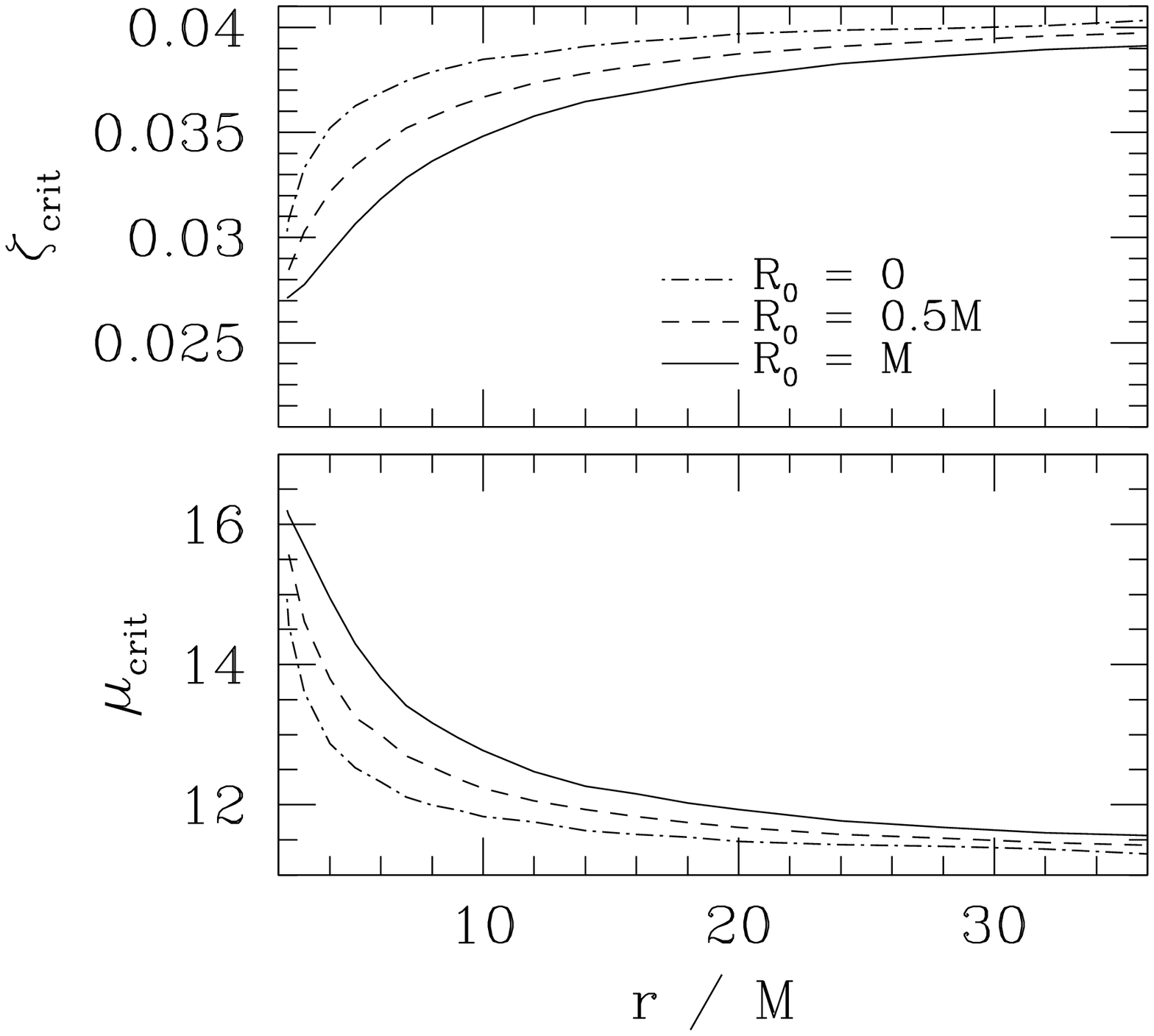} \\
\epsfxsize=2.85in
\leavevmode
(c)\epsffile{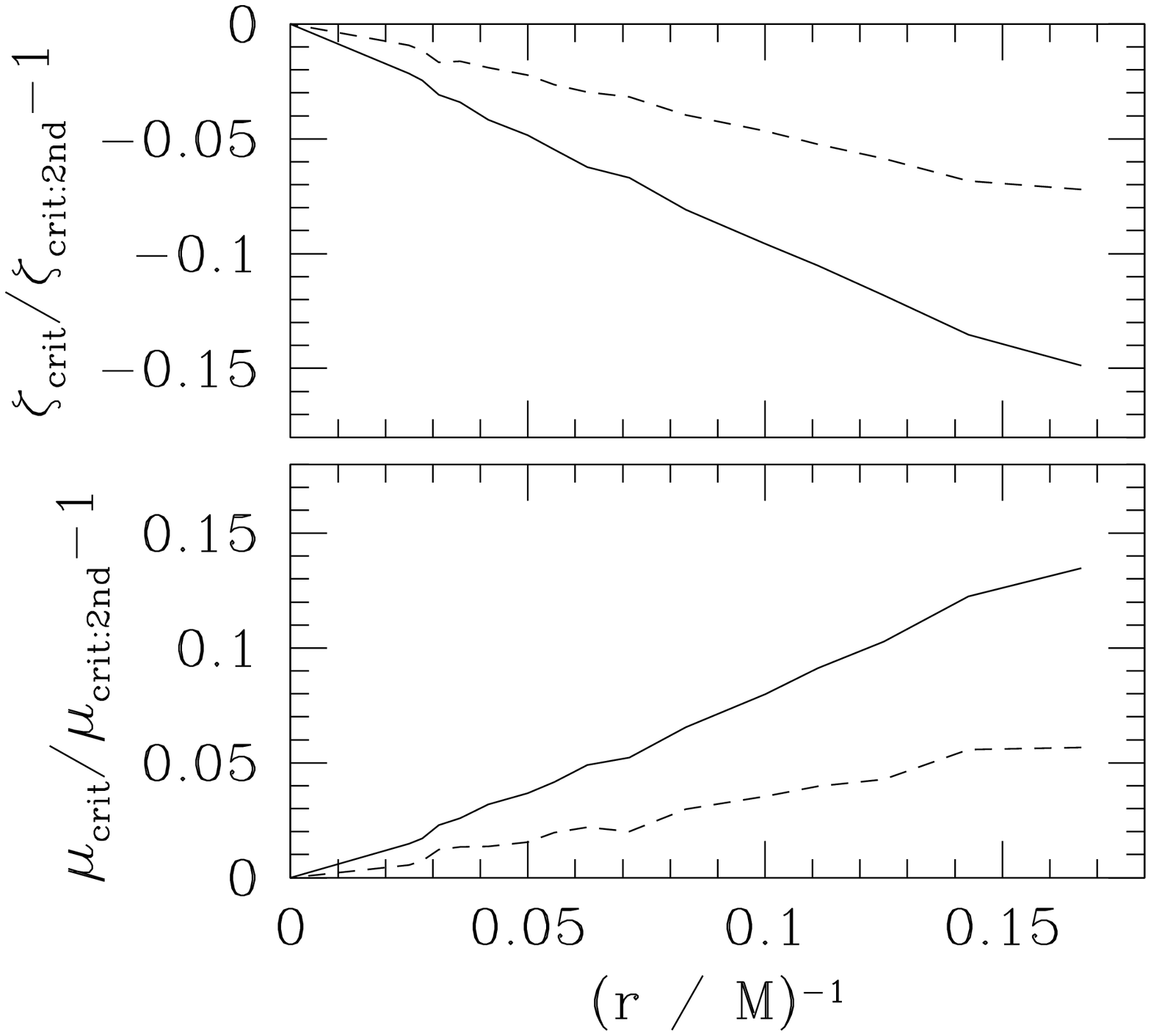}
\epsfxsize=2.85in
\leavevmode
~~(d)\epsffile{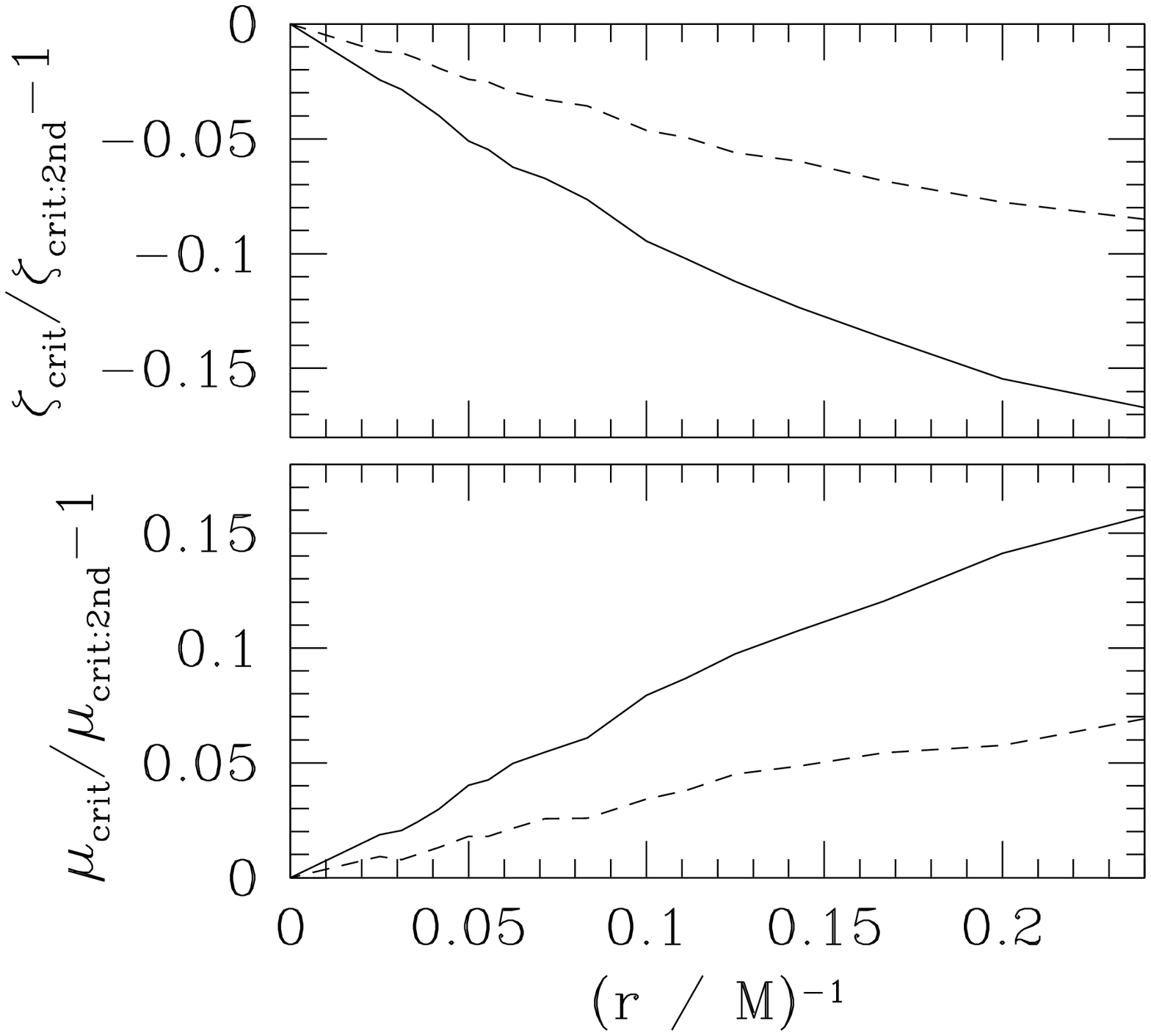} \\
\end{center}
\vspace{-2mm}
\caption{(a) $\zeta_{\rm crit}$ and $\mu_{\rm crit}$ as functions of $r$ 
for $R_0/M=0$, 0.5, 1, and 2, and for $a=0$. 
(b) the same as (a) but for $R_0/M=0$, 0.5, and 1 and for $a=0.9M$.
(c) $\zeta_{\rm crit}/\zeta_{\rm crit:2nd}-1$ as a function of
$M/r$ for $R_0/M=$0.5 (dashed curve) and 1 (solid curve), and for $a=0$.
(d) the same as (c) but for $a=0.9M$.
Here, $n=1$. $\zeta_{\rm crit:2nd}$ is $\zeta_{\rm crit}$
with $R_0=0$ and equal to that in the second-order tidal approximation.
We note that a star with $\zeta > \zeta_{\rm crit}$ 
(or $\mu < \mu_{\rm crit}$) for a given value of $r/M$
is unstable against tidal disruption. 
\label{FIG2} }
\end{figure}

\begin{figure}[tbh]
\vspace*{-6mm}
\begin{center}
\epsfxsize=2.85in
\leavevmode
(a)\epsffile{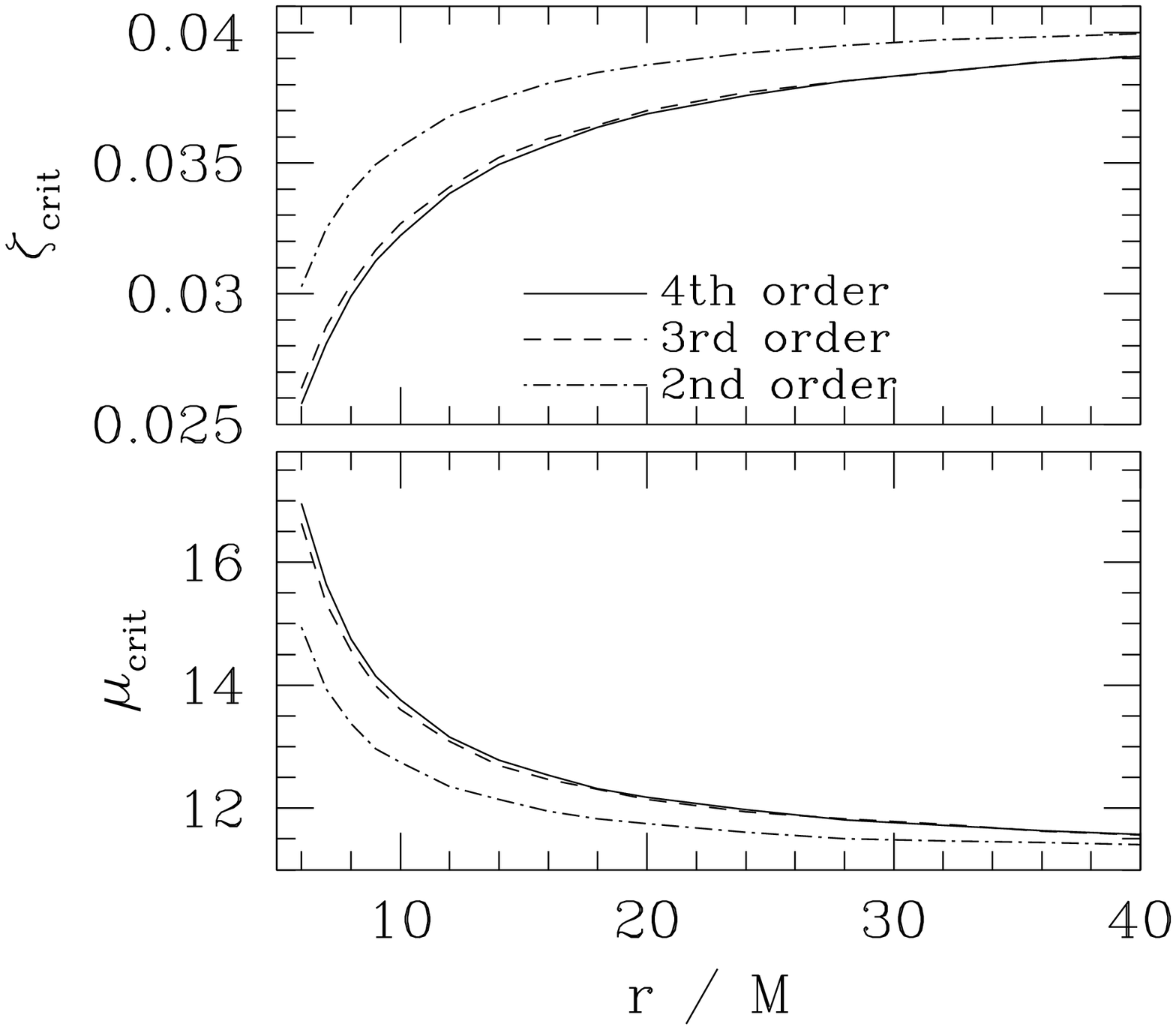}
\epsfxsize=2.85in
\leavevmode
~~(b)\epsffile{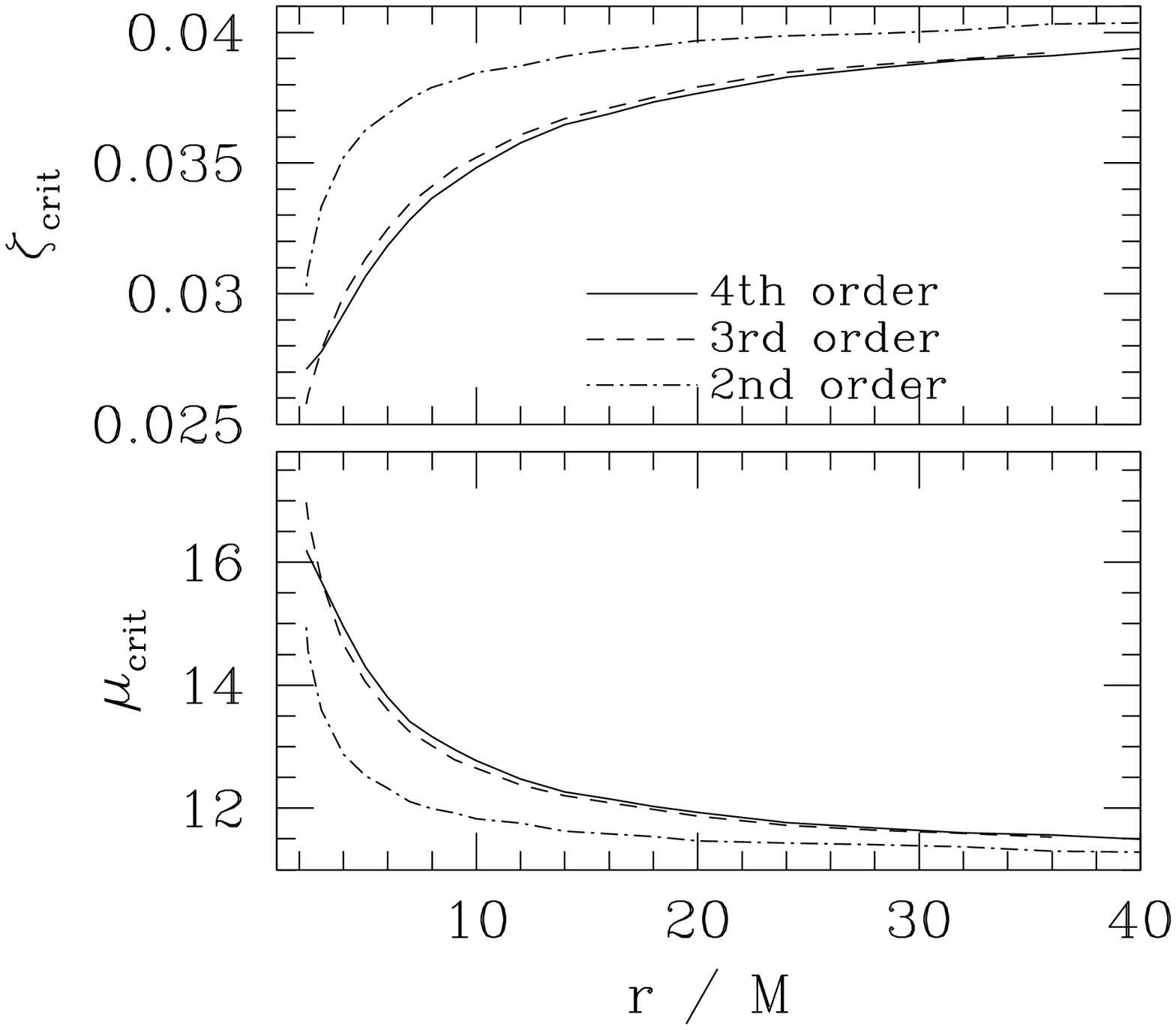}
\end{center}
\vspace*{-2mm}
\caption{$\zeta_{\rm crit}$ and $\mu_{\rm crit}$ as functions of $r$ 
(a) for $a=0$ and (b) for $a=0.9M$.
For both cases, $n=1$ and $R_0=M$. 
The solid, dashed, and dotted-dashed curves denote the results 
in the fourth-, third-, and second-order tidal approximations,
respectively. 
\label{FIG3} }
\end{figure}

\begin{figure}[p]
\vspace*{-6mm}
\begin{center}
\epsfxsize=2.85in
\leavevmode
(a)\epsffile{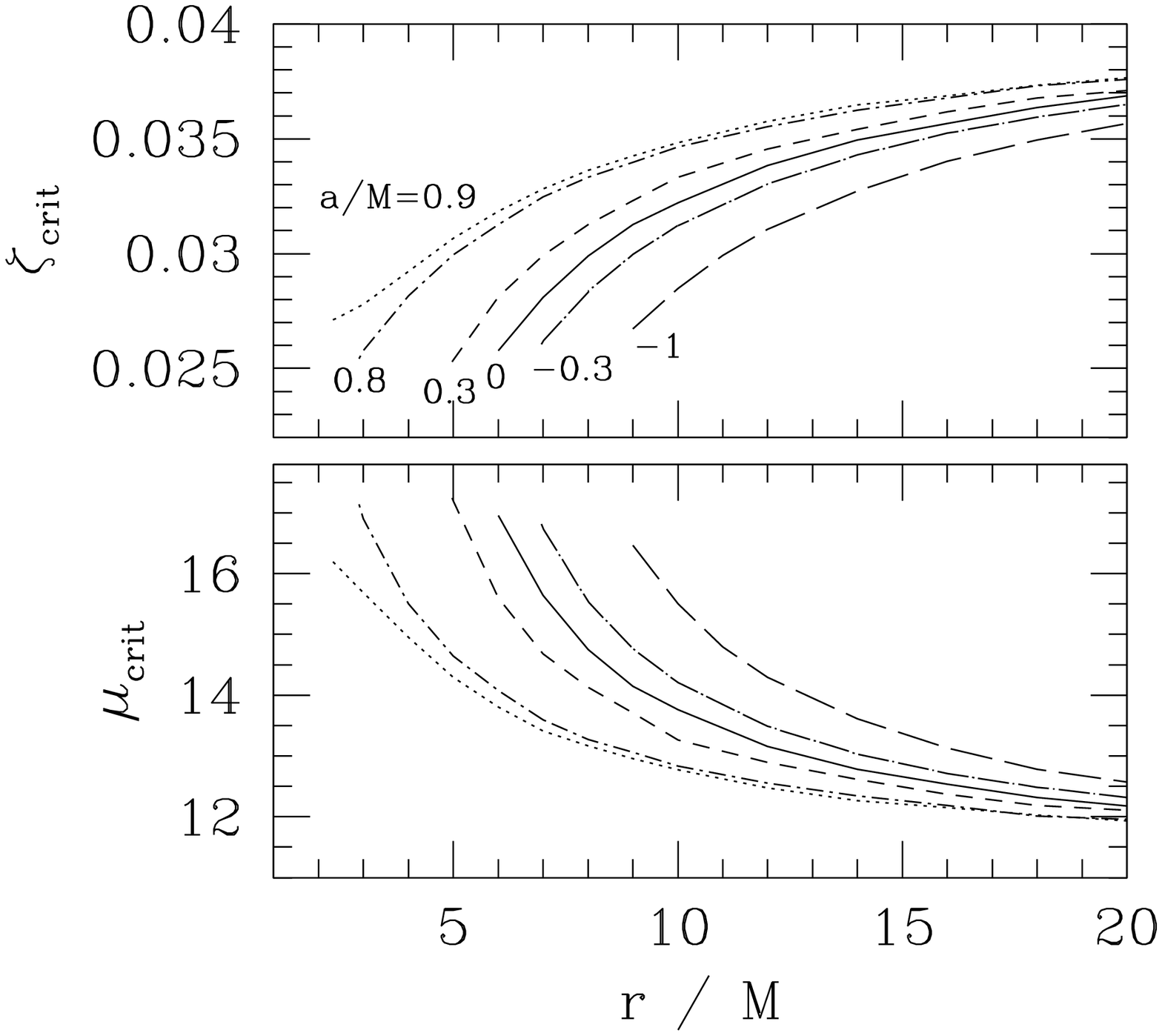}
\epsfxsize=2.85in
\leavevmode
~~(b)\epsffile{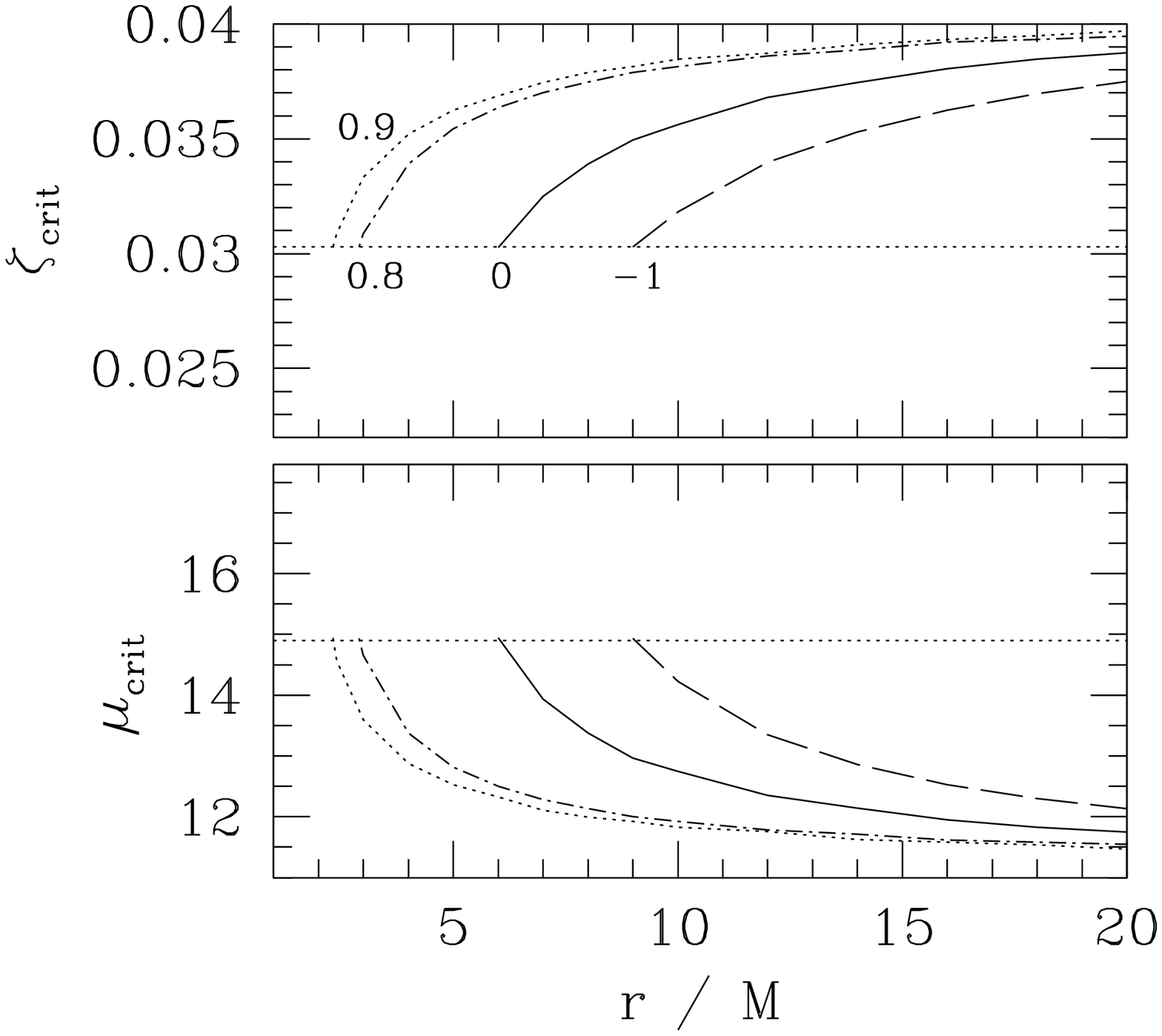}
\end{center}
\vspace*{-2mm}
\caption{$\zeta_{\rm crit}$ and $\mu_{\rm crit}$ as functions of $r$ 
for $n=1$ and various values of $a/M$. 
(a) The dotted, dotted-dashed, dashed, solid, dotted-long-dashed,
and long-dashed curves
denote the results in the fourth-order tidal approximation
with $R_0=M$ for $a/M=0.9$, 0.8, 0.3, 0, $-0.3$, and $-1$, respectively.
(b) The dotted, dotted-dashed, solid, and long-dashed curves
denote the results in the second-order tidal approximation
for $a/M=0.9$, 0.8, 0, and $-1$, respectively.
Note that in the second-order tidal approximation, 
$\zeta_{\rm crit}$ and $\mu_{\rm crit}$ at ISCOs are 
about 0.0303 and 14.9 irrespective of the value of $a$ 
(dotted horizontal lines). 
\label{FIG4} }
\end{figure}

\begin{figure}[p]
\vspace*{-6mm}
\begin{center}
\epsfxsize=2.4in
\leavevmode
(a)\epsffile{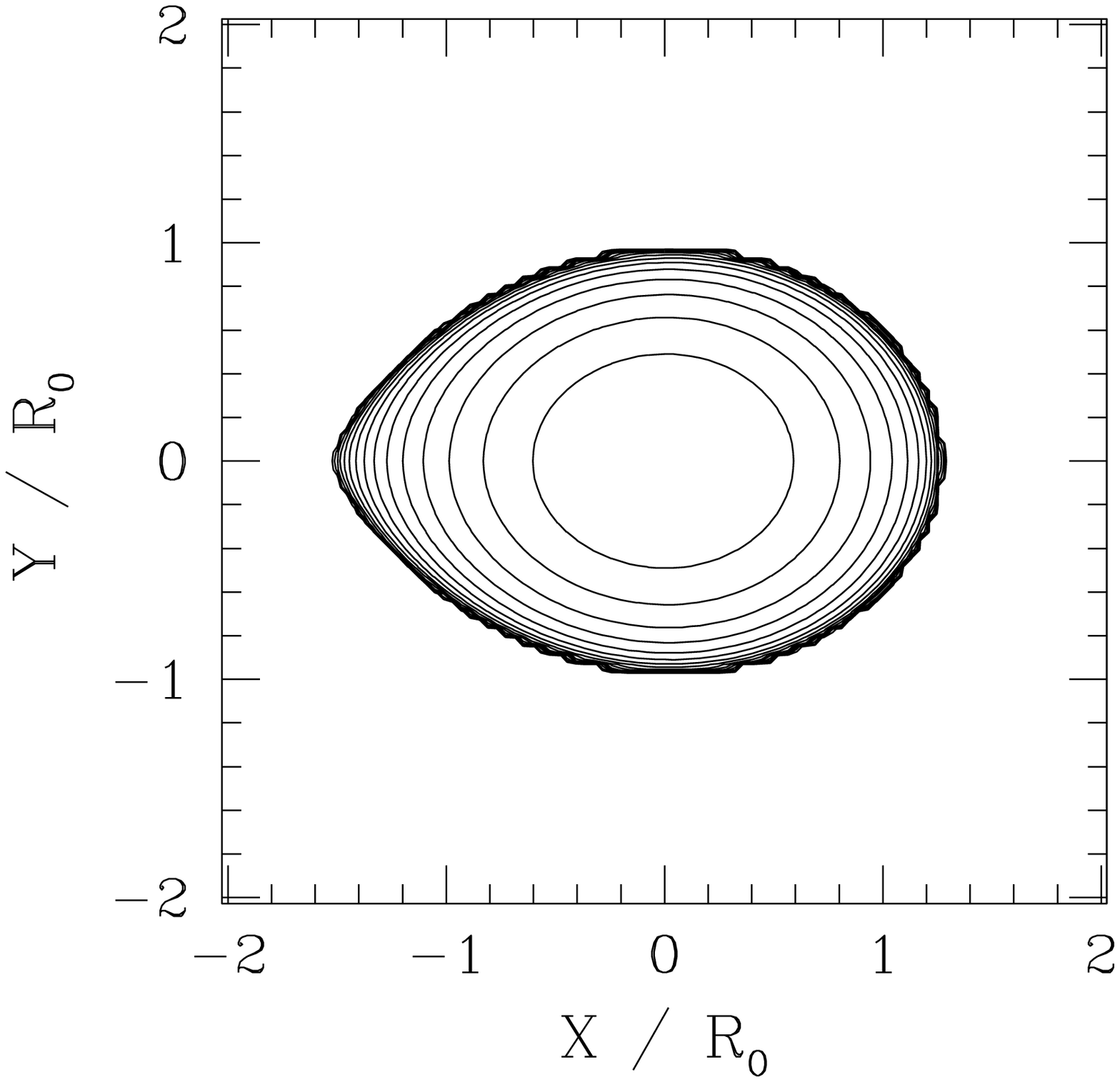}
\epsfxsize=2.4in
\leavevmode
~~(b)\epsffile{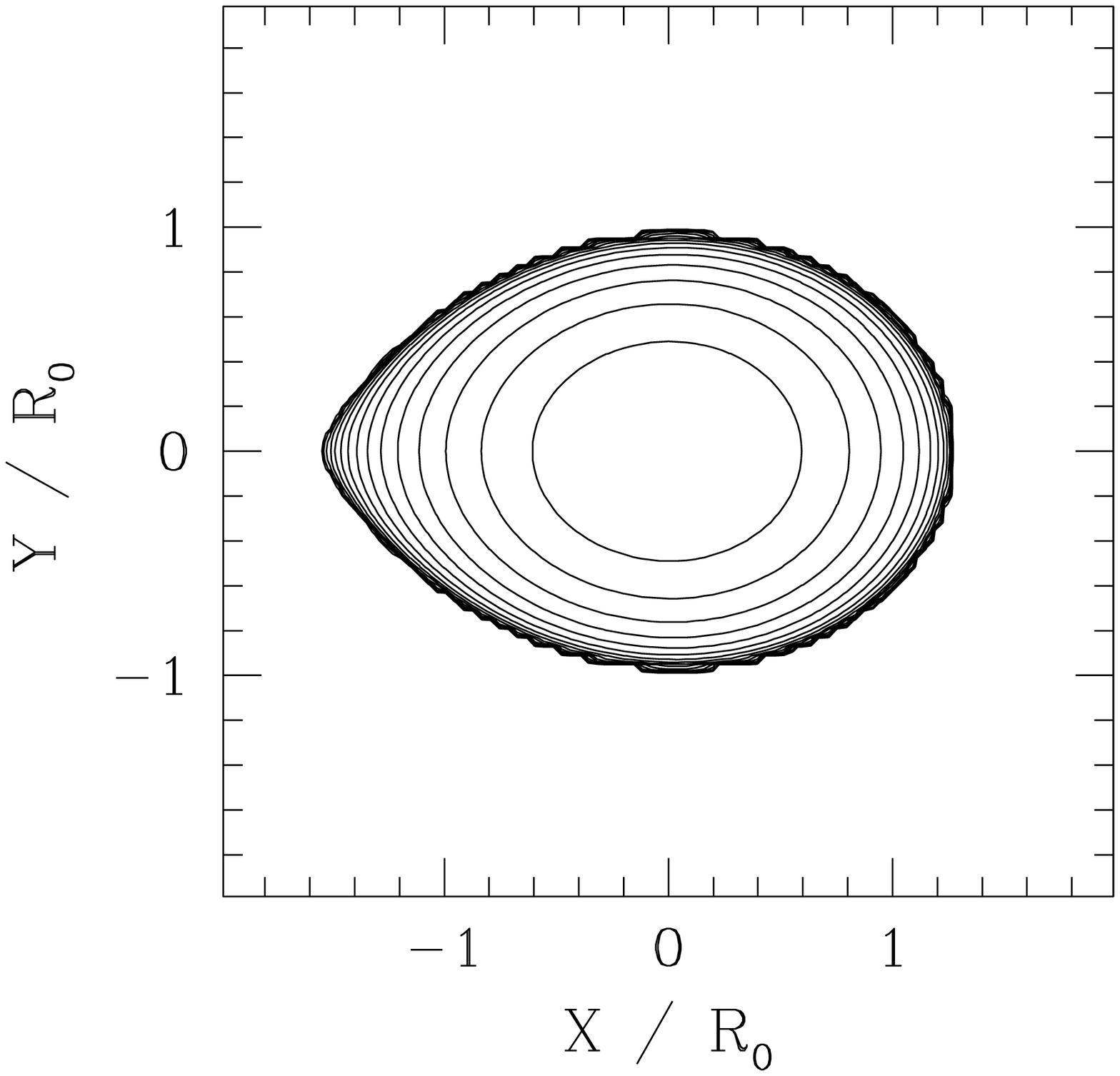} \\
\epsfxsize=2.4in
\leavevmode
(c)\epsffile{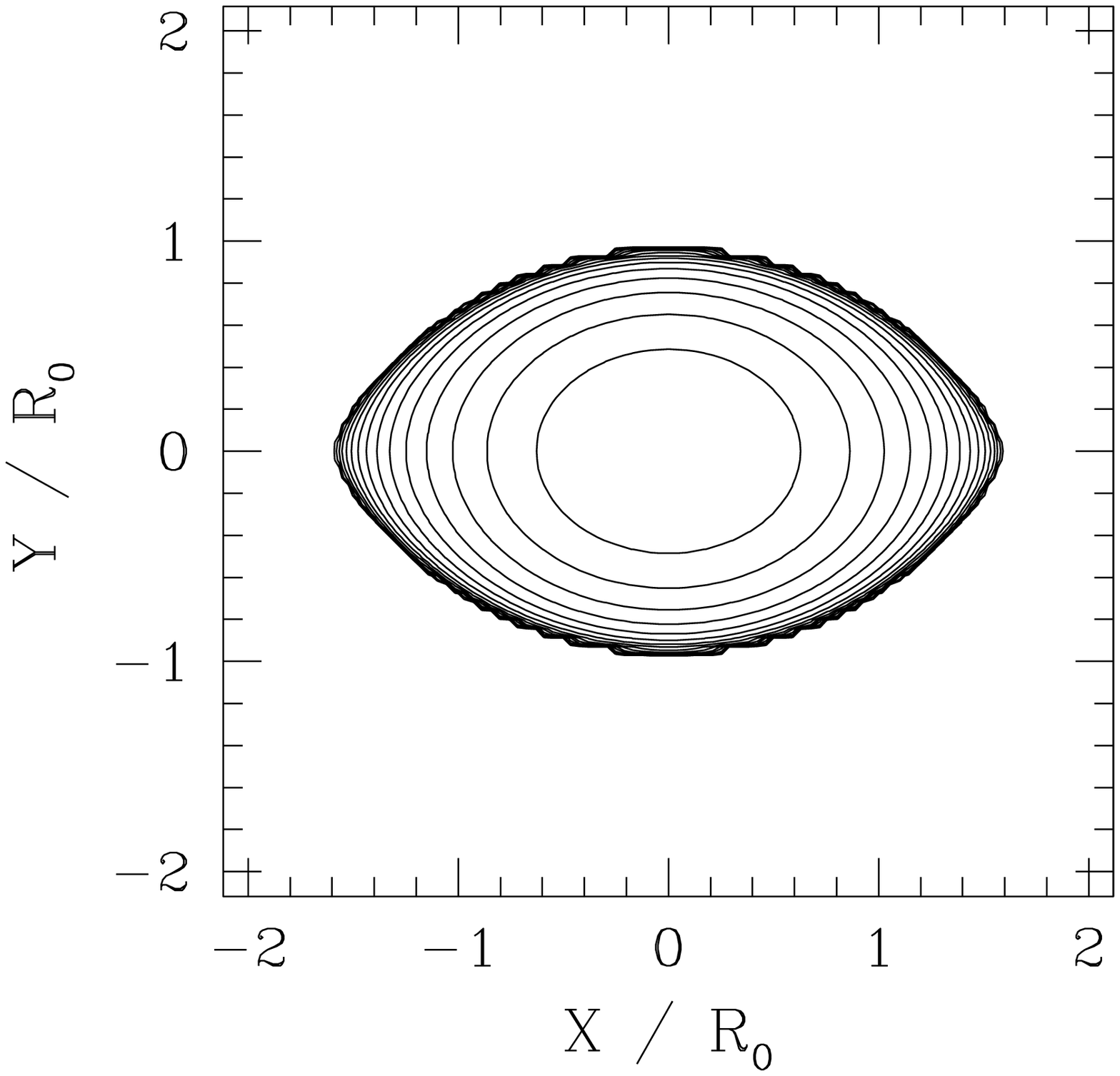}
\epsfxsize=2.4in
\leavevmode
~~(d)\epsffile{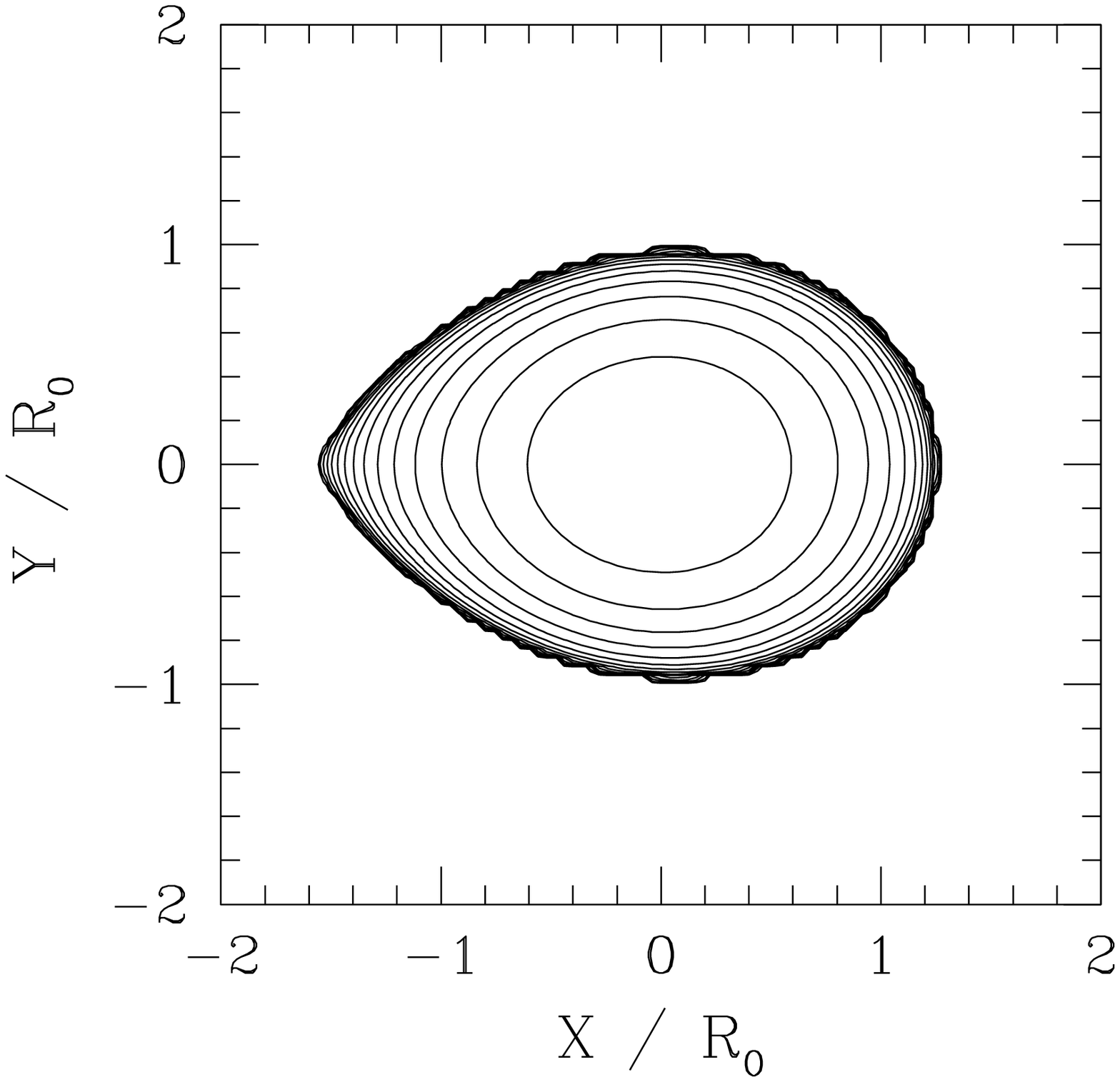}
\end{center}
\vspace*{-2mm}
\caption{Density contour curves at Roche limits
in the (a) fourth-, (b) third-, and (c) second-order tidal approximations 
for $r=6M$, $R_0=M$, $a=0$, and $n=1$.
(d) density contour curves at a Roche limit
in the fourth-order approximation for $r=2.4M$, $R_0=M$, $a=0.9M$,
and $n=1$. The contour curves are drawn for $\rho/\rho_c=10^{-0.2j}$ 
for $j=0,1,2,\cdots,15$. 
$X$ and $Y$ denote $\tilde x^1$ and $\tilde x^3$, respectively. 
\label{FIG5} }
\end{figure}

In Fig. 2, we show $\zeta_{\rm crit}$ and
$\mu_{\rm crit} \equiv Q_{\rm crit}(r/R_0)^3$ as functions of $r/M$
(a) for $n=1$, $a=0$, and $R_0/M=0$, 0.5, 1, and 2, and 
(b) for $n=1$, $a=0.9M$, and $R_0/M=0$, 0.5 and 1.
Stars with $\zeta < \zeta_{\rm crit}$ and 
$\mu > \mu_{\rm crit}$ for a given value of $r/M$ 
are stable against tidal disruption. 
The computations for $R_0 \not=0$ were performed taking into account the
tidal potential up to fourth order. 
Note that the values of $\zeta_{\rm crit}$ and $\mu_{\rm crit}$
in the third- and fourth-order approximations with $R_0=0$
are equal to those in the second-order approximation
with an arbitrary value of $R_0$.
Thus, ``$R_0=0$'' implies that the computations were performed
in the second-order tidal approximation.
With increasing $R_0/M$, $\zeta_{\rm crit}$ ($\mu_{\rm crit}$) at a
given orbital radius decreases (increases). 
This illustrates that the tidal force in the fourth-order approximation
is stronger than in the second-order one. 

As in previous works \cite{Fishbone,shibata} carried out by the 
second-order tidal approximation, $\zeta_{\rm crit}$ ($\mu_{\rm crit}$)
decreases (increases) with decreasing $r/M$. 
This implies that with the decrease of $r/M$, the tidal force is enhanced. 
At $r=r_{\rm ISCO}$, 
$\zeta_{\rm crit}$ and $\mu_{\rm crit}$ become minimum and maximum for
given values of $a$ and $R_0$. We denote them by 
$\zeta_{\rm crit:min}$ and $\mu_{\rm crit:max}$.
On the other hand, for $r \rightarrow \infty$, 
$\zeta_{\rm crit}$ and $\mu_{\rm crit}$ become maximum and minimum for
given values of $a$ and $R_0$, and they agree with 
the values in the Newtonian limit \cite{IS}. 
For given values of $a$ and $R_0$, a star at an ISCO 
has to be tidally disrupted whenever 
$\zeta > \zeta_{\rm crit} \geq \zeta_{\rm crit:min}$ or 
$\mu < \mu_{\rm crit} \leq \mu_{\rm crit:max}$. 
In other words, for $\zeta < \zeta_{\rm crit:min}$
and $\mu > \mu_{\rm crit:max}$, the star in a circular orbit 
is not tidally disrupted outside the ISCOs. 

In Figs. 2(c) and (d), $\zeta_{\rm crit}/\zeta_{\rm crit:2nd}-1$ 
and $\mu_{\rm crit}/\mu_{\rm crit:2nd}-1$
as functions of $M/r$ are shown for the same parameters as
in Figs. 2(a) and (b), respectively. 
Here, $\zeta_{\rm crit:2nd}$ and $\mu_{\rm crit:2nd}$ denote
$\zeta_{\rm crit}$ and $\mu_{\rm crit}$ determined in the second-order
tidal approximation. We note that in the second-order tidal 
approximation, $\zeta_{\rm crit}$ and $\mu_{\rm crit}$ 
are independent of $R_0$. It is found that for $M/r(\alt 0.1)$, 
$\zeta_{\rm crit}/\zeta_{\rm crit:2nd}-1$ and
$\mu_{\rm crit}/\mu_{\rm crit:2nd}-1$ are approximately linear
functions of $M/r$. Thus, for the large values of $r/M$, the third-order
tidal potential dominantly modifies the values of $\zeta_{\rm crit}$ 
and $\mu_{\rm crit}$. The numerical results show that the following
relations approximately hold: 
\beqn
&&\zeta_{\rm crit} \approx
\zeta_{\rm crit:2nd}\biggl(1 - C_1 {R_0 \over r}\biggr), \label{fit0}
\\
&&\mu_{\rm crit} \approx \mu_{\rm crit:2nd}
\biggl(1 + C_2 {R_0 \over r}\biggr),
\label{fit}
\eeqn
Here, the coefficients of the
correction factors $C_1$ and $C_2$ are $\approx 0.95$ and $\approx 0.80$ 
irrespective of the value of $a$ for $n=1$. The error is within 
$\sim 0.05$ for both coefficients. 
For $M/r \agt 0.1$, the fourth-order tidal potential becomes important. 
However, Eqs. (\ref{fit0}) and (\ref{fit}) approximately hold even for
$r \agt 6M$. For $r \alt 6M$, the correction factors in these fitting
formulae give overestimated values.

To clarify the importance of the higher-order terms of $R_0/r$
in the tidal potential, in Fig. 3,
we show $\zeta_{\rm crit}$ and $\mu_{\rm crit}$ as functions of $r/M$
(a) for $n=1$, $a=0$, and $R_0=M$, and 
(b) for $n=1$, $a=0.9M$, and $R_0=M$ 
with the second-, third-, and fourth-order tidal potentials 
(dotted, dashed, and solid curves, respectively).
With the higher-order tidal potential,
the value of $\zeta_{\rm crit}$ decreases. Namely, 
the minimum allowed value of $\rho_c$ increases. This also illustrates that
in the second-order approximations, the tidal force is underestimated. 
For $r/M(=r/R_0)=6~(10)$, the values of $\zeta_{\rm crit}$ in the
third- and fourth-order approximations are about 13 (8) and 15 (10)\%
smaller than that in the second-order one.
This shows that the standard tidal approximation taken up to the
second-order term provides a result with the error 
of $\sim 10(R_0/M)\%$ for close orbits with $R_0/r \agt 0.1$. 
However, the convergence appears to be very good if the third- and 
fourth-order terms are included for $R_0 \alt M$.
In other words, the fourth-order term plays a quantitatively minor role. 
All these results agree qualitative with a Newtonian analysis \cite{IS}. 

\begin{table}[b]
\begin{center}
\begin{tabular}{cccccccc} \hline
$n$ & $\zeta_{\rm crit:2nd}$(ISCO) & $\mu_{\rm crit:2nd}$(ISCO) 
& $\zeta_{\rm crit}$($r\rightarrow \infty$) 
& $\mu_{\rm crit}$($r\rightarrow \infty$) 
& $M_c(M_{\odot})$ & $C_1$ & $C_2$ \\ \hline
0   & 0.0664 & 20.1 & 0.0901 & 14.8 & 4.54 & --- & --- \\ 
0.5 & 0.0480 & 19.4 & 0.0651 & 14.5 & 4.45 & 0.7  & 0 \\ 
1.0 & 0.0303 & 14.9 & 0.0411 & 11.1 & 3.90 & 0.95 & 0.80 \\ 
1.5 & 0.0171 & 13.7 & 0.0232 & 10.1 & 3.74 & 0.97 & 0.97 \\
\end{tabular}
\caption{Values of $\zeta_{\rm crit:2nd}$, $\mu_{\rm crit:2nd}$, 
$M_c$, $C_1$, and $C_2$ for each value of $n$. 
The values for $n=0$ were determined by Fishbone [13]. 
$\zeta_{\rm crit:2nd}$ and $\mu_{\rm crit:2nd}$ 
do not depend on the value of $a$, and 
$C_1$ and $C_2$ depend very weakly on it. The value of $M_c$ 
shown here is that for $a=0$ and $r=r_{\rm ISCO}=6M$, and for $a\not=0$, 
$M_c(a)=M_c(a=0)(6M/r_{\rm ISCO})^{3/2}$. 
For the values of $C_1$ and $C_2$, the numerical error is
within 5\% for $n=1$ and 1.5, and $\sim 10\%$ for $n=0.5$. 
}
\end{center}
\end{table}

In Fig. 4(a), we show $\zeta_{\rm crit}$ and $\mu_{\rm crit}$
as functions of $r/M$ 
for $n=1$, $R_0=M$, and $a/M=0.9$, 0.8, 0.3, 0, $-0.3$, and $-1$ in 
the fourth-order tidal approximation. 
For comparison, the results in the second-order
approximation are shown in Fig. 4(b). 
These figures clarify effects of the black hole spin 
on the Roche limit.  For each curve of a given value of 
$a$, the minimum value of $r$ is equal to $r_{\rm ISCO}$. 
In the second-order tidal approximation, the values of 
$\zeta_{\rm crit}$ and $\mu_{\rm crit}$ at the ISCO are independent of $a$ 
\cite{Fishbone} and are about 0.0303 and 14.9, respectively (cf. Table II).
In the third- and fourth-order approximations, on the other hand,
these values depend slightly on the value of $a$.
As mentioned before, the value of $\zeta_{\rm crit}$ ($\mu_{\rm crit}$)
in the fourth-order tidal approximation 
is smaller (larger) than that of $\zeta_{\rm crit:2nd}$
($\mu_{\rm crit:2nd}$).
The magnitude of the difference between $\zeta_{\rm crit}$
and $\zeta_{\rm crit:2nd}$ is slightly smaller
for a large value of $a \agt 0.9M$ and $a <0$.
The effects of the third- and fourth-order
terms in the tidal potential are strongest for $0.3 \alt a/M \alt 0.8$. 

In Figs. 5(a)--(d), the density contour curves in the equatorial plane
($\tilde x^1$-$\tilde x^3$ plane) at the Roche limit 
are displayed. Figures 5(a)--(c) are those 
in the fourth-, third-, and second-order tidal approximations 
for $r=6M$, $R_0=M$, $a=0$, and $n=1$, and Fig. 5(d)
is in the fourth-order approximation 
for $r=2.4M$, $R_0=M$, $a=0.9M$, and $n=1$. 
In the second-order approximation, the 
density is symmetric with respect to the $\tilde x^1=0$ plane,
and hence, the asymmetry is induced by the third-order term. 
At the tidal disruption, $x_s \equiv p q_s \approx 1.6$, 1.55, and
$1.5 R_0$ in the second-, third-, and fourth-order approximations.
On the other hand, the length of the minor axes is $\sim R_0$
independent of the order of the approximation.
Thus, in the higher-order approximations, 
the ellipticity at the Roche limit is slightly smaller. 
Comparing Figs. 5(a) and (d), it is found that 
the density configurations near ISCOs for
different values of $a$ are very similar irrespective of $r$. 

\begin{figure}[tbh]
\vspace*{-6mm}
\begin{center}
\epsfxsize=2.85in
\leavevmode
(a)\epsffile{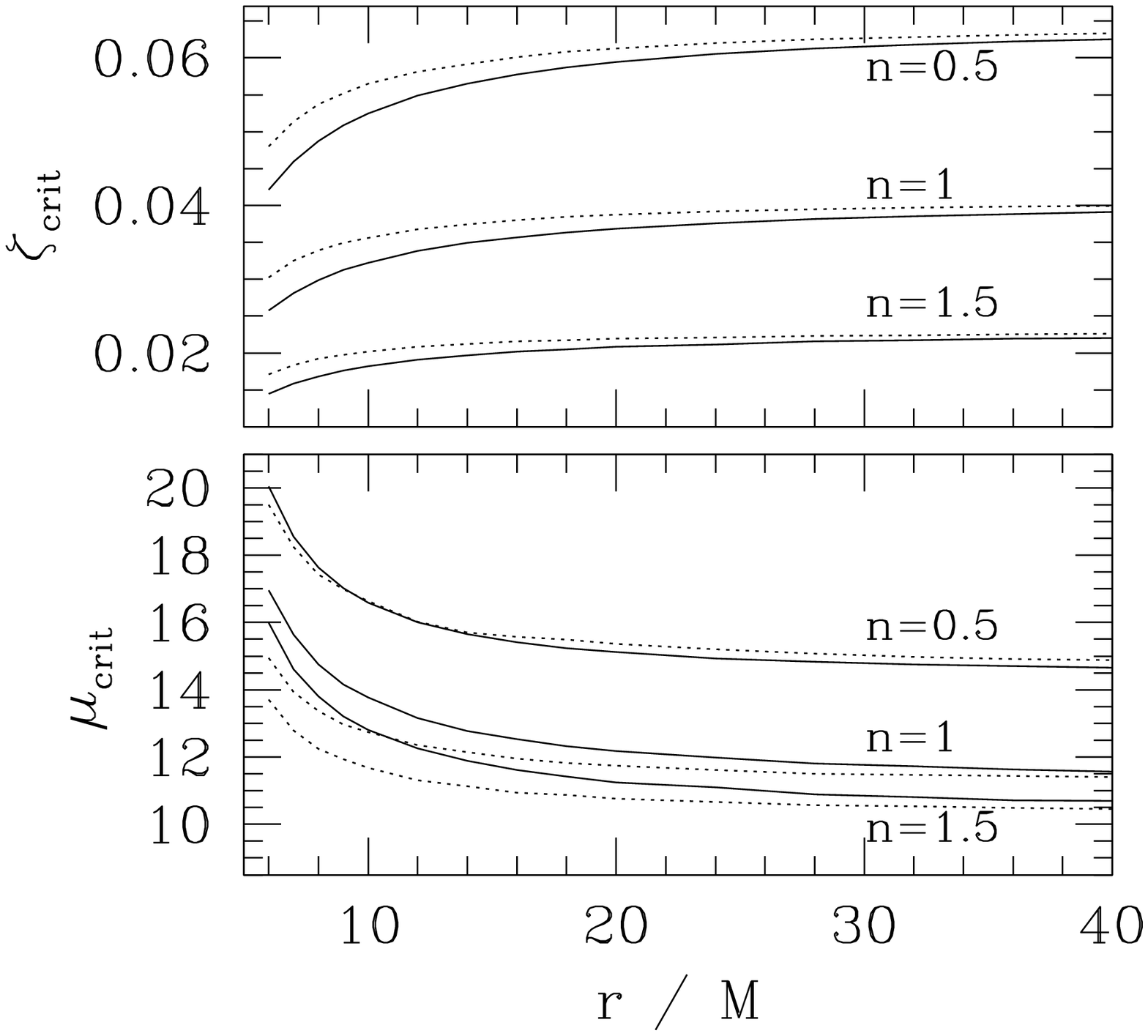}
\epsfxsize=2.85in
\leavevmode
~~(b)\epsffile{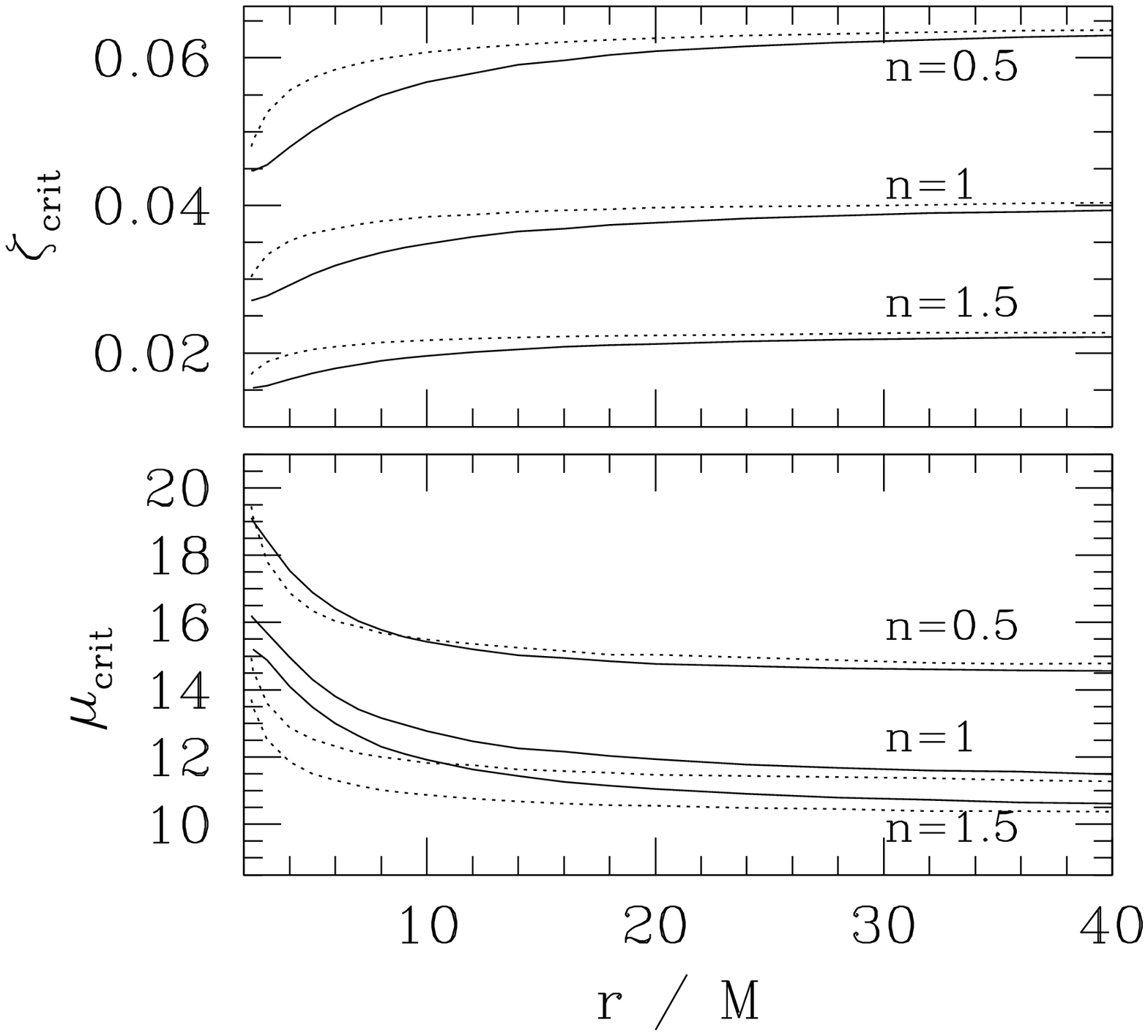}
\end{center}
\vspace*{-2mm}
\caption{$\zeta_{\rm crit}$ and $\mu_{\rm crit}$ as functions of $r$ 
(a) for $a=0$ and (b) for $a=0.9M$. For both figures, 
the results with $n=0.5$, 1, and 1.5 are shown. 
The solid and dotted curves for each value of $n$ denote the
results in the fourth-order tidal approximation 
with $R_0=M$ and in the second-order tidal approximation, respectively. 
\label{FIG6} }
\end{figure}

To clarify the dependence of the Roche limit on the equations of state,
we show 
$\zeta_{\rm crit}$ and $\mu_{\rm crit}$ for $n=0.5$, 1, and 1.5 and 
(a) for $a=0$ and (b) for $a=0.9M$ in  Fig. 6. 
The solid and dotted curves for each value of $n$ denote the results in the 
fourth-order tidal approximation
with $R_0=M$ and in the second-order tidal approximation, respectively. 
Recall that for $\zeta > \zeta_{\rm crit}$ or $\mu < \mu_{\rm crit}$,
a star is stable against tidal disruption. 
This implies that for given values of {\it mass and radius} ($R_0 \alt M$),
a star with {\it softer} equations of state 
is stronger against tidal disruption than that with the stiffer one. 
On the other hand, $\zeta_{\rm crit}$ is smaller for softer equations
of state. This implies that for given values of
{\it central density and radius} ($R_0 \alt M$), 
a star with {\it stiffer} equations of state
is stronger against tidal disruption \footnote{Figure \ref{FIG6}
suggests that these statements may not be correct for a very large value of 
$R_0 \gg M$. However, for such a large value of $R_0$, the tidal 
approximation is not applied, and thus, this point is not clear.}

From the analysis of $\zeta_{\rm crit}/\zeta_{\rm crit:2nd}-1$ and 
$\mu_{\rm crit}/\mu_{\rm crit:2nd}-1$ as functions of $M/r$,
the coefficients $C_1$ and $C_2$ are computed for $a=0$ and $0.9M$.
We find that they depend very weakly on $a$ and 
$C_1 \sim 0.7$ and $C_2 \sim 0$ for $n=0.5$, and 
$C_1 \approx 0.97$ and $C_2 \approx 0.97$ for $n=1.5$. 
For $n=0.5$, $C_1$ and $C_2$ are not determined very accurately, and
hence, the error would be $\sim \pm 0.1$. This is because
for very stiff equations of state, the density steeply decreases 
near the stellar surface and it is not easy to accurately apply
the condition for determining the Roche limit (cf. Eq. (\ref{eq187})).
Nevertheless, $C_1$ and $C_2$ for $n=0.5$ are much smaller than those
for $n=1$ and 1.5, and therefore, we can conclude that 
$C_1$ and $C_2$ are larger for softer equations of state. 
This implies that the third- and fourth-order terms in
the tidal potential play a more important role in 
softer equations of state.
For $n=0.5$, $C_2$ is approximately zero within the numerical error.
This can be explained as follows. 
The central density at the Roche limit 
in the fourth-order approximation is always larger than that 
in the second-order one. On the other hand, the volume of a star at
the Roche limit in the fourth-order approximation is smaller
than in the second-order one. These two effects seem 
to approximately cancel for $n=0.5$. 

\begin{figure}[tbh]
\vspace*{-6mm}
\begin{center}
\epsfxsize=2.85in
\leavevmode
(a)\epsffile{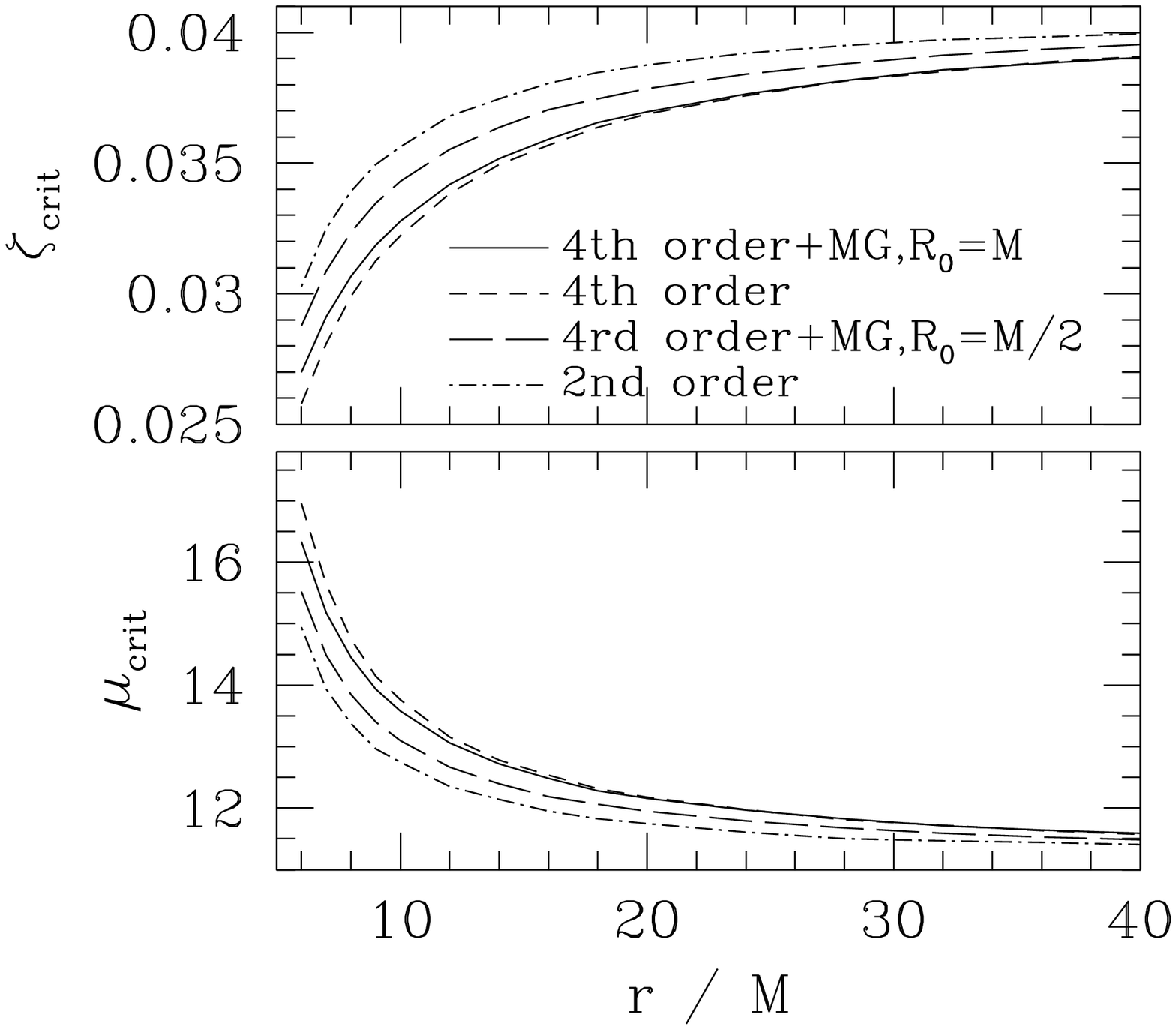}
\epsfxsize=2.85in
\leavevmode
~~(b)\epsffile{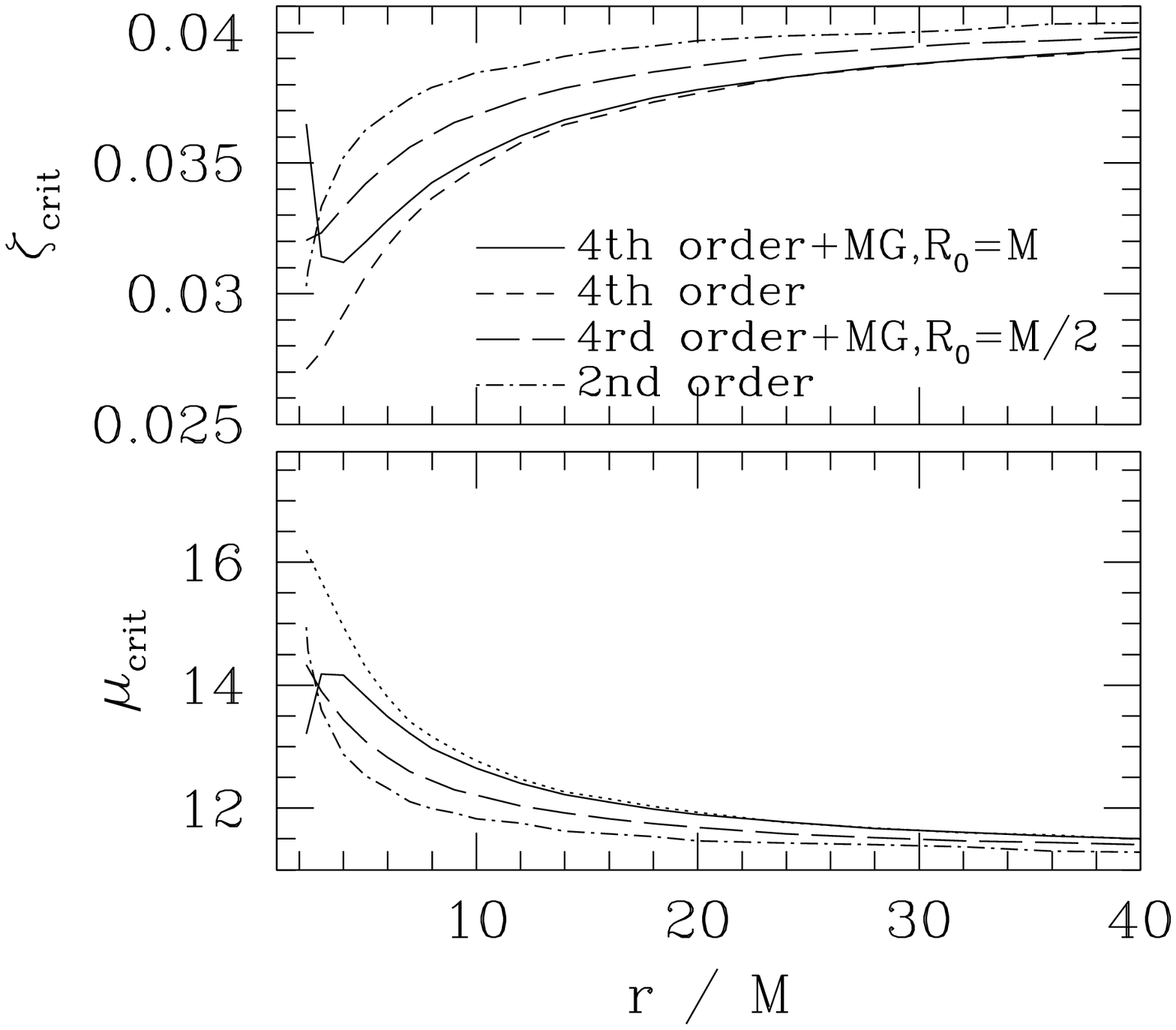}
\end{center}
\vspace*{-2mm}
\caption{$\zeta_{\rm crit}$ and $\mu_{\rm crit}$ as functions of $r$ 
in the presence of the gravitomagnetic (GM) term
(a) for $a=0$ and (b) for $a=0.9M$. 
For both figures, the results with $n=1$ and with $R_0/M=0.5$ and 1
are shown. 
The solid, dashed, long-dashed, and dotted-dashed curves denote the 
results in the fourth-order tidal approximation with the
GM term and for $R_0=M$,
in the fourth-order tidal approximation with no GM term
and for $R_0=M$, and 
in the fourth-order tidal approximation with the GM term
and for $R_0=0.5M$, and 
in the second-order tidal approximation with no GM term, 
respectively.  
\label{FIG7} }
\end{figure}

Computations were also performed including the gravitomagnetic 
terms associated with $A_i$ for $n=1$, $R_0/M=0.5, 1$, and $a/M=0$, 0.9. 
In Fig. 7, we display  
$\zeta_{\rm crit}$ and $\mu_{\rm crit}$ as functions of $r/M$.
This figure shows that the gravitomagnetic terms
decrease $\zeta_{\rm crit}$ and increase $\mu_{\rm crit}$ 
for a given value of $r$, respectively.  
This implies that the gravitomagnetic tidal force 
reduces the magnitude of the tidal force. 
This result is reasonable since the gravitomagnetic
force in the Fermi normal coordinate 
is mainly associated with the coupling 
between the star's spin and orbital motion, and the 
spin-orbit coupling force has a repulsive nature in the case
that their axes are parallel \cite{SO}. 
The order of magnitude of this term is smaller by the order of $r^{-2}$
as described in the end of Sec. IV A. 
Thus, it plays an important role only for very small orbital radii 
$r \alt 5M$ as in the fourth-order term of $\phi_{\rm tidal}$. 
The numerical results 
show that at $r=6M$, $\zeta_{\rm crit}$ and $\mu_{\rm crit}$ 
change only by $\sim 4(R_0/M)\%$ for $a=0$ and by $\sim 3(R_0/M)\%$
for $a=0.9M$. However, for rapidly spinning black holes, 
this term can significantly modify the Roche limit near the ISCOs.
Indeed, for $a=0.9M$ with $r \sim r_{\rm ISCO}\approx 2.32M$, their values
are changed by 20--$30\%$ for $R_0 \approx M$. 

\subsection{Application to neutron star-black hole binaries}

Now, we apply the results obtained in the
previous subsections to binary systems of a neutron star
and a black hole. Using the result for
$\mu_{\rm crit} =\mu_{\rm crit:2nd}(1+C_2R_0/r)=(m/M)_{\rm crit}(r/R_0)^3$,
the condition for the mass of a black hole that can tidally disrupt
a neutron star outside the ISCO is derived as 
\beqn
M \alt M_c(a) \biggl({R_0 \over 10~{\rm km}}\biggr)^{3/2}
\biggl({m \over 1.4M_{\odot}}\biggr)^{-1/2}
\biggl(1+ C_2 {R_0 \over r}\biggr)^{1/2}, \label{masscrit}
\eeqn
where $M_c(a)$ is a function of $a$.
$M_{c}$ is estimated for $r=r_{\rm ISCO}$, and 
thus, it denotes the maximum mass
of a black hole that can tidally disrupt neutron stars of
a given set of $m=1.4M_{\odot}$, $R_0=10$ km, and $r_{\rm ISCO}/M$. 
The dependence of $M_c$ on $a$ 
only results from the dependence of $r_{\rm ISCO}/M$ on $a$ since 
$\mu_{\rm crit:2nd}$ at the ISCO is independent of $a$ 
($\mu_{\rm crit:2nd}=19.4$, 
14.9, and 13.7 at the ISCOs for $n=0.5$, 1, and 1.5; cf. Table II). 

For $a=0$, $M_c(a=0) \approx 4.45M_{\odot}$, $3.90M_{\odot}$,
and $3.74M_{\odot}$ for $n=0.5$, 1, and 1.5, respectively. Note that
$M_c(a=0) \approx 4.54M_{\odot}$ and $\mu_{\rm crit:2nd}=20.1$ 
at ISCOs for $n=0$ \cite{Fishbone,shibata}, and thus, they 
are close to the values for $n=0.5$. 
$M_c(a)$ for $a\not=0$ can be computed by $M_c(a=0)(6M/r_{\rm ISCO})^{3/2}$. 
For a rapidly rotating black hole with $a=0.99M$ ($0.9M$), 
$r_{\rm ISCO}/M \approx 1.4545$ ($2.3209$), and hence, 
$M_c \approx 37.3M_{\odot}$, $32.7M_{\odot}$, and $31.3M_{\odot}$  
($18.5M_{\odot}$, $16.2M_{\odot}$, and $15.5M_{\odot}$)
for $n=0.5$, 1, and 1.5, respectively. 
Thus, for the large value of $a/M=0.9$--1, 
neutron stars of $R_0 \approx 10$ km
are tidally disrupted for a wide mass range of 
the black hole with $M \alt 15$--$40M_{\odot}$.
If the typical radius of neutron stars is $R_0=15$ km, the value of $M_c$ 
increases by a factor of $\approx 1.84$, and $M_c$ becomes 
$\approx 30$--$75M_{\odot}$ for $a/M=0.9$--1. 

In the second-order tidal approximation, $M_c$ for given
values of $R_0$ and $m$ is smaller for the larger value of $n$.  It is
interesting to point out that difference between $M_c$ for
$n=0.5$ and for $n=1$ is fairly large $\sim 14\%$ although
the differences for $n=1$ and $n=1.5$ and for $n=0$ and $n=0.5$
are only $\sim 4\%$ and $\sim 2\%$, respectively. 
This suggests that the critical mass ratio $Q_{\rm crit}$ for orbits
close to the ISCO depends sensitively on the equations of state in a stiff
region of $n=0.5$--1.

For $n=1$ and 1.5, the higher-order terms in the tidal potential increases the
critical mass for the tidal disruption. 
For parameters $M=4.5M_{\odot}$ and $R_0=10~{\rm km}$, we
obtain $R_0 \approx 1.5M$.
In this case, the critical mass with $a=0$ and $r_{\rm ISCO}=6M$
increases by a factor of $\approx 1+C_2/8$.
For $M=15M_{\odot}$ and $R_0=10$ km, $R_0 \approx 4M/9$.
In this case, the critical mass with $a=0.9M$ and $r_{\rm ISCO}\approx 2.32M$
increases by a factor of $\approx 1+C_2/10$. Thus, the critical mass
for orbits close to the ISCO 
increases by $\sim 5\%$ beyond the value of $M_c$ for $n=1$ and 1.5
due to the effect of the higher-order tidal potential, and hence, 
the dependence of the critical mass on $n$ would be weaker in reality. 
Nevertheless, the critical mass near the ISCO still 
depends fairly strongly on $n$ for $n=0.5$--1 which are plausible
polytropic indices for neutron stars \cite{ST}.  In 
contrast, $\zeta_{\rm crit}$ is smaller for softer equations of state,
and the higher-order terms in the tidal potential make this feature 
stronger. 

\begin{figure}[tbh]
\vspace*{-6mm}
\begin{center}
\epsfxsize=2.85in
\leavevmode
(a)\epsffile{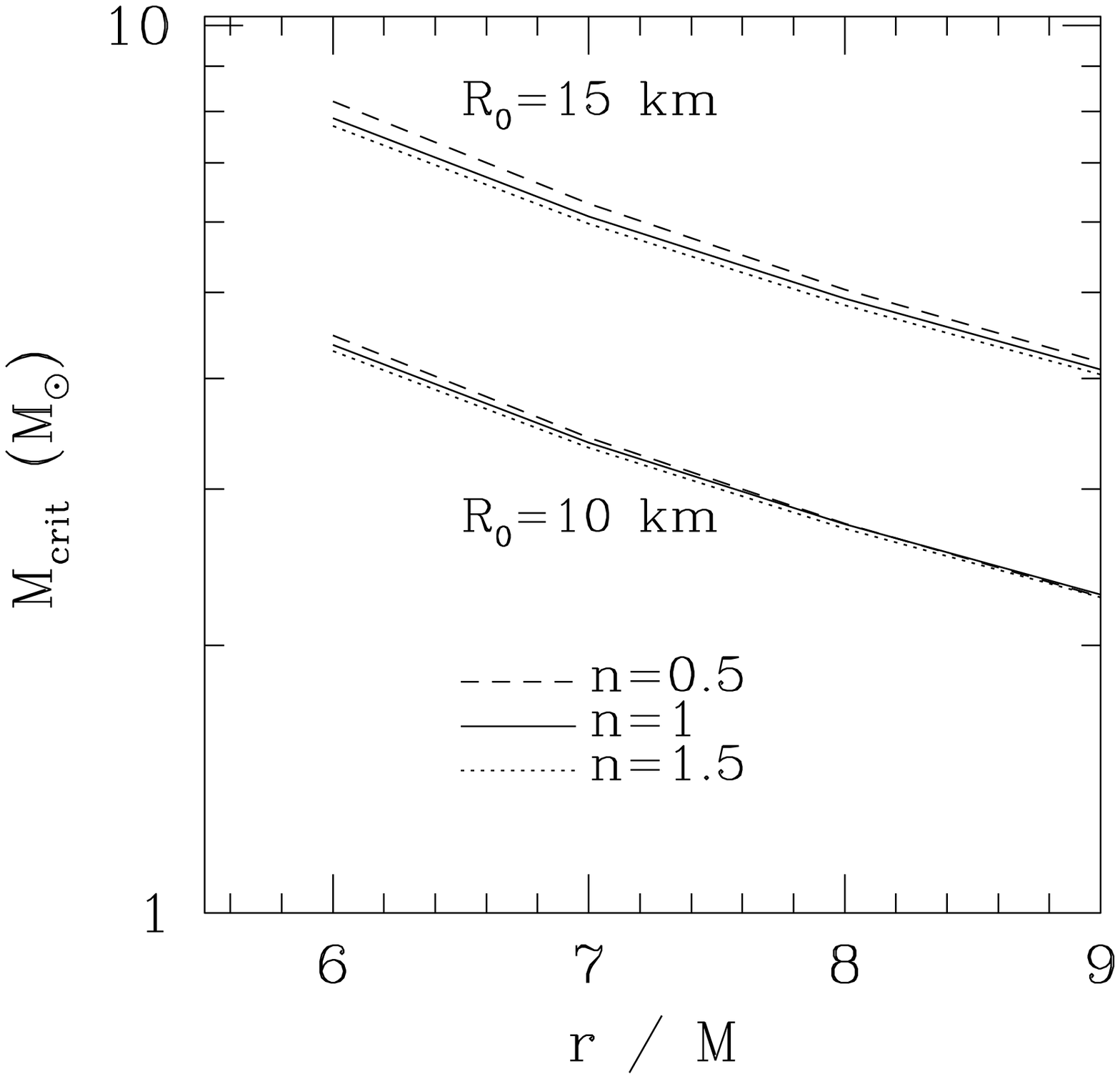}
\epsfxsize=2.85in
\leavevmode
~~(b)\epsffile{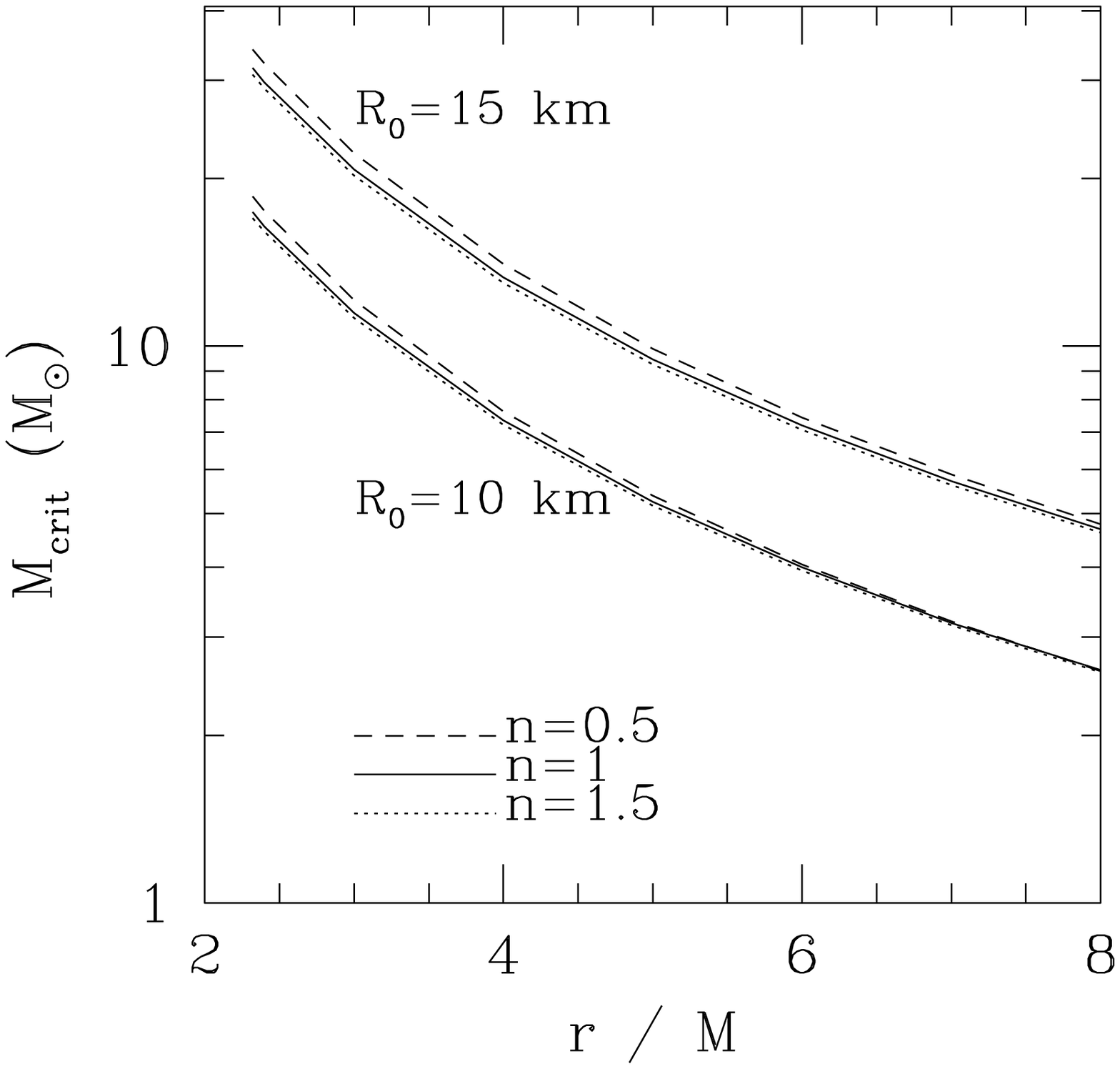}
\end{center}
\vspace*{-2mm}
\caption{Critical mass $M_{\rm crit}$ of a black hole for the tidal
disruption of a neutron star of mass $1.4M_{\odot}$
and radius $R_0=10$ km and 15 km as a function of $r/M$ 
(a) for $a=0$ and (b) for $a=0.9M$. 
The dashed, solid, and dotted curves denote the
results for $n=0.5$, 1, and 1.5, respectively. 
For $M < M_{\rm crit}$ at a given value of $r/M$,
the neutron star is unstable against tidal disruption. 
\label{FIG8} }
\end{figure}

To clarify its dependence on the orbital radius,
in Fig. \ref{FIG8}, we display the critical mass of a black hole 
as a function of $r/M$. Here, the critical mass in the
fourth-order approximation is defined by 
\beqn
M_{\rm crit}=\mu_{\rm crit}^{1/2} R_0^{3/2} m^{-1/2} 
\biggl({r \over M}\biggr)^{-3/2}. 
\eeqn
For $M < M_{\rm crit}$ at a given orbital radius $r/M$,
the neutron star is tidally disrupted.
In Fig. \ref{FIG8}, we show $M_{\rm crit}$ for $a=0$ and $0.9M$,
for $R_0=10$ and 15 km, and for $n=0.5$, 1, and 1.5. 
It is found that if we assume that the mass of black holes is
larger than $3M_{\odot}$, the tidal disruption can happen only for
close orbits with $r/M \alt 7$ for $R_0=10$ km and
for $r/M \alt 10$ for $R_0=15$ km.
Based on an observational results for black hole binaries in
our galaxy and in the LMC, typical mass of black holes is 
in the range between 6 and $8M_{\odot}$ \cite{BH}. This suggests
that the tidal disruption of neutron stars may happen frequently
only if (i) the radius of neutron stars is fairly large as $\sim 15$ km or
(ii) the typical spin parameter of black holes is fairly large as
$a/M \agt 0.6$. 

As mentioned above, the values of $M_{\rm crit}$ 
are very close each other for $n=1$ and 1.5 
irrespective of the orbital radius $r/M$, and difference in 
the value of $M_{\rm crit}$ for $n=0.5$ and 1 is remarkable 
for $r \sim r_{\rm ISCO}$. It is interesting to note that 
the value of $M_{\rm crit}$ depends very weakly on the
equations of state for orbits not very close to the ISCO 
(or in other words for $M_{\rm crit} \alt 4M_{\odot}$). 
The reason is that with increasing $r/M$, the value of $M_{\rm crit}$
steeply decreases as $M_{\rm crit}/R_0 \propto (M/r)^{3/2}$,
and hence, for a given value of $R_0$, 
the importance of the correction due to the higher-order terms associated
with the term $R_0/r=(R_0/M)(M/r)$ in determining $M_{\rm crit}$
is enhanced for $n=1$ and 1.5  \footnote{
The tidal approximation cannot be used 
for $M_{\rm crit} \alt 2M_{\odot}$ because 
$M_{\rm crit}$ and $r$ should be much larger than $m$ and $R_0$, 
respectively. Thus, one should focus only on the results for
$M \agt M_{\rm crit}$ in Fig. \ref{FIG8}. 
}. 
Nevertheless, $M_{\rm crit}$ for small values of $r/M \alt 5$ 
(or for large black hole mass with $M \agt 4M_{\odot}$) 
still depends on the equations of state. 
This suggests that by determining the tidal disruption limit of binaries 
of a neutron star and a massive and rapidly spinning
black hole in an observation, 
the equations of state of neutron stars may be constrained. 

An observation of binaries composed 
of a stellar-mass black hole and a neutron star 
will be possible in the near future using laser 
interferometric gravitational wave detectors such as LIGO \cite{KIP}. 
For the orbital separation $r \agt 10M$, the 
binary adiabatically evolves due to emission of gravitational waves.
In such inspiral phase, the masses of the black hole and the neutron
star as well as the black hole spin will be determined from
the chirp signal of gravitational waves by using matched
filtering techniques \cite{CF}. If the mass of the
black hole is smaller than the maximum value of $M_{\rm crit}$,
the tidal disruption of the neutron star
will happen near an ISCO. It is reasonable to expect that
the chirp signal of gravitational waves quickly shutdowns at the
tidal disruption. This suggests that from the signal of gravitational
waves, the radius and the orbital frequency at the
tidal disruption may be identified. With increasing 
observational samples, $M_{\rm crit}$ (or $Q_{\rm crit}$) may 
be determined. If that becomes possible,
the equations of state of neutron stars will be constrained. 

Unfortunately, the present formulation is not yet appropriate
for an accurate determination of the values of 
$M_{\rm crit}$ and $Q_{\rm crit}$ for a binary of a neutron star and
a black hole, since $m$ and $R_0$ are not much smaller than $M$ and $r$. 
Although it is reasonable to expect that 
dependence of $M_{\rm crit}$ and $Q_{\rm crit}$ on the equations of
state is qualitatively unchanged even in a more accurate computation, 
the values would be systematically modified by 10--20\% (see Appendix A). 
Thus, $M_{\rm crit}$ and $Q_{\rm crit}$ should be determined 
more accurately in terms of fully general relativistic computations
in the future. 

\section{Summary and discussion}

As an extension of previous works by Fishbone \cite{Fishbone}
and Marck \cite{Marck}, we have derived the tidal potential 
induced by a black hole up to the fourth order in $R/r$ 
in the Fermi normal coordinate system 
using the method developed by Manasse and Misner \cite{MM}. 
The new tidal potential is incorporated into the
Newtonian equations of motion for a star orbiting the black hole. 
Using the new formulation, 
we determined the tidal disruption limit (Roche limit) 
for corotating Newtonian stars in equatorial circular orbits 
around a black hole. It is found that the third- and fourth-order terms 
in the tidal potential always amplify the tidal force and 
modify the Roche limit for close orbits.
In particular, the third-order term plays a quantitatively
important role for orbits with $R/r \agt 0.1$. 
To this time, tidal problems for a star orbiting a black hole 
have been widely studied taking into account only the second-order 
terms of the black hole tidal field 
(e.g., \cite{Fishbone,CL,novikov2,shibata}). 
The present results illustrate that for close orbits with $R/r \agt 0.1$, 
the second-order approximation might not provide quantitatively accurate 
results. 

For a specific illustration of the importance of the higher-order terms in 
the tidal potential as well as the dependence of the Roche limit
on the equations of state of the star, numerical computations 
are performed for plausible parameters 
of binaries of a black hole and a neutron star. Since neutron stars 
are general relativistic objects and the mass ratio $Q$ between two 
stars are not very small, the present framework might not be 
yet appropriate for such studies. However, it is still possible to 
extract many qualitative features on the tidal disruption limit.
The following is the summary of the results: 
(i) general relativistic corrections amplify 
the tidal potential with the decrease of the orbital separation
irrespective of the order of the tidal approximation; 
(ii) as found by Fishbone \cite{Fishbone},
in the second-order tidal approximation, 
the minimum value of $\zeta_{\rm crit}$ and 
the maximum value of $\mu_{\rm crit}$ are independent of $a$. 
In the third- and fourth-order approximations, they depend on $a$; 
(iii) because of the third- and fourth-order
terms in the tidal potential, the critical mass ratio $Q_{\rm crit}$ for the 
tidal disruption is changed by $\sim 10$--15\% for close orbits near ISCOs
with $R_0 \sim M$; 
(iv) with the increase of the value of spin parameter $a$,
the magnitude of the tidal potential decreases 
for a given value of $r/M$, and as a result, $\zeta_{\rm crit}$
increases and $\mu_{\rm crit}$ decreases. 
This property is independent of the order of the tidal approximation; 
(v) for given values of central density and stellar radius ($R_0 \alt M$), 
a neutron star with stiffer equations of state 
is stronger against tidal disruption. On the other hand, 
for given values of stellar mass and radius ($R_0 \alt M$), 
a neutron star with softer equations of state 
is stronger against tidal disruption. 
The maximum value of $\mu_{\rm crit}$ depends sensitively on 
stiffness of the equations of state for $n=0.5$--1. 

As mentioned above, the present formulation is not yet appropriate 
for an accurate determination of the tidal disruption limit of a neutron star
by stellar-mass black holes, 
since the effects of mass, spin, and multipole moments of 
the neutron star to the orbital motion \cite{papapetrou,dixon} as well as 
its general relativistic self-gravity are neglected in the analysis.
In Appendix A, we estimate the order of magnitude of the error
associated with such neglected effects. 
To more accurately determine the tidal disruption limit, 
fully general relativistic computation is obviously necessary. 
We expect that our result presented in this paper could be a guideline
for the future computation.
In particular, we note that
for the case that the mass and radius of a neutron star
are much smaller than the black hole mass and orbital radius, respectively, 
the present formulation provides an 
accurate result for the Roche limit.
The fully general relativistic results computed in the future
should be compared with our numerical results to check the accuracy. 

So far, we have focused on binaries of a black hole and a neutron star.
The results with $n=1.5$ may be used for determining the Roche 
limit of a white dwarf near the ISCO of a massive black hole.
For typical values of mass $\sim 0.7M_{\odot}$ and radius $\sim 10^4$ km
of a white dwarf with $n \approx 1.5$ \cite{ST,weidemann},
Eq. (\ref{masscrit}) is written as
\beqn
M \alt 
1.67 \times 10^5 M_{\odot} \biggl({R_0 \over 10^4~{\rm km}}\biggr)^{3/2}
\biggl({m \over 0.7 M_{\odot}}\biggr)^{-1/2}
\biggl({r \over 6M}\biggr)^{-3/2}
\biggl({\mu_{\rm crit} \over 13.7}\biggr)^{1/2}. 
\label{WD}
\eeqn
In this case, $R_0/r \ll 1$, and therefore, 
the third- and fourth-order terms in the tidal potential 
are not important and can be neglected. 
Equation (\ref{WD}) implies that an intermediate-mass black hole
of $M \sim 10^3M_{\odot}$ will tidally disrupt a typical white 
dwarf in a circular orbit at $r \sim 160M$. 
To tidally disrupt a typical white dwarf in a circular orbit, 
a supermassive black hole of $M \sim 10^6 M_{\odot}$ has to be
rapidly rotating with $a \agt 0.95M$ for which $r_{\rm ISCO}\alt 2M$. 

White dwarfs orbiting a supermassive black hole in galactic centers
are likely to have highly elliptic orbits with $E \approx 1$ or
parabolic orbits with $E=1$ (e.g., \cite{HB,shibata94}). Even in this case,
the values of $M_{\rm crit}$, which are determined for the
circular orbits with $r \approx r_{\rm ISCO}$, may be used for determining 
the upper mass of a black hole, $M_{\rm max}$, for which
a white dwarf with $E \approx 1$ is tidally disrupted. 
The reason is that the tidal disruption limit will be determined
by the tidal force at the periastron radius and thus 
$M_{\rm max}$ will be determined by the radius of 
the marginally bound orbits. For highly elliptic or parabolic 
orbits in general relativity, the marginally bound orbits  
become the so-called zoom-whirl orbits \cite{GK}, which 
have a nearly circular trajectory near the periastron with 
the orbital radius \cite{ST}
\beqn
r=r_{\rm mb}=2M-a+2M\sqrt{1-a/M}. 
\eeqn
Therefore, the analysis in assumption of the circular orbits is 
approximately applicable. 
Here, the circular orbits with $r=r_{\rm mb}$ is unstable. Nevertheless, 
we can estimate the Roche limit by the analysis presented in Sec V
mathematically.

The result is that $\mu_{\rm crit:2nd} \approx 20.7$ for $n=1.5$
irrespective of value of $a$. Here, we note that for such
circular orbits with $E=1$ and $r=r_{\rm mb}$,
$B/r=1$ independent of $a$, and so is 
$\mu_{\rm crit:2nd}$. Replacing $r_{\rm ISCO}$ to $r_{\rm mb}$ and
adopting the new value of $\mu_{\rm crit:2nd}$, we can approximately estimate 
$M_{\rm max}$ of the black hole for the tidal disruption 
of a white dwarf in highly elliptic or parabolic orbits as 
\beqn
M_{\rm max} \approx 
3.77 \times 10^5 M_{\odot} \biggl({R_0 \over 10^4~{\rm km}}\biggr)^{3/2}
\biggl({m \over 0.7 M_{\odot}}\biggr)^{-1/2}
\biggl({r_{\rm mb} \over 4M}\biggr)^{-3/2}. \label{WD2}
\eeqn
Therefore, a black hole of $M \alt 3.8\times 10^{5}M_{\odot}$ and $a=0$
($M \alt 3\times 10^6M_{\odot}$ and $a \approx M$) can 
tidally disrupt white dwarfs of $m \approx 0.7M_{\odot}$ and
$R_0 \approx 10^4$ km. 

In this paper, we have only studied the tidal disruption limit for a star
in equatorial circular orbits. 
The formulation derived in this paper can be used for the 
hydrodynamic tidal problem of an ordinary star or a white dwarf 
in parabolic orbits around a supermassive black hole \cite{novikov2}. 
In Appendix B, we write the tidal potential for a parabolic orbit 
in the equatorial plane, which may be used for 
an extension of the works presented in \cite{novikov2}.
To confirm the prediction (\ref{WD2}) for the tidal disruption of
white dwarfs by a supermassive black hole, we plan to perform 
numerical simulations.

\acknowledgments

We are grateful Kip Thorne for pointing out the importance of 
higher-order terms in the tidal potentials.
His suggestion motivates this work. 
Algebraic manipulations in Secs. III, IV, and
Appendix B were performed by Mathematica. 
This work was supported by Japanese Monbukagaku-Sho 
Grant Nos. 14047207, 15037204, 15740142, and 16029202. 
YM was supported by the NASA Center for Gravitational Wave 
Astronomy at University of Texas at Brownsville (NAG5-13396), 
and grant NSF-PHY-0140326. 

\appendix

\section{Order of magnitude of the error} 

In the analysis for the tidal disruption limit of a star 
by black holes in terms of the present tidal approximation,
we neglect the effects associated with the mass, spin, and
multipole moments of the companion star to its orbital motion.
Here, we estimate the
order of magnitude of the error due to neglecting these effects.

Neglecting the mass of the star results in the error of the
orbital angular velocity by a factor of $m/M$. This 
error is included in the centrifugal force associated
with the corotating velocity field (cf. Eq. (\ref{coro})).
Since the order of magnitude of the centrifugal force is 
nearly identical with the tidal force, the values of 
$\zeta_{\rm crit}$ and $\mu_{\rm crit}$ which characterize 
the tidal disruption limit would be modified by a factor
$\alt m/2M \sim 1$--20\% for $m=1.4M_{\odot}$ and
$M=M_{\rm crit}\sim 4$--50$M_{\odot}$. 

An equation of motion for an extended body was considered in \cite{dixon} 
ignoring the self-gravity of the body. 
If we take into account the spin and quadrupole moment
of the body, we have the equations of motion 
\begin{eqnarray}
{D\over ds}p_\alpha &=& 
{1\over 2}v^\beta S^{\gamma\delta}R_{\alpha\beta\gamma\delta}
+{1\over 6}J^{\beta\gamma\delta\epsilon}
\nabla_\alpha R_{\beta\gamma\delta\epsilon} \label{eq:dixon} \,, 
\end{eqnarray}
where $p^\alpha$ is the momentum vector, 
$s$ an affine parameter (see Eq.171 in \cite{dixon}), 
$S^{\alpha\beta}$ the spin tensor, 
and $J^{\alpha\beta\gamma\delta}$ 
the quadrupole mass distribution of the extended body. 
In the post-Newtonian approximation, 
two terms in the right-hand side of Eq. (\ref{eq:dixon}) are estimated as
\beqn
\delta F^s_{\alpha}&\equiv&
{1\over 2}v^\beta S^{\gamma\delta}R_{\alpha\beta\gamma\delta}
\sim {m M R^2\Omega^2 \over r^2} \sim {m M^2 R^2 \over r^5}\,, \\
\delta F^q_{\alpha}&\equiv&
J^{\beta\gamma\delta\epsilon}
\nabla_\alpha R_{\beta\gamma\delta\epsilon} 
\sim {m M^2 R^4 \over r^7} \,. 
\eeqn
where we assume that the orbital angular velocity
$\Omega$ is equal to the spin angular velocity observed by a comoving flame, 
and that the quadrupole moment is induced by the
black hole tidal field. 
The ratio of these terms to the Newtonian force is 
\begin{eqnarray}
{\delta |F^{s}_{\alpha}| \over Mm/r^2} &\sim& {M R^2 \over r^3}\,, \\ 
{\delta |F^{q}_{\alpha}| \over Mm/r^2} &\sim& {M R^4 \over r^5} \,. 
\end{eqnarray}
Thus, these corrections modify the orbital angular velocity
by a factor of $\sim MR^2/r^3$ and $\sim MR^4/r^5$, respectively. 
They are likely to be much smaller than $m/M$ for neutron star binaries. 
We note that for the irrotational velocity field, the correction
due to the spin is approximately zero. 

With the modification of the orbital motion,
the location of the ISCO will be modified.
By this effect, $\zeta_{\rm crit:min}$ and $\mu_{\rm crit:max}$
will be also modified. 

\section{Tidal tensors for equatorial parabolic orbits}

In Sec. IV, we derive the formulation of
the tidal problem for a star in equatorial
circular orbits. Another interesting case is a parabolic encounter
of an ordinary star or a white dwarf with a supermassive black hole. 
Here, we write the tidal potential for equatorial
parabolic orbits.

For equatorial parabolic orbits with $E=1$ and $L > L_{\rm crit}$
where $L_{\rm crit}$ is a critical value which depends on $a$ and
$L_{\rm crit}=4M$ for $a=0$, 
the first integrals of the geodesic equations are written as
\beqn
u^t&=&{1 \over \Delta r^2}\Big[(r^2+a^2)r^2 - 2Mar\ell\Big],\\
u^r&=& \pm V_1,\\
u^{\varphi}&=&{-2M\ell + L r \over \Delta r}, 
\eeqn
and $u^{\theta}=0$. Here,
\beq
V_1\equiv \sqrt{\Big(1-{a\ell \over r^2}\Big)^2
-{\Delta \over r^2} V_2^2}, ~~~
V_2\equiv \sqrt{1+{\ell^2 \over r^2}},
\eeq
and $\ell \equiv L-a$. 
Then, the nonzero components of the tidal tensor
in the tilde frame are 
\beqn
&&\tilde C_{11}={M \over r^3}\biggl(1 - 3{r^2 + \ell^2 \over r^2}\biggr),\\
&&\tilde C_{22}={M \over r^3}\biggl(1 + 3{\ell^2 \over r^2}\biggr),\\
&&\tilde C_{33}={M \over r^3},\\
&&\tilde B_{131}=\tilde B_{311}=-\tilde B_{232}=-\tilde B_{322}
=-{1 \over 2}\tilde B_{113}={1 \over 2} \tilde B_{223}
=-{3M \ell V_2 \over 2r^4},\\
&&\tilde C_{111}={3M \over r^4 V_2}\Bigl[2+{3\ell^2-4a\ell \over r^2}
-{5a\ell^3 \over r^4}\Bigr],\\
&&\tilde C_{131}=\tilde C_{311}=\tilde C_{113}
=\pm {M \ell V_1 \over r^5 V_2}\Bigl[4+{5\ell^2 \over r^2}\Bigr],\\
&&\tilde C_{122}=\tilde C_{212}=\tilde C_{221}
=-{M \over r^4 V_2}\Bigl[3+{7\ell^2 -11 a\ell \over r^2}
-{15a\ell^3 \over r^4}\Bigr],\\
&&\tilde C_{133}=\tilde C_{313}=\tilde C_{331}
=-{M \over r^4 V_2}\Bigl[3+{2\ell^2 - a\ell\over r^2} \Bigr],\\
&&\tilde C_{322}=\tilde C_{232}=\tilde C_{223}
=\mp {M \ell V_1 \over r^5 V_2}\Bigl[1+{5\ell^2 \over r^2}\Bigr],\\
&&\tilde C_{333}=\mp {3M \ell V_1 \over r^5 V_2},\\
&& \tilde C_{1111}={3M \over r^5 V_2^2}
\Bigl[-8+{2M \over r}-{12\ell^2 -32 a\ell +4a^2 \over r^2}
+{5M\ell^2\over r^3}+{40a\ell^3 -35a^2 \ell^2 \over r^4}
+{3 \ell^4 M \over r^5}-{35 a^2\ell^4 \over r^6}\Bigr],\\
&& \tilde C_{1113}=\tilde C_{1131}=\tilde C_{1311}=\tilde C_{3111}
=\mp {3M \over 2r^{6} V_2^2}\Bigl[16\ell - 4a
+{20\ell^3 - 35a\ell^2 \over r^2}-{35 a\ell^4 \over r^4}\Bigr] V_1 ,\\
&& \tilde C_{1122}=\tilde C_{1212}=\tilde C_{1221}
=\tilde C_{2112}=\tilde C_{2121}=\tilde C_{2211}\nonumber \\
&&=-{M \over 2 r^5 V_2^2}
\Bigl[-24 + {11 M \over r} -{55 \ell^2 - 148a\ell +31a^2 \over r^2}
+ {28M \ell^2 \over r^3}
-{5\ell^4 - 200 a\ell^3 + 215a^2\ell^2 \over r^4}+{17M \ell^4 \over r^5}
-{210 a^2\ell^4 \over r^6}\Bigr],\\
&& \tilde C_{1133}=\tilde C_{1313}=\tilde C_{1331}
=\tilde C_{3113}=\tilde C_{3131}=\tilde C_{3311} \nonumber \\
&&=-{M \over 2 r^5}
\Bigl[-24 + {11 M \over r} - {5 \ell^2 - 20a\ell + 5a^2 \over r^2}
+ {76 M \ell^2 \over r^3}
-{35 \ell^4 +70a\ell^3+35a^2\ell^2\over r^4}
+{75 M \ell^4 \over r^5}\Bigr],\\
&& \tilde C_{1223}=\tilde C_{1232}=\tilde C_{1322}
=\tilde C_{2123}=\tilde C_{2132}=\tilde C_{3122}
=\tilde C_{2213}=\tilde C_{2312}=\tilde C_{3212}
=\tilde C_{2231}=\tilde C_{2321}=\tilde C_{3221} \nonumber \\
&&=\pm {M  \over 2r^{6} V_2^2}
\Bigl[14\ell - a +{40\ell^3 -80 a \ell^2 \over r^2}
-{105 a\ell^4 \over r^4}\Bigr] V_1,\\
&& \tilde C_{1333}=\tilde C_{3133}=\tilde C_{3313}=\tilde C_{3331}
=\pm {3 M \over 2 r^{6} V_2^2}\Bigl[14\ell - a
+{10\ell^3-5a\ell^2 \over r^2}\Bigr]V_1,\\
&& \tilde C_{2222}={3M \over r^5}\Bigl[
-3 + {2M \over r} -{5\ell^2 -20a\ell +5a^2\over r^2}
+{10M \ell^2 \over r^3}-{35 a^2\ell^2 \over r^4}\Bigr],\\
&& \tilde C_{2233}=\tilde C_{2323}=\tilde C_{2332}
=\tilde C_{3223}=\tilde C_{3232}=\tilde C_{3322} \nonumber \\
&&=-{M \over 2 r^5 V_2^2}
\Bigl[-6 + {4 M \over r} -{12 \ell^2 -18a\ell +6a^2 \over r^2}
+ {25 M \ell^2 \over r^3}
-{15 \ell^4 +15 a^2\ell^2 \over r^4}+{96 M \ell^4 \over r^5} \nonumber \\
&& \hskip 3cm-{35 \ell^6 +70a\ell^5+35a^2\ell^4 \over r^6}
+{75 M \ell^6 \over r^7}\Bigr],\\
&& \tilde C_{3333}={3M \over r^5 V_2^2}\Bigl[
-3 + {2M \over r} -{4 \ell^2 + 2a\ell + a^2 \over r^2}
+ {13M \ell^2 \over r^3}
-{5 \ell^4 +10a\ell^3 +5a^2\ell^2 \over r^4}
+{11 M \ell^4 \over r^5}\Bigr]. 
\eeqn
Components in the Fermi normal coordinates are calculated
by operating rotational matrices associated with $\Psi$
as in Sec. IV. Here, evolution
equation of the rotation angle is written as
\beq
{d \Psi \over d \tau}={L \over r^2 + \ell^2}. 
\eeq
If computations are performed in the tilde frame,
$\Psi$ is not necessary for computing the tidal potential.
Instead, $d \Psi/d \tau$ and $d^2\Psi/d \tau^2$
appear in computing the inertial forces in the
equations of motion.


\begin{thebibliography}{99}

\bibitem{ayal} S. Ayal, M. Livio, and T. Piran, Astrophys. J. {\bf 545}, 772
(2000). 

\bibitem{obs} T. Bogdanovic, M. Eracleous, S. Mahadevan, S. Sigurdsson, and
P. Laguna, Astrophys. J. {\bf 610}, 707 (2004). 

\bibitem{fryer} C. L. Fryer, S. E. Woosley, M. Herant, and
M. B. Davies, Astrophys. J. {\bf 520}, 650 (1999). 

\bibitem{cutler} C. Cutler et al., Phys. Rev. Lett. {\bf 70}, 2984 (1993). 

\bibitem{micheal} M. Vallisneri, Phys. Rev. Lett. {\bf 84}, 3519 (2000).  

\bibitem{EK} C. R. Evans and C. S. Kochanek, Astrophys. J. {\bf 346},
L13 (1989).

\bibitem{UE} K. Uryu and Y. Eriguchi, Mon. Not. R. astr. Soc. {\bf 303},
329 (1999). 

\bibitem{Max} H.-T. Janka, T. Eberl, M. Ruffert, and
C. L. Fryer, Astrophys. J. {\bf 527}, L39 (1999). 

\bibitem{CL} B. Carter and J.-P. Luminet, 
Astron. Astrophys. {\bf 121}, 97 (1983); 
Mon. Not. R. astr. Soc. {\bf 212}, 23 (1985). 

\bibitem{Novikov0} A. Khokhlov, I. D. Novikov, and C. J. Pethick, 
Astrophys. J. {\bf 418}, 163 (1993): {\em ibid} {\bf 418}, 181 (1993). 

\bibitem{MLB} J. A. Marck, A. Lioure, and S. Bonazzola, Astron.
Astrophys. {\bf 306}, 666 (1996). 

\bibitem{Laguna} P. Laguna, W. A. Miller, W. H. Zurek, and M. B. Davies,
Astrophys. J. {\bf 410}, L83 (1993). 

\bibitem{Fishbone} L. G. Fishbone, Astrophys. J. {\bf 185}, 43 (1973). 

\bibitem{mash} B. Mashhoon, Astrophys. J. {\bf 197}, 705 (1975). 

\bibitem{Marck} J.-A. Marck, Proc. R. Soc. Lond. A {\bf 385}, 431 (1983).

\bibitem{novikov2} 
V. P. Frolov, A. M. Khokhlov, I. D. Novikov, and 
C. J. Pethick, Astrophys. J. {\bf 432}, 680 (1994): 
P. Diener, V. P. Frolov, A. M. Khokhlov, I. D. Novikov, and 
C. J. Pethick, Astrophys. J. {\bf 479}, 164 (1997).

\bibitem{shibata} M. Shibata, Prog. Theor. Phys. {\bf 96}, 917 (1996). 

\bibitem{IS} M. Ishii and M. Shibata, Prog. Theor. Phys. {\bf 112},
399 (2004). 

\bibitem{MM} F. K. Manasse and C. W. Misner, J. Math. Phys. {\bf 4}, 735
(1963).

\bibitem{BPT} J. M. Bardeen, W. H. Press, and S. A. Teukolsky,
Astrophys. J. {\bf 178}, 347 (1972).

\bibitem{miller} M. Miller, gr-qc/0106017. 

\bibitem{bss} T. W. Baumgarte, M. L. Skoge, and S. L. Shapiro, gr-qc/0405077. 

\bibitem{Wald} R. M. Wald, {\em General relativity} (The University
of Chicago Press, Chicago and London, 1984). 

\bibitem{chandra} S. Chandrasekhar, {\em The Mathematical
Theory of Black Holes}, Oxford Science Publications
(Oxford University Press, New York, 1983), chapter 6. 

\bibitem{carter} B. Carter, Phys. Rev. {\bf 174}, 1559 (1968). 

\bibitem{Membrane} K. S. Thorne, R. H. Price, and D. A. Macdonald,
{\em The Membrane Paradigm} (Yale University Press, New Haven and London,
1986). 

\bibitem{MTW} 
C. W. Misner, K. S. Thorne, and J. A. Wheeler, {\em Gravitation}
(Freeman, San Francisco, 1973). 

\bibitem{shibapn} M. Shibata, Prog. Theor. Phys. {\bf 96}, 317 (1996): 
Phys. Rev. D {\bf 55}, 6019 (1997). 

\bibitem{ST} S. L. Shapiro and S. A. Teukolsky, {\em Black
Holes, White Dwarfs, and Neutron Stars}, Wiley Interscience
(New York, 1983), chapter 12. 

\bibitem{SO} E.g., L. E. Kidder, C. M. Will, and A. G. Wiseman,
Phys. Rev. D {\bf 47}, R4183 (1993) and references therein. 

\bibitem{KIP}
C. Cutler and K. S. Thorne, in {\em Proceedings of the 16th
International Conference on General Relativity and Gravitation},
edited by N. T. Bishop and S. D. Maharaj (World Scientific, 2002), p.72. 

\bibitem{CF} C. Cutler and E. E. Flanagan, Phys. Rev. D {\bf 49}, 2658 
(1994).

\bibitem{BH} J. E. McClintock and R. A. Remillard, in {\em
Compact Stellar X-ray Sources}, edited by W. H. G. Lewin and van
der Klis (Cambridge University Press, Cambridge, 2005) to be published
(astro-ph/0306213).

\bibitem{papapetrou} A. Papapetrou, Proc. R. Soc. London Ser. {\bf A209},
248 (1951). 

\bibitem{dixon} W. G. Dixon, in {\em Isolated Gravitating Systems
in General Relativity}, edited by J. Ehlers (North-Holland,
Amsterdam, 1979), pp. 156. 

\bibitem{weidemann} V. Weidemann, Annu. Rev. Astron. Astrophys.
{\bf 28}, 103 (1990). 

\bibitem{HB} D. Hils and P. L. Bender, Astrophys. J. Lett. {\bf 445}, 
L7 (1995). 

\bibitem{shibata94} M. Shibata, Phys. Rev. D {\bf 50}, 6297 (1994). 

\bibitem{GK} K. Glampedakis and D. Kennefick, Phys. Rev. D {\bf 66},
044002 (2002). 

\end{thebibliography}
\end{document}